\documentclass[aps,prd,preprint,superscriptaddress,showpacs,floatfix]{revtex4}
\usepackage{amssymb}
\usepackage{amsmath}
\RequirePackage{xspace}

\usepackage{relsize}
\def\babar{\mbox{\slshape B\kern-0.1em{\smaller A}\kern-0.1em
    B\kern-0.1em{\smaller A\kern-0.2em R}}}

\def\mil    {\ensuremath{\!\cdot\!10^{-3}}}

\def\stot  {\ensuremath{\sigma_{TOT} }}
\def\sine  {\ensuremath{\sigma_{INE } }}
\def\sela {\ensuremath{\sigma_{ELA} }}
\def\sqel  {\ensuremath{\sigma_{QEL} }}

\def\mup        {\ensuremath{\mu^+}\xspace}
\def\mun        {\ensuremath{\mu^-}\xspace} % muon negative (\mum is taken)

\def\nub        {\ensuremath{\overline{\nu}}\xspace}

\def\nub        {\ensuremath{\overline{\nu}}\xspace}

\def\nue        {\ensuremath{\nu_e}\xspace}
\def\nueb       {\ensuremath{\nub_e}\xspace}

\def\num        {\ensuremath{\nu_\mu}\xspace}
\def\numb       {\ensuremath{\nub_\mu}\xspace}

\def\piz   {\ensuremath{\pi^0}\xspace}

\def\pip   {\ensuremath{\pi^+}\xspace}
\def\pim   {\ensuremath{\pi^-}\xspace}

\def\pipm  {\ensuremath{\pi^\pm}\xspace}

\def\Kbar  {\kern 0.2em\overline{\kern -0.2em K}{}\xspace}

\def\Kz    {\ensuremath{K^0}\xspace}
\def\Kzb   {\ensuremath{\Kbar^0}\xspace}
\def\KzKzb {\ensuremath{\Kz \kern -0.16em \Kzb}\xspace}
\def\Kp    {\ensuremath{K^+}\xspace}
\def\Km    {\ensuremath{K^-}\xspace}
\def\Kpm   {\ensuremath{K^\pm}\xspace}

\def\KpKm  {\ensuremath{\Kp \kern -0.16em \Km}\xspace}
\def\KS    {\ensuremath{K^0_{\scriptscriptstyle S}}\xspace} 
\def\KL    {\ensuremath{K^0_{\scriptscriptstyle L}}\xspace}

\def\Dbar    {\kern 0.2em\overline{\kern -0.2em D}{}\xspace}

\def\Dz      {\ensuremath{D^0}\xspace}
\def\Dzb     {\ensuremath{\Dbar^0}\xspace}
\def\DzDzb   {\ensuremath{\Dz {\kern -0.16em \Dzb}}\xspace}
\def\Dp      {\ensuremath{D^+}\xspace}
\def\Dm      {\ensuremath{D^-}\xspace}

\def\DpDm    {\ensuremath{\Dp {\kern -0.16em \Dm}}\xspace}

\def\Bbar    {\kern 0.18em\overline{\kern -0.18em B}{}\xspace}

\def\Bz      {\ensuremath{B^0}\xspace}
\def\Bzb     {\ensuremath{\Bbar^0}\xspace}
\def\BzBzb   {\ensuremath{\Bz {\kern -0.16em \Bzb}}\xspace}
\def\Bu      {\ensuremath{B^+}\xspace}
\def\Bub     {\ensuremath{B^-}\xspace}

\def\BpBm    {\ensuremath{\Bu {\kern -0.16em \Bub}}\xspace}

\mathchardef\Upsilon="7107
\def\Y#1S{\ensuremath{\Upsilon{(#1S)}}\xspace}% no space before {...}!

\mathchardef\Deltares="7101
\mathchardef\Xi="7104
\mathchardef\Lambda="7103
\mathchardef\Sigma="7106
\mathchardef\Omega="710A

\def\Deltabar{\kern 0.25em\overline{\kern -0.25em \Deltares}{}\xspace}
\def\Lbar{\kern 0.2em\overline{\kern -0.2em\Lambda\kern 0.05em}\kern-0.05em{}\xspace}
\def\Sigbar{\kern 0.2em\overline{\kern -0.2em \Sigma}{}\xspace}
\def\Xibar{\kern 0.2em\overline{\kern -0.2em \Xi}{}\xspace}
\def\Obar{\kern 0.2em\overline{\kern -0.2em \Omega}{}\xspace}
\def\Nbar{\kern 0.2em\overline{\kern -0.2em N}{}\xspace}
\def\Xb{\kern 0.2em\overline{\kern -0.2em X}{}\xspace}

\newcommand{\tev}{\ensuremath{\mathrm{\,Te\kern -0.1em V}}\xspace}
\newcommand{\gev}{\ensuremath{\mathrm{\,Ge\kern -0.1em V}}\xspace}
\newcommand{\mev}{\ensuremath{\mathrm{\,Me\kern -0.1em V}}\xspace}
\newcommand{\kev}{\ensuremath{\mathrm{\,ke\kern -0.1em V}}\xspace}
\newcommand{\ev}{\ensuremath{\mathrm{\,e\kern -0.1em V}}\xspace}
\newcommand{\gevc}{\ensuremath{{\mathrm{\,Ge\kern -0.1em V\!/}c}}\xspace}
\newcommand{\mevc}{\ensuremath{{\mathrm{\,Me\kern -0.1em V\!/}c}}\xspace}
\newcommand{\gevcc}{\ensuremath{{\mathrm{\,Ge\kern -0.1em V\!/}c^2}}\xspace}
\newcommand{\mevcc}{\ensuremath{{\mathrm{\,Me\kern -0.1em V\!/}c^2}}\xspace}
\def\cm   {\ensuremath{\rm \,cm}\xspace}

\def\mus  {\ensuremath{\rm \,\mus}\xspace}

\def\mus        {\ensuremath{\,\mu{\rm s}}\xspace}    %% microsecond
\def\mrad{\ensuremath{\rm \,mrad}\xspace}               %% milliradian

\def\to                 {\ensuremath{\rightarrow}\xspace}

\def\pep2{PEP-II}

\def\gsim{{~\raise.15em\hbox{$>$}\kern-.85em
          \lower.35em\hbox{$\sim$}~}\xspace}
\def\lsim{{~\raise.15em\hbox{$<$}\kern-.85em
          \lower.35em\hbox{$\sim$}~}\xspace}

\xspace

\def\jetset74   {\mbox{\tt Jetset \hspace{-0.5em}7.\hspace{-0.2em}4}\xspace}

\usepackage{graphicx}
\begin{document}

%\preprint{}

\title{The Neutrino Flux prediction at MiniBooNE}



%\author{}
%\thanks{}
%\affiliation{}
%\input{authors_new}
\newcommand{\bama}{University of Alabama; Tuscaloosa, AL 35487}
\newcommand{\bucknell}{Bucknell University; Lewisburg, PA 17837}
\newcommand{\cinci}{University of Cincinnati; Cincinnati, OH 45221}
\newcommand{\colorado}{University of Colorado; Boulder, CO 80309}
\newcommand{\columbia}{Columbia University; New York, NY 10027}
\newcommand{\embry}{Embry-Riddle Aeronautical University; Prescott, AZ 86301}
\newcommand{\fnal}{Fermi National Accelerator Laboratory; Batavia, IL 60510}
\newcommand{\florida}{University of Florida; Gainesville, FL 32611}
\newcommand{\indiana}{Indiana University; Bloomington, IN 47405}
\newcommand{\lanl}{Los Alamos National Laboratory; Los Alamos, NM 87545}
\newcommand{\lsu}{Louisiana State University; Baton Rouge, LA 70803}
\newcommand{\umich}{University of Michigan; Ann Arbor, MI 48109}
\newcommand{\princeton}{Princeton University; Princeton, NJ 08544}
\newcommand{\marys}{Saint Mary's University of Minnesota; Winona, MN 55987}
\newcommand{\vtech}{Virginia Polytechnic Institute \& State University; Blacksburg, VA 24061}
\newcommand{\yale}{Yale University; New Haven, CT 06520}
\newcommand{\beijing}{Institute of High Energy Physics; Beijing 100049, China}
\newcommand{\hope}{Hope College; Holland, MI 49423}
\newcommand{\iit}{Illinois Institute of Technology; Chicago, IL 60616}
\newcommand{\imsa}{Illinois Mathematics and Science Academy; Aurora IL 60506}
\newcommand{\massit}{Massachusetts Institute of Technology; Cambridge, MA 02139}
\newcommand{\caltech}{California Institute of Technology; Pasadena, CA 91125}
\newcommand{\bu}{Boston University; Boston, MA 02215}
\newcommand{\valencia}{IFIC, Universidad de Valencia and CSIC; 46071 Valencia, Spain}
\newcommand{\ubc}{University of British Columbia, Vancouver, BC V6T 1Z1, Canada}
\newcommand{\imperial}{Imperial College; London SW7 2AZ, United Kingdom}

\affiliation{\bama}
\affiliation{\bucknell}
\affiliation{\cinci}
\affiliation{\colorado}
\affiliation{\columbia}
\affiliation{\embry}
\affiliation{\fnal}
\affiliation{\florida}
\affiliation{\indiana}
\affiliation{\lanl}
\affiliation{\lsu}
\affiliation{\umich}
\affiliation{\princeton}
\affiliation{\marys}
\affiliation{\vtech}
\affiliation{\yale}

\author{A.~A. Aguilar-Arevalo}\affiliation{\columbia}  
\author{C.~E.~Anderson}\affiliation{\yale} 	     
\author{A.~O.~Bazarko}\affiliation{\princeton} 	     
\author{S.~J.~Brice}\affiliation{\fnal} 	     
\author{B.~C.~Brown}\affiliation{\fnal} 	     
\author{L.~Bugel}\affiliation{\columbia} 		     
\author{J.~Cao}\altaffiliation{Present Address: \beijing}\affiliation{\umich}
\author{L.~Coney}\altaffiliation{Present Address: \hope}\affiliation{\columbia}
\author{J.~M.~Conrad}\affiliation{\columbia} 	     
\author{D.~C.~Cox}\affiliation{\indiana} 	     
\author{A.~Curioni}\affiliation{\yale} 	     
\author{Z.~Djurcic}\affiliation{\columbia} 	     
\author{D.~A.~Finley}\affiliation{\fnal} 	     
\author{B.~T.~Fleming}\affiliation{\fnal}\affiliation{\yale}
\author{R.~Ford}\affiliation{\fnal} 		     
\author{F.~G.~Garcia}\affiliation{\fnal} 	     
\author{G.~T.~Garvey}\affiliation{\lanl} 	     
\author{C.~Green}\affiliation{\lanl}\affiliation{\fnal}
\author{J.~A.~Green}\affiliation{\indiana}\affiliation{\lanl}
\author{T.~L.~Hart}\altaffiliation{Present Address: \iit}\affiliation{\colorado}
\author{E.~Hawker}\altaffiliation{Present Address: \imsa}\affiliation{\lanl}\affiliation{\cinci}
\author{R.~Imlay}\affiliation{\lsu} 	     
\author{R.~A. ~Johnson}\affiliation{\cinci}	     
\author{G.~Karagiorgi}\affiliation{\columbia} 	     
\author{P.~Kasper}\affiliation{\fnal} 	     
\author{T.~Katori}\affiliation{\indiana} 	     
\author{T.~Kobilarcik}\affiliation{\fnal} 	     
\author{I.~Kourbanis}\affiliation{\fnal} 	     
\author{S.~Koutsoliotas}\affiliation{\bucknell} 	     
\author{E.~M.~Laird}\affiliation{\princeton} 	     
\author{S.~K.~Linden}\affiliation{\yale} 	     
\author{J.~M.~Link}\affiliation{\columbia}\affiliation{\vtech}
\author{Y.~Liu}\affiliation{\umich} 		     
\author{Y.~Liu}\affiliation{\bama} 		     
\author{W.~C.~Louis}\affiliation{\lanl} 	     
\author{K.~B.~M.~Mahn}\affiliation{\columbia} 	     
\author{W.~Marsh}\affiliation{\fnal} 		     
\author{P.~S.~Martin}\affiliation{\fnal} 	     
\author{G.~McGregor}\affiliation{\lanl} 	     
\author{W.~Metcalf}\affiliation{\lsu} 	     
\author{P.~D.~Meyers}\affiliation{\princeton} 	     
\author{F.~Mills}\affiliation{\fnal} 		     
\author{G.~B.~Mills}\affiliation{\lanl} 	     
\author{J.~Monroe}\altaffiliation{Present Address: \massit}\affiliation{\columbia}
\author{C.~D.~Moore}\affiliation{\fnal} 	     
\author{R.~H.~Nelson}\affiliation{\colorado} 	     
\author{V.~T.~Nguyen}\affiliation{\columbia} 	     
\author{P.~Nienaber}\affiliation{\marys} 	     
\author{J.~A.~Nowak}\affiliation{\lsu} 	     
\author{S.~Ouedraogo}\affiliation{\lsu} 	     
\author{R.~B.~Patterson}\altaffiliation{Present Address: \caltech}\affiliation{\princeton}
\author{D.~Perevalov}\affiliation{\bama} 	     
\author{C.~C.~Polly}\affiliation{\indiana} 	     
\author{E.~Prebys}\affiliation{\fnal} 	     
\author{J.~L.~Raaf}\altaffiliation{Present Address: \bu}\affiliation{\cinci}
\author{H.~Ray}\affiliation{\lanl}\affiliation{\florida}
\author{B.~P.~Roe}\affiliation{\umich} 	     
\author{A.~D.~Russell}\affiliation{\fnal} 	     
\author{V.~Sandberg}\affiliation{\lanl} 	     
\author{R.~Schirato}\affiliation{\lanl} 	     
\author{D.~Schmitz}\affiliation{\columbia} 	     
\author{M.~H.~Shaevitz}\affiliation{\columbia} 	     
\author{F.~C.~Shoemaker}\affiliation{\princeton} 	     
\author{D.~Smith}\affiliation{\embry}
\author{M.~Soderberg}\affiliation{\yale} 	     
\author{M.~Sorel}\altaffiliation{Present Address: \valencia}\affiliation{\columbia}
\author{P.~Spentzouris}\affiliation{\fnal} 	     
\author{I.~Stancu}\affiliation{\bama} 	     
\author{R.~J.~Stefanski}\affiliation{\fnal} 	     
\author{M.~Sung}\affiliation{\lsu} 		     
\author{H.~A.~Tanaka}\altaffiliation{Present Address: \ubc}\affiliation{\princeton}
\author{R.~Tayloe}\affiliation{\indiana} 	     
\author{M.~Tzanov}\affiliation{\colorado} 	     
\author{R.~Van~de~Water}\affiliation{\lanl} 	     
\author{M.~O.~Wascko}\altaffiliation{Present Address: \imperial}\affiliation{\lsu}
\author{D.~H.~White}\affiliation{\lanl} 	     
\author{M.~J.~Wilking}\affiliation{\colorado} 	     
\author{H.~J.~Yang}\affiliation{\umich} 	     
\author{G.~P.~Zeller}\affiliation{\columbia}\affiliation{\lanl}
\author{E.~D.~Zimmerman}\affiliation{\colorado}

\collaboration{MiniBooNE Collaboration}\noaffiliation
%\pacs{14.60.Lm, 29.27.-a, 14.60.P, 13.15.+g }

%\collaboration{MiniBooNE}

\date{\today}

\begin{abstract}
% Text of abstract
The Booster Neutrino Experiment (MiniBooNE) searches for $\num\to\nue$ oscillations
using the $\mathcal{O}(1 \gev)$ neutrino beam produced by the Booster synchrotron
at the Fermi National Accelerator Laboratory (FNAL).
The Booster delivers protons with 8 $\gev$ kinetic energy ($8.89\gevc$ momentum)  to a 
beryllium target, producing neutrinos from the decay of secondary particles in the beam line.
We describe the Monte Carlo simulation methods used to estimate the flux of neutrinos
from the beamline incident on the MiniBooNE detector for both polarities of the focusing
horn. The simulation uses the Geant4
framework for propagating particles, accounting for electromagnetic processes
and hadronic interactions in the beamline materials, as well as the decay of particles. The absolute double differential cross sections of pion and kaon production in the simulation have
 been tuned to match external measurements, as have the hadronic cross sections for nucleons 
and pions. The statistical precision of the flux predictions is enhanced through reweighting 
and resampling techniques. Systematic errors in the flux estimation have been determined by 
varying parameters within their uncertainties, accounting for correlations where appropriate. 
\end{abstract}


\maketitle

%\input{introduction}
\section{Introduction}
The Booster Neutrino Experiment (MiniBooNE) at Fermilab searches for 
the oscillation of muon neutrinos $(\num)$ to electron neutrinos $(\nue)$ 
indicated by the LSND experiment\cite{lsnd}\cite{oscillation}. The 
neutrino beam is produced by
the Booster Neutrino Beamline (BNB), where protons with 8 $\gev$ kinetic 
energy ($8.89\gevc$ momentum) are extracted
from the Fermilab Booster synchrotron and directed towards a beryllium
target. Secondary mesons produced by the interaction of the protons
in the target decay to produce a neutrino beam with an average energy of 
$\sim 800\mev$. 
Neutrino interactions are observed in a 6.1-meter-radius, 
spherical detector situated 541 meters from the center of the 
target. The detector is composed of 800 metric tons of mineral oil that serves as both the target for neutrino interactions and the medium in which charged 
particles produced in neutrino interactions radiate Cherenkov and 
scintillation 
photons. The photons are detected on an array of 1520 photomultipliers, and the resulting
spatial and temporal patterns of light are used to identify and reconstruct the interactions.  
Understanding both the spectrum and 
composition of the neutrino beam is critical to the neutrino oscillation 
analysis, which searches for an excess of $\nue$ events over a background of 
both non-oscillation sources of $\nue$ in the beamline and misidentified
$\num$ interactions. 

%In this paper, we describe the Geant4-based Monte Carlo simulation
%of the BNB used to predict the neutrino flux at MiniBooNE. The simulation
%tracks the $8\gev$ primary protons from the Booster into the target, accounting
%for interactions that produce additional particles and recording particle
%decays relevant for neutrino production. We start by describing the elements
%of the BNB in Section \ref{sec:bnb} and discuss the Geant4-based software
%infrastructure for tracking particles and their interactions, as well
%as statistical enhancement techniques used to reduce statistical errors
%in the flux prediction in Section \ref{sec:g4}. Sections \ref{sec:secondary}
%and \ref{sec:hadronic} discuss measurements of hadron production and 
%scattering cross sections, whose results are incorporated directly into the 
%simulation. In Section \ref{sec:neutrinoflux}, we present the neutrino flux 
%predictions which result from the simulation, followed by Section \ref{sec:systematic},
%where we discuss the effect of various systematic uncertaintainties. 

The neutrino oscillation analysis at MiniBooNE utilizes
observed data to constrain the uncertainties in the expected event rates
of certain key processes. These constraints typically reduce the uncertainties 
that would result from a direct estimation using solely
 the predicted neutrino flux and cross sections. Within the $\num\to\nue$ 
oscillation analysis,  the  observed rate of $\num$ 
charged current quasi-elastic events and neutral current $\piz$ events are used
directly in the estimation of the number and spectrum of background and 
expected neutrino oscillation events. The Monte Carlo-based flux prediction 
described here is one input to this process. In this article, we focus on the 
flux prediction itself, which is based on external data, without regard to the
observed neutrino rates at MiniBooNE. Predictions for both polarities of the 
focusing horn are presented. Detailed comparisons of the observed and 
predicted neutrino event rates, as well as descriptions of the use of the 
predicted neutrino fluxes in various analyses, are described in publications 
relating to the analysis of the neutrino data itself, {\em e.g.},  
References \cite{oscillation,ccqe,jocelyn,ryan}.

\section{The Booster Neutrino Beamline}
\label{sec:bnb}
\begin{figure}[hp]
\begin{center}
\includegraphics[width=200mm,angle=90]{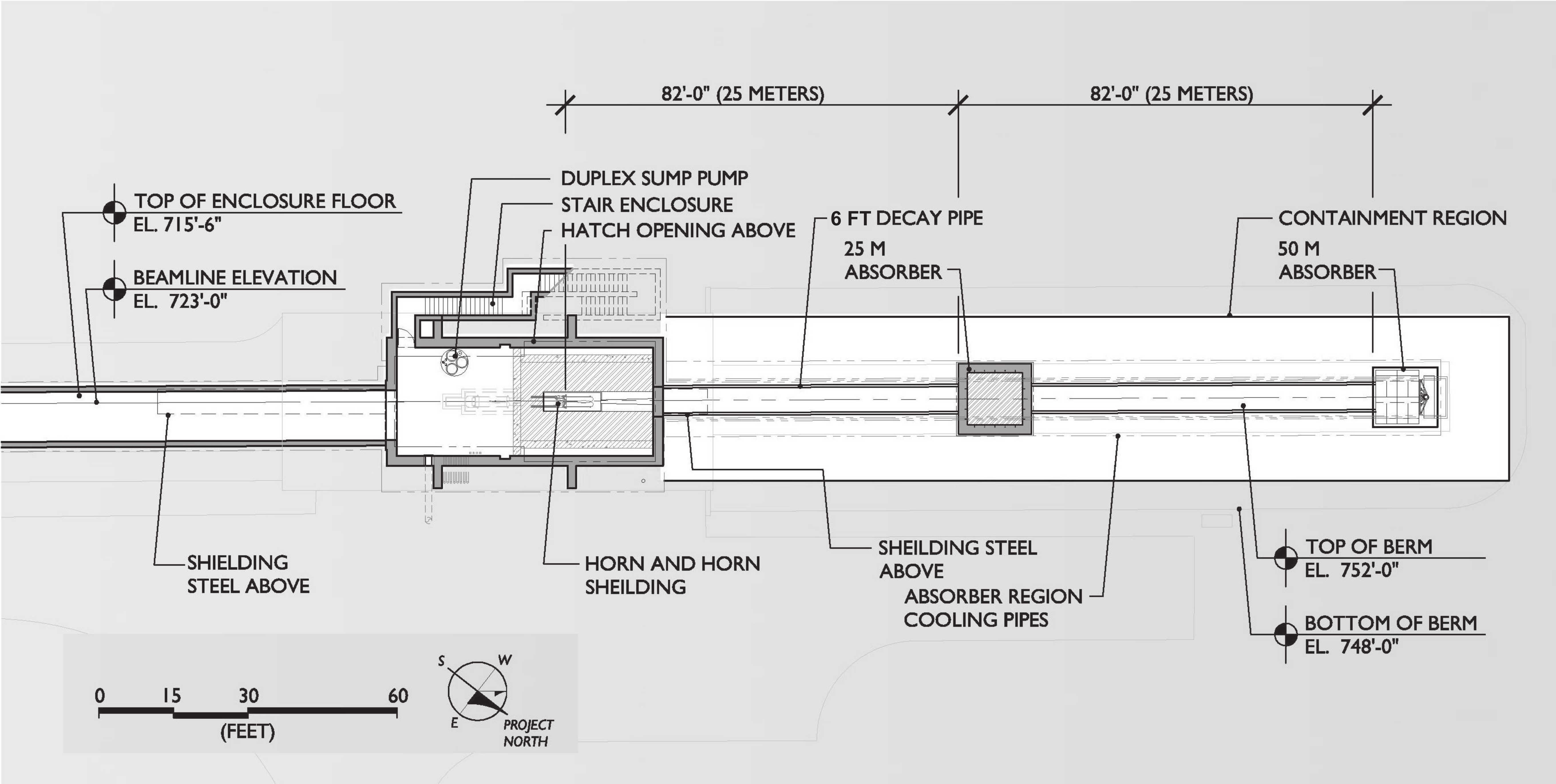}
\caption{\label{fig:bnb}Overall layout of the BNB. The primary
proton beam, extracted from the Booster, enters the target hall from the left.
Upon exiting the target hall, particles encounter a 50-meter-long decay region, terminating in the beam
stop on the  right.}
\end{center}
\end{figure}

The Booster Neutrino Beamline (BNB) produces neutrinos using
$8.89\gevc$ momentum  protons from the Booster synchrotron that
are incident on a beryllium target. The layout of the BNB is shown in 
Figure \ref{fig:bnb}.
The target is embedded within a pulsed electromagnet (the ``horn'')  that produces a 
toroidal magnetic
field to focus positive secondary particles and defocus negative
secondary particles emerging from proton-beryllium interactions. These secondary
particles enter a 50-meter-long decay region, resulting in a neutrino-enhanced beam. 
The polarity of the horn can be reversed to focus negative secondary particles and 
produce an antineutrino-enhanced beam.
The axis of the beam, defined by
the center of the decay pipe, is displaced vertically from the center of the
MiniBooNE detector by 1.9 meters.

The particle production is dominated by pions, though there is significant
kaon production as well.  Neutrinos also result from the
decay of muons whose primary source is the decay of pions produced in the target. This results
in a significant flux of $\nue/\nueb$ in neutrino/antineutrino mode, while the corresponding flux of 
$\numb/\num$ is small compared to the
$\num/\numb$ which result directly from the decay of the pions.
A beam stop at the end of the decay region absorbs particles apart from the neutrinos. 
The predicted composition of the neutrino beam is described in Section \ref{sec:neutrinoflux}. 
A detailed description of the BNB can be found in the 
Technical Design Report for the BNB \cite{bnbtdr}. This section describes the beamline geometry and
components relevant for the neutrino flux prediction.

\subsection{FNAL Booster and Proton Extraction}

\begin{figure}
\begin{center}
\includegraphics[width=80mm,angle=270]{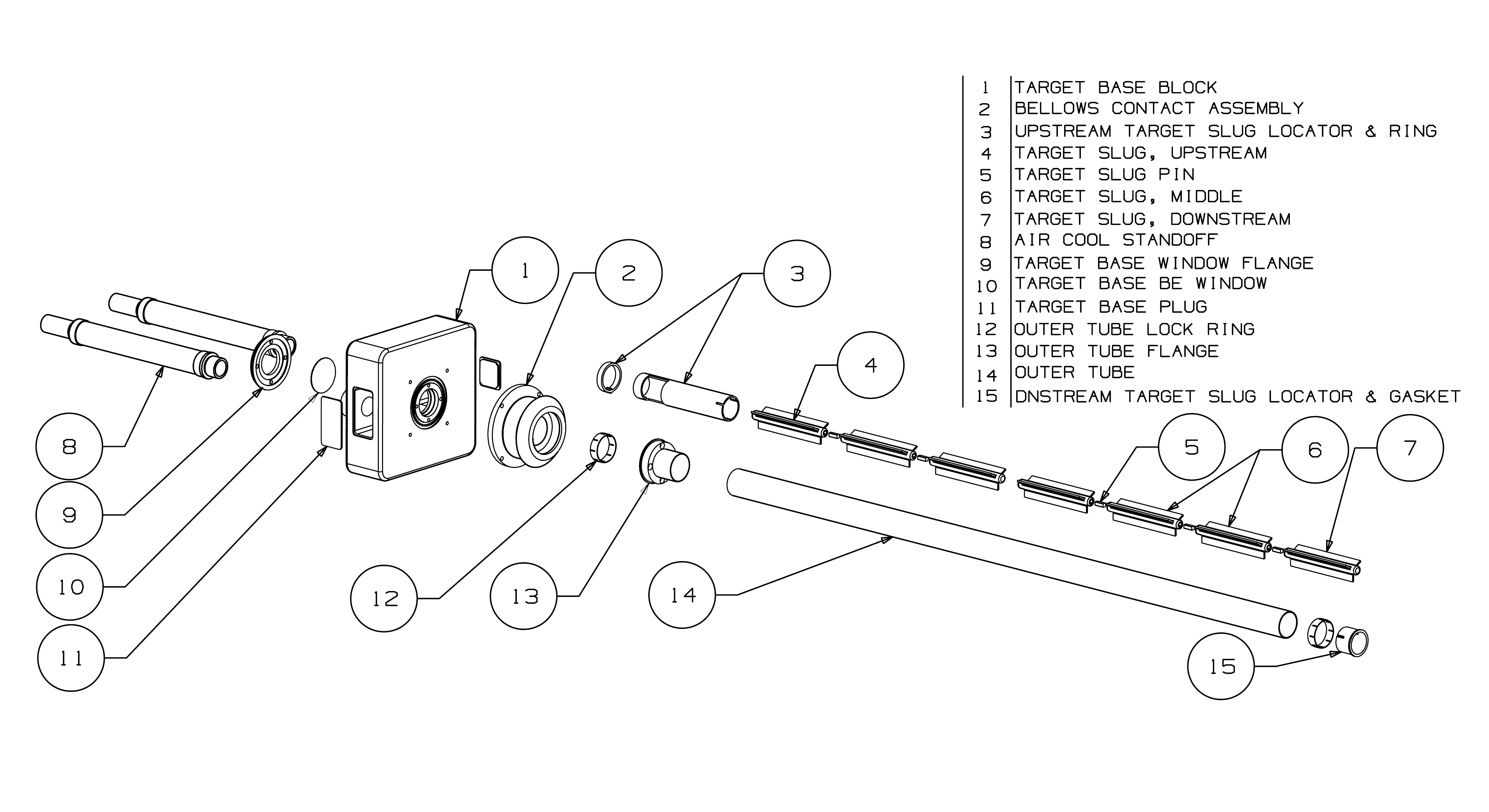}\\
\includegraphics[width=80mm,angle=270]{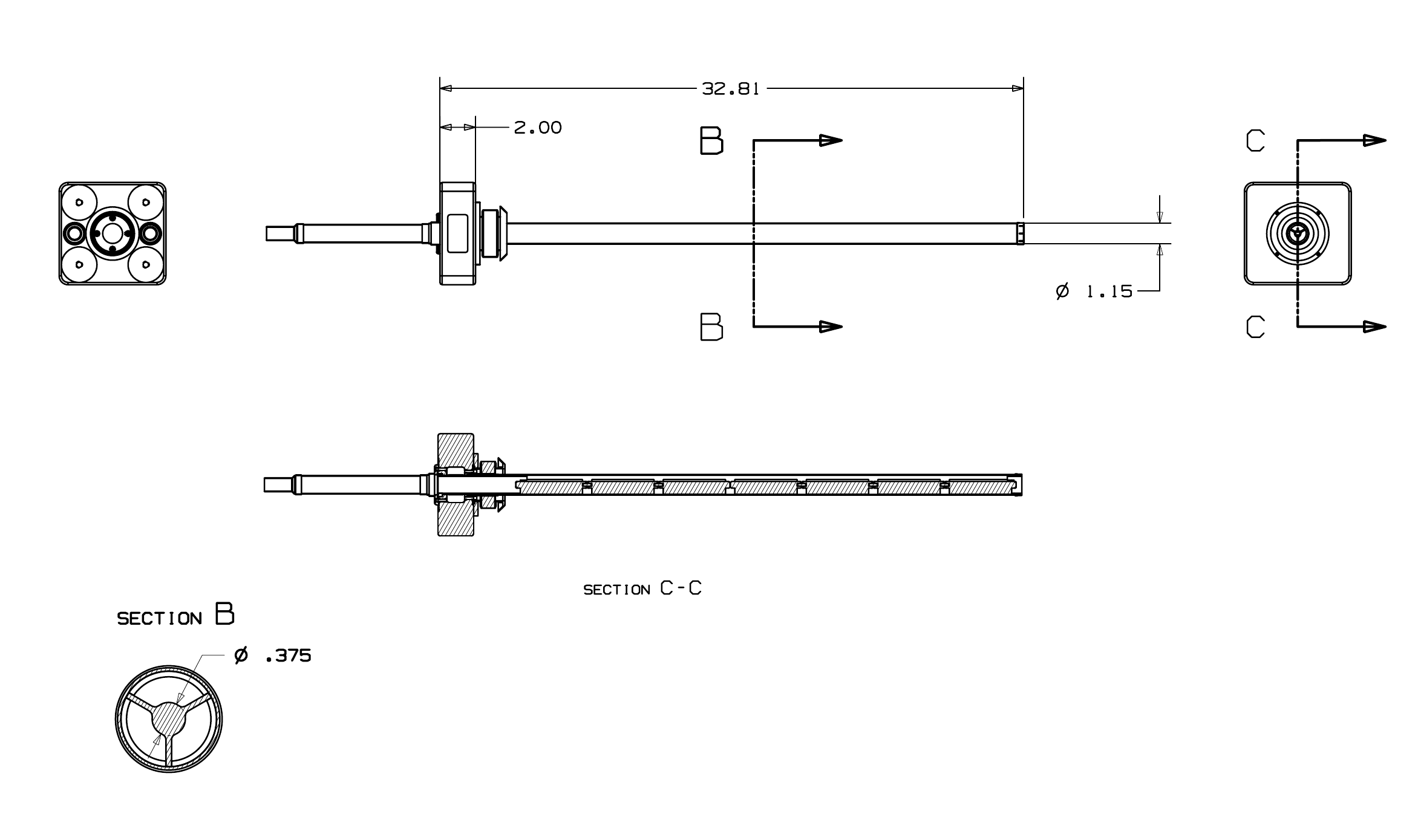}
\caption{\label{fig:target}The MiniBooNE target assembly. The top shows an exploded
view of the components, while the bottom shows the assembled configuration. The proton beam enters from the left in both figures, striking the finned beryllium slugs. Dimensions are in
inches.}
\end{center}
\end{figure}

The FNAL Booster is a 474-meter-circumference synchrotron operating at 15Hz. 
Protons from the Fermilab LINAC are injected at 400 MeV and accelerated to 8 
GeV kinetic energy ($8.89\gevc$ momentum).  The Booster has a harmonic 
number of 84, of which 81 buckets are filled.  The beam is extracted into the 
BNB using a fast-rising kicker that extracts all of the particles in a single 
turn.  The resulting structure is a series of 81 bunches of protons each 
$\sim 2$ ns wide and 19 ns apart.

Upon leaving the Booster, the proton beam is transported through a lattice of 
focusing and defocusing quadrupole (FODO) and dipole magnets.  A switch magnet 
steers the beam to the Main Injector or to the BNB. The BNB is also a FODO that
terminates with a triplet that focuses the beam on the target.  The design and 
measured optics of BNB are in agreement \cite{tom,moorepac2003}.

The maximum allowable average repetition rate for delivery of protons to the 
BNB is 5 Hz (with a maximum of 11 pulses in a row at 15 Hz) and 
$5\times 10^{12}$ protons-per-pulse. The 5 Hz limit is set by the design of 
the horn (described below) and its power supply. As of January 2008, over 
$10^{21}$ protons have been delivered to the BNB, with a typical up time of 
greater than $90\%$ during normal operations. The neutrino oscillation
results in neutrino mode were published using $5.6\times10^{20}$ 
protons-on-target delivered prior to 2006, when the polarity of the horn was 
reversed to collect antineutrino mode data \cite{nubarloi}.

\begin{figure}
\begin{center}
\includegraphics[height=48 mm]{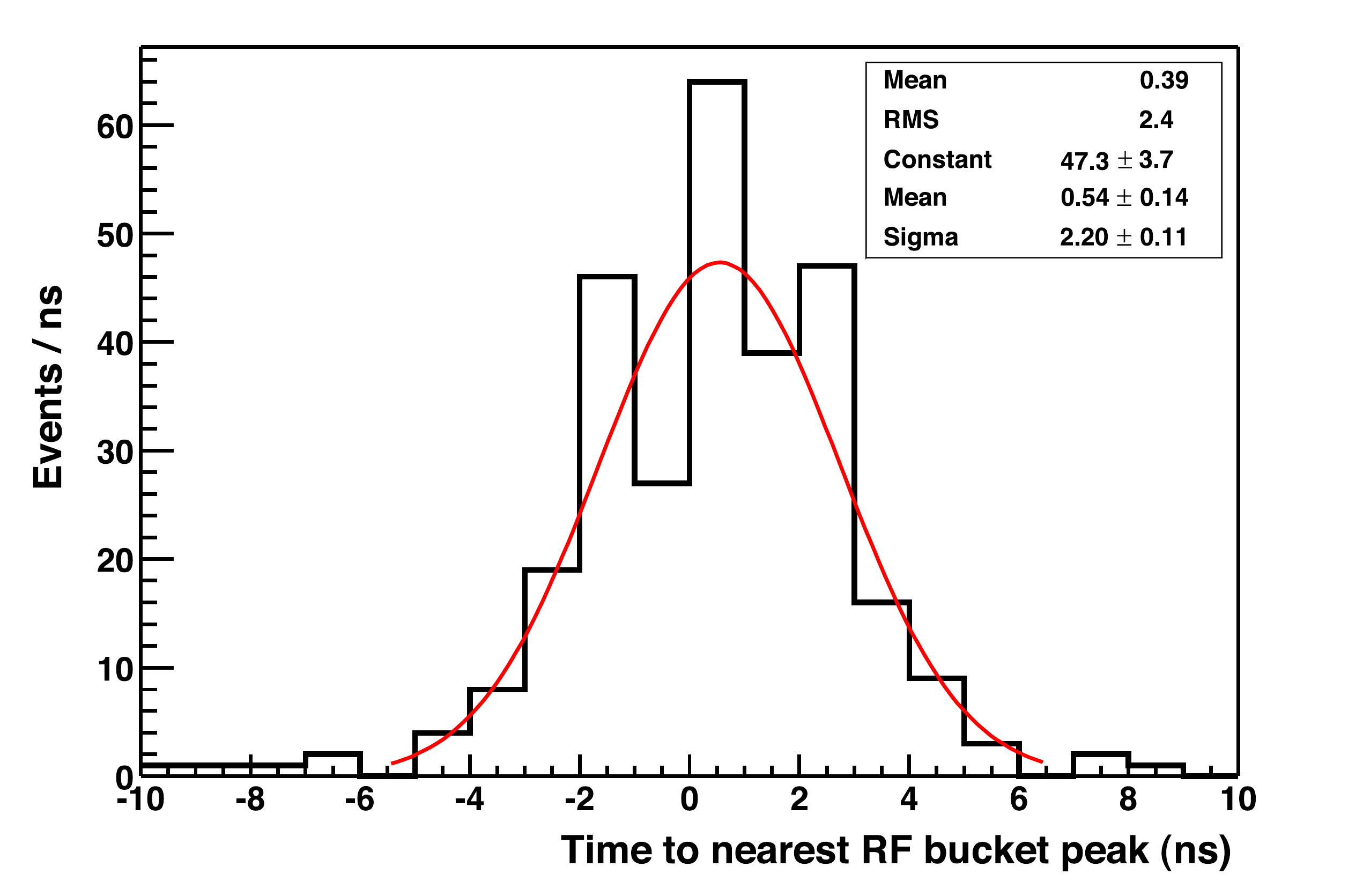}
\includegraphics[width=48 mm,angle=90]{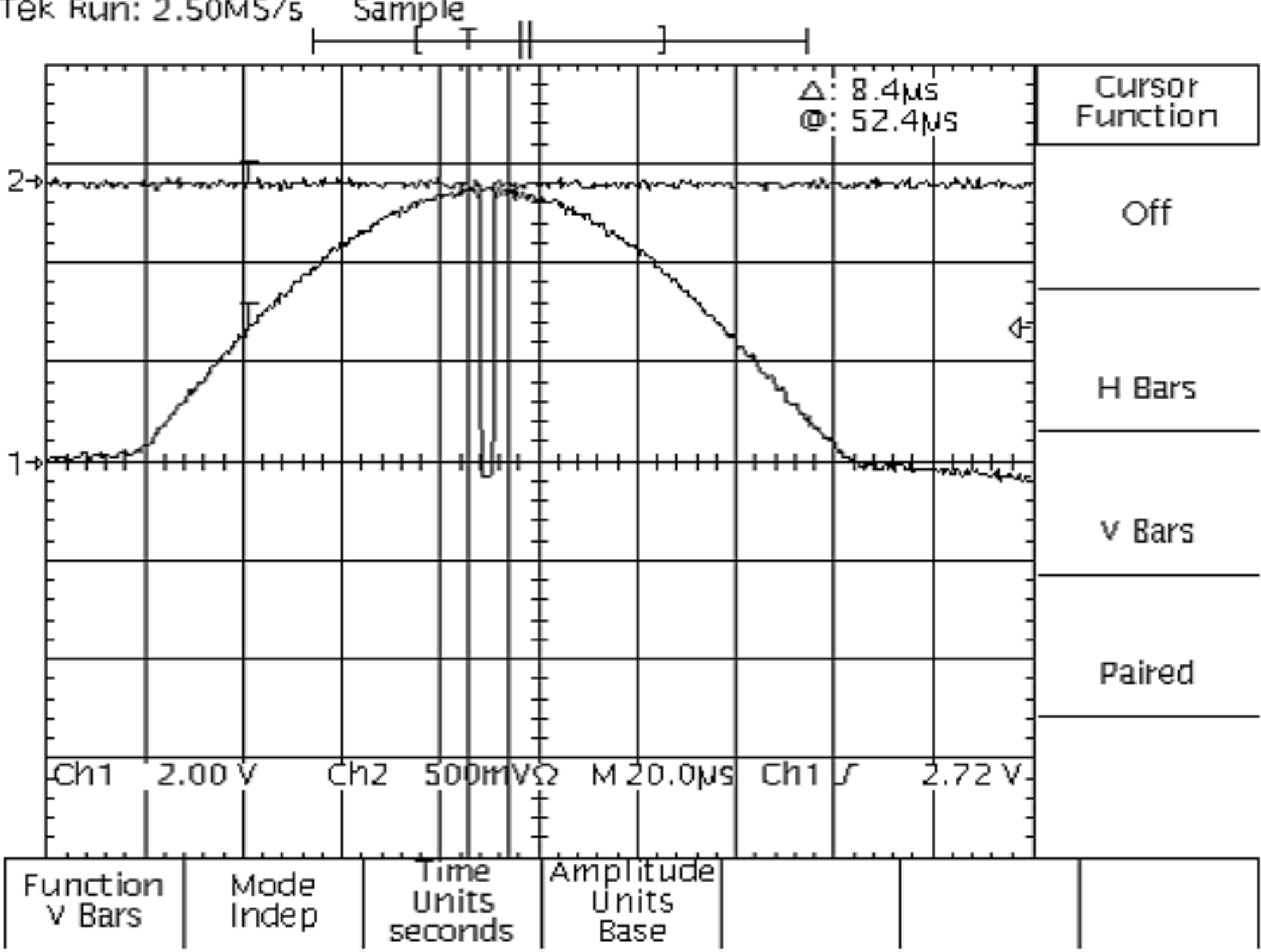}
\caption{\label{fig:rwmhorn} 
Left: Neutrino event times relative to the nearest RF bucket (measured by the RWM) corrected for expected time-of-flight.
Right: An oscilloscope trace showing the coincidence
of the beam delivery with the horn pulse. The top trace (labeled ``2'' on the 
left) is a discriminated signal from the resistive 
wall monitor (RWM), indicating the arrival of the beam pulse. The bottom 
trace (labeled ``1'' on the left) 
is the horn pulse. The horizontal divisions are 20 $\mus$ each. 
}
\end{center}
\end{figure}

\subsection{Target}
The target consists of seven identical cylindrical slugs of beryllium arranged
to produce a cylinder 71.1 cm long and 0.51 cm in radius. The target is
contained within a beryllium sleeve 0.9 cm thick with an inner radius of
1.37 cm. Each target slug is supported within
the sleeve by three ``fins'' (also beryllium) which extend radially out from 
the target to the sleeve. The volume of air within the sleeve is circulated to 
provide cooling for the target when the beamline is in operation. The target
and associated assembly are shown in Figure \ref{fig:target}, where the
top figure shows an ``exploded'' view of the various components (with the
downstream end of the target on the right), and the bottom shows
the components in assembled form. The choice of beryllium as the target
material was motivated by residual radioactivity issues in the event that
the target assembly needed to be replaced, as well as energy loss 
considerations that allow the air-cooling system to be sufficient.

Upstream of the target, the primary proton beam is monitored using four 
systems: two toroids measuring its intensity (protons-per-pulse), beam 
position monitors (BPM) and a multi-wire chamber determining the beam width 
and position, and a resistive wall monitor (RWM) measuring both the time and
 intensity of the beam spills. The vacuum of the beam pipe extends to about 5
feet upstream of the target, minimizing upstream proton interactions.

The toroids are continuously calibrated at 5 Hz  with their absolute 
calibrations verified twice a year. The calibrations have shown minimal 
deviation $(<0.5\%)$. The proton flux measured in the two toroids agree to 
within $2\%$, compatible with the expected systematic uncertainties. The BPMs 
are split-plate devices that measure the difference of charge induced on two 
plates.  By measuring the change in beam position at several locations without 
intervening optics, the BPMs are found to be accurate to 0.1 mm (standard deviation). 
% !! I presented this but never wrote it up !!  
The multi-wire is a wire chamber 
with 48  horizontal and 48 vertical wires and 0.5 mm pitch.  The profile of the beam
is measured using the secondary emission induced by the beam on the wires.  
%The present multi-wire has a short in the readout cable, 
%rendering it unusable (should I just drop this?).

The RWM is located upstream of the target to monitor the time and intensity
of the proton pulses prior to striking the target. 
While the data from the RWM did not directly
enter the $\num\to\nue$ analysis, it allowed many useful cross checks, such as those  shown
in Figure \ref{fig:rwmhorn}. The left figure shows a comparison of the production times of 
neutrinos observed in the MiniBooNE detector estimated based on the vertex and time
reconstructed by the detector and subtracting the time-of-flight. 
This time is then compared to the nearest bucket as measured by the RWM. The distribution 
indicates that neutrino events can be matched not only to pulses from the Booster, but to 
a specific bucket within the pulse. The tails of the distribution result from
the resolution of the vertex reconstruction of the neutrino event in the 
detector, which is needed to determine the time of the event and correct for 
the time-of-flight.
The right plot demonstrates the synchronization of the horn pulse 
(described in Section \ref{sec:horn}) with the delivery of the beam as measured by the RWM.

\begin{figure}
\begin{center}
\includegraphics[width=110mm]{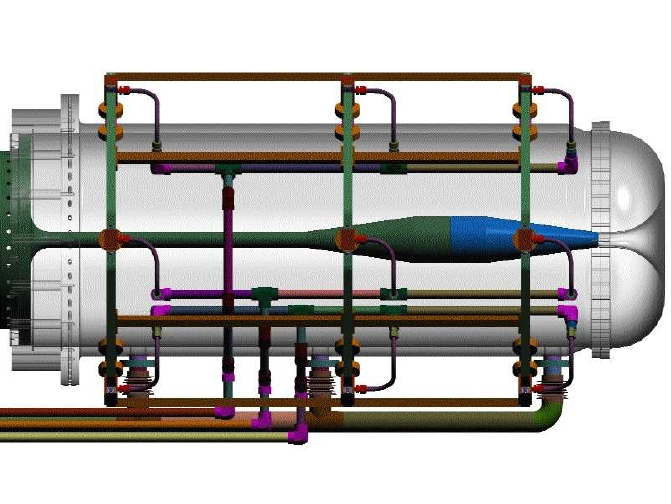}
\caption{\label{fig:horn}The MiniBooNE pulsed horn system. The outer conductor (gray)
is transparent to show the inner conductor components running along the center (dark green and blue).
The target assembly is inserted into the inner conductor from the left side. In neutrino-focusing mode, the (positive) current flows from left-to-right along the inner conductor, returning along the outer conductor. The plumbing associated with the water cooling system is also shown. }
\vskip 0.2 cm
\includegraphics[width=80mm]{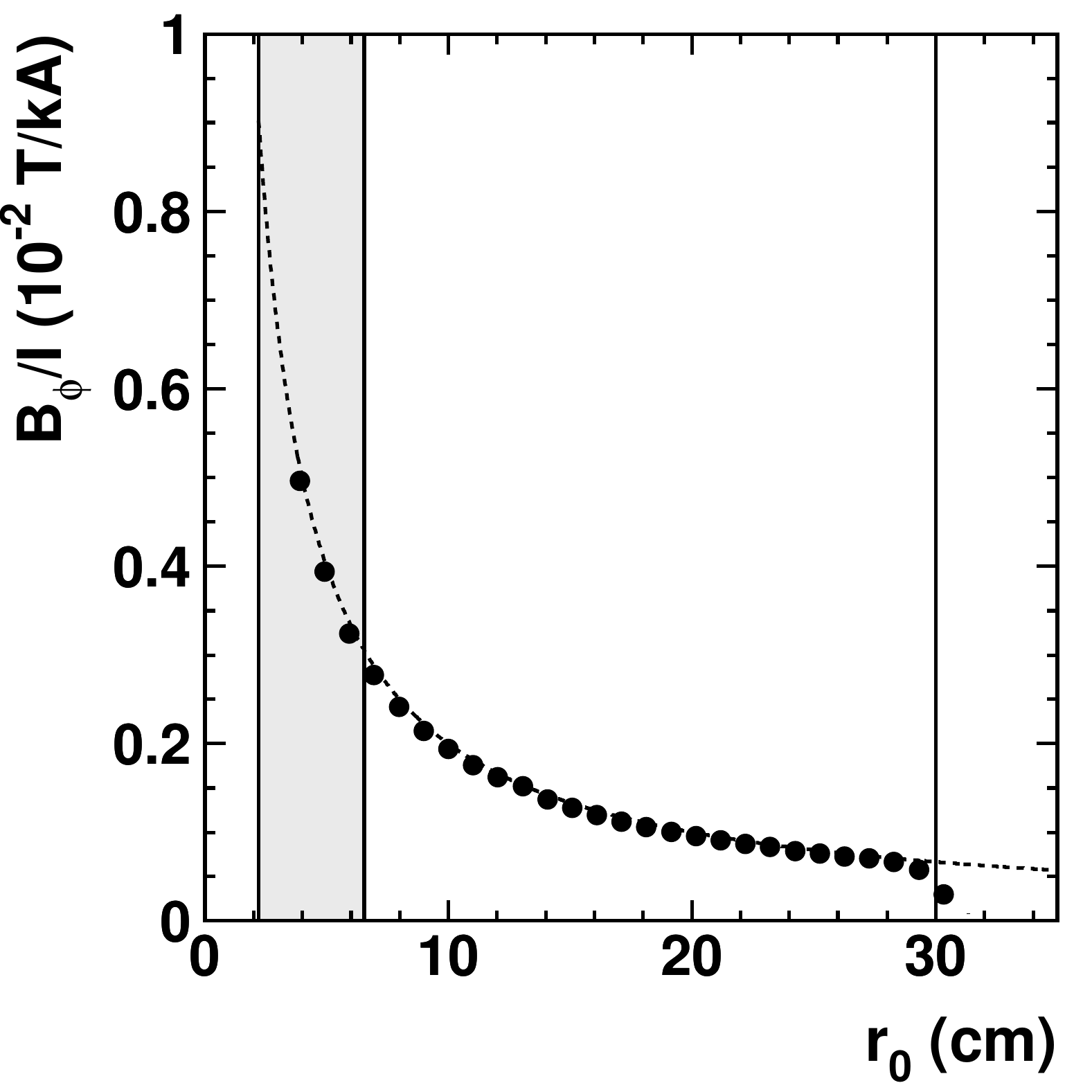}
\caption{\label{fig:hornfield}Measurements of the azimuthal magnetic field within the
horn. The points show the measured magnetic field, while the line shows the expected
$1/R$ dependence. The black lines indicate the minimum and maximum radii of the inner
conductor.}
\end{center}
\end{figure}

\subsection{Horn Electromagnet}
\label{sec:horn}
The horn, shown in Figure \ref{fig:horn}, is a pulsed toroidal electromagnet
composed of an aluminum alloy (6061 T6). 
The current in the horn is a 143 $\mus$-long pulse with a nominal peak of 
170 kA coinciding with the arrival of the proton beam at the target.
The actual operating values are typically $174$ kA in both neutrino and
antineutrino mode with $\sim 1$ kA variations. In neutrino mode, the
flow of current (in the positive sense) runs along the inner conductor (containing
the target), which folds outwards at a length of 185 cm to return via the 
the outer conducting cylinder of the horn at 30 cm radius. 
Within the horn cavity, defined by 
the volume between the outer and inner conducting cylinders, the 
pulse creates a magnetic field that falls as $1/R$, where $R$ is the 
distance from the cylindrical symmetry axis of the horn. The largest field
values of $1.5$ Tesla are obtained where the inner conductor is narrowest
(2.2 cm radius). The effects of time-varying fields within the cavity of the horn are found to 
be negligible. The expected field properties of the horn have been verified by measuring
the current induced in a wire coil inserted into the portals of the horn. Figure \ref{fig:hornfield}
shows the measured $R$ dependence of the azimuthal magnetic field compared with
the expected $1/R$ dependence. The ``skin effect'', in which the time-varying 
currents traveling on the surface of the conductor penetrate into the conductor, 
results in electromagnetic fields within the conductor itself.

During operation, the horn is cooled by a closed water 
system which sprays water onto the inner conductor via portholes in 
the outer cylinder. The target assembly is rigidly fixed to the upstream 
face of the horn, although the target is electrically isolated from its current path.
At the time of writing, two horns have been in operation in the BNB. The first
operated for 96 million pulses before failing, while the second is still in 
operation as of this writing after over 130 million pulses \cite{michel}.

\subsection{Collimator}
A concrete collimator is located downstream of the target/horn assembly and serves
as the entrance into the decay region. The collimator absorbs particles that would 
not otherwise contribute to the neutrino flux and is 214 cm long, with an upstream
aperture of 30 cm radius that grows to 35.5 cm on the downstream end.  By absorbing
these particles, the collimator reduces radiation elsewhere in the beamline.The upstream
end of the collimator is located 259 cm from the upstream face of the target. Simulations
indicate that the collimator does not limit the neutrino flux at the MiniBooNE detector.

\subsection{Decay Region and Absorber}
The air-filled cylindrical decay region following the collimator is 3 feet 
in radius and extends for 45 meters, terminated by the beam stop 50 meters from
the upstream face of the target. Survey data indicate that the constructed
decay region is 49.87 meters, including the collimator region. The wall 
of the decay region is a corrugated steel pipe surrounded
by packed dolomite gravel ($\mbox{CaMg(CO}_2\mbox{)}_{3}$, $\rho=2.24 \mbox{g}/\cm^3$).  
The beam stop itself is made of steel and concrete, within which is an array
of gas proportional counters that detects muons penetrating the beam stop. 

To allow potential systematic studies, a set of ten steel absorbing plates are positioned above the decay 
pipe at 25 meters. When lowered into the decay region, the steel absorbers reduce the effective decay path from 
50 to  25 meters. This has the effect of reducing the overall flux, but preferentially reducing the 
decay of the longer-lived muons, a major source of non-oscillation $\nue$ background. The 25 meter
absorber was not deployed during the neutrino running for the $\num\to\nue$ oscillation analysis.

\subsection{Little Muon Counter}
The Little Muon Counter (LMC) is an off-axis spectrometer that measures the rate and spectrum of
muons produced at a $7^\circ$ angle to the beam axis in the decay pipe pointing back to the alcove for the 25 meter absorber. The detector consists of a 40 foot
drift pipe extending from the decay region at $7^\circ$ leading to an enclosure. The kinematics of the two-body
pion and kaon decay are such that kaons produce a momentum distribution peaked at  $1.8 \gevc$ at this angle,
whereas pions produce muons at lower momentum. Muons sent down the drift pipe to the enclosure encounter
an iron collimator with a tungsten core that further restricts the angular acceptance of the counter and reduces backgrounds.
Following the collimator, the muon momentum is determined by a spectrometer consisting of a dipole magnet and planes of 
scintillating fiber trackers. Finally, a range stack consisting of alternating scintillator and tungsten layers allow high
energy muons to be distinguished from pions and other particles based on  the number of tungsten
planes penetrated by the particle. Further details on the LMC can be found in Reference \cite{boblmc}.
\section{Geant4-based Monte Carlo Simulation}
\label{sec:g4}

The properties of the MiniBooNE neutrino flux are determined using a
Geant4-based Monte Carlo simulation \cite{geant4}. The simulation can be divided
roughly into five steps:
\begin{itemize}
\item The definition of the beamline geometry, specified by the shape, location, and
composition of the components of the BNB, through which the 
primary protons and all other particles  propagate (Section \ref{sec:geometry}).
\item The generation of the primary protons according to the expected
beam optics properties  upstream of the target (Section \ref{sec:primaryoptics}).
\item The simulation of particles produced in the primary $\mbox{p}$-Be interactions,
including the elastic and quasi-elastic scattering of protons in the target. Custom
tables for the production of protons, neutrons, $\pipm$, 
$\Kpm$ and $\Kz$ in these interactions have been developed to accommodate
production models based on external data (Section \ref{sec:pbeint}). 
\item The propagation of the particles using the Geant4 framework, accounting for energy loss and
electromagnetic and hadronic processes that alter the kinematics of the
particles as described in Section \ref{sec:tracking}. Hadronic interactions and decay 
processes may also annihilate the particle in the tracking process  and create new 
particles to be tracked. 
Within the horn, the effect of the expected magnetic field on the trajectory 
of the particles is accounted for (Section \ref{sec:g4field}).
\item The identification of decay processes that result in neutrinos. The simulation
of the decays is handled by a custom
decay model, described in Section \ref{sec:decays}, outside of the Geant4 framework. The
decay model reflects  the latest branching fraction measurements and simulates polarization 
effects and kinematic distributions resulting from decay form factors. A number 
of techniques to enhance the statistical precision of the flux prediction are employed 
(Section \ref{sec:statenhance}).
\end{itemize}

%The initial properties of each incident proton are randomly chosen according to the measured
%position and angular distributions of the beam.  A detailed model
%has been developed for proton, neutron, and pion interaction cross
%sections on both aluminum and beryllium, as described in Section
%\ref{sec:hadronic}.  In addition, for inelastic proton-beryllium
%interactions, we use a custom model for secondary particle production
%based primarily on fits to external data and input from other hadronic
%interaction packages where external data is not available, as described in Section \ref{sec:secondary}. %The primary source of neutrinos in the experiment are the decays of
%these secondary particles.  In order to determine the characteristics of the flux with 
%sufficient statistical precision, a  series of enhancement techniques that include resampling
%the kinematics of these decays (described in Section \ref{sec:statenhance})
%are used to boost reduce the statistical uncertainties.
 
\subsection{Geant4 Description of the BNB Geometry:}
\label{sec:geometry}
The Geant4 Monte Carlo geometry consists of the last 50 meters of the Booster
beamline, the target hall, and the 50 meter meson decay volume. The geometry
description is defined to match the actual constructed beamline as closely
as possible; differences from the specifications are noted here. Since the generation of 
primary protons in the simulation starts immediately upstream
of the target (see Section \ref{sec:primaryoptics}), the geometry description 
of the beamline leading to the target is simplified.
Each section is simulated with concrete walls surrounded by a
uniform bed of dolomite.  The entire structure is filled with air at standard
temperature and pressure.

\begin{figure}
\begin{center}
\includegraphics[width=130mm]{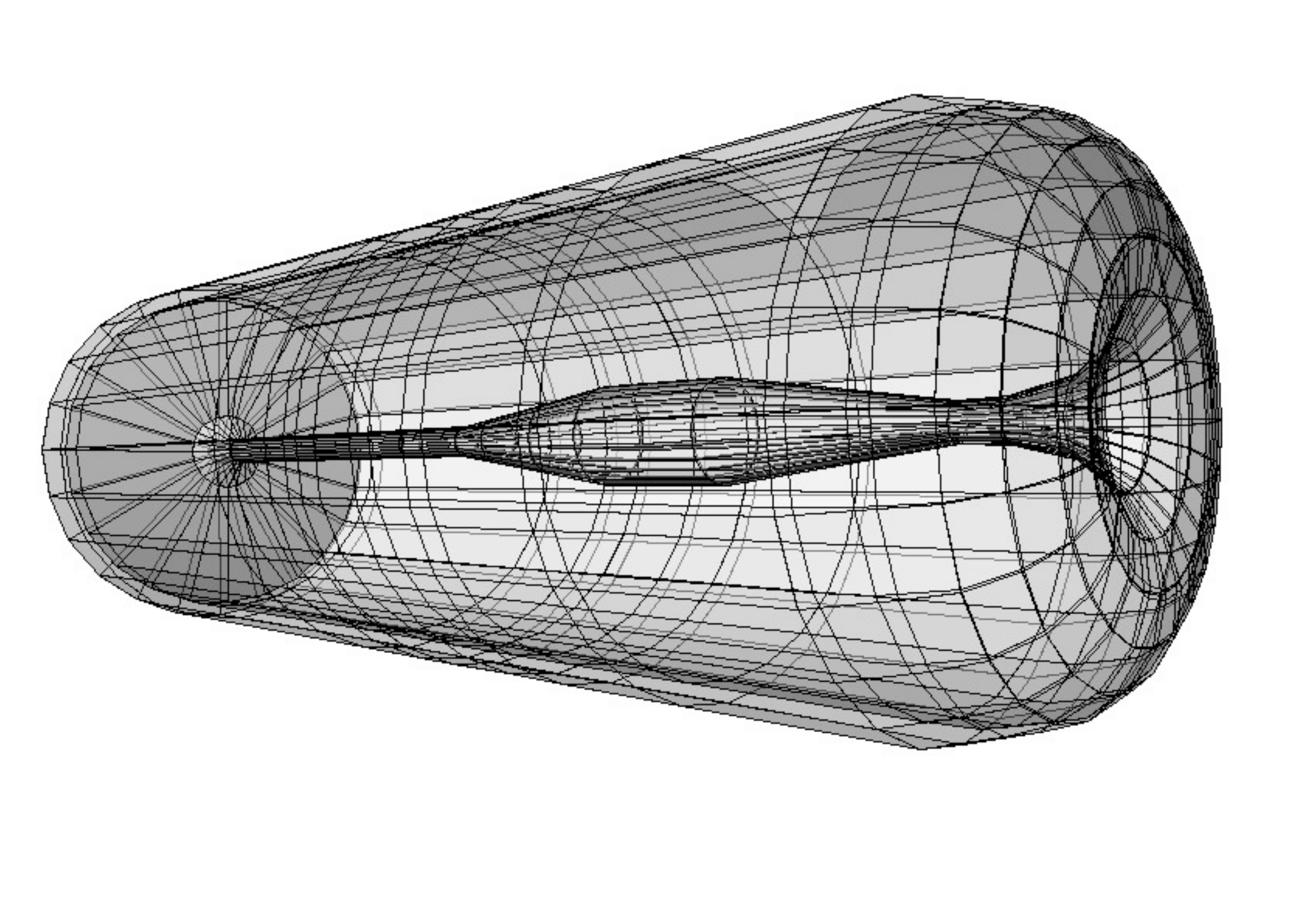}
\caption{\label{fig:g4horn}The MiniBooNE horn as simulated in the Geant4
beam Monte Carlo.}
\end{center}
\end{figure}

The simulated target hall contains the target, horn, and secondary beam
collimator.  In addition to the concrete walls, the target hall is
lined with 1.28 meters of steel shielding.  The seven slugs of the target and the target
sleeve, together with the fins which support them within the sleeve,
compose the simulated target geometry.
The horn is constructed using an aluminum Geant polycone that
specifies the inner and outer radius at 14 different points along the
direction of the beam.  A polycone of air is placed inside of the aluminum
polycone to set the thickness of the inner and outer conductors.  A
graphical representation of the Monte Carlo horn is shown in Figure
\ref{fig:g4horn}.

%Just downstream of the horn is the secondary beam
%collimator, which is a circular hole in the target hall
%steel shielding with an initial radius of 30 cm that linearly
%expands to a final radius of 35.5 cm.

The meson decay region is simulated as
two 20-meter-long decay pipes separated by the 25 meter absorber enclosure, followed by
the beam stop.  The decay pipes are made of concrete with an inner
diameter of 6 feet and an outer diameter of 10 feet.  In contrast, the actual
decay pipe is corrugated steel and surrounded by dolomite; this simplification
is not expected to affect the flux prediction.
%The simulated  25-meter absorber volume is  a concrete box with a steel floor and 10 steel 
%absorber
%modules hanging just above the beamline.  The beam stop is composed of
%11 feet of steel and 3 feet of concrete along the direction of the beam.

\subsection{\label{sec:g4field}Horn Magnetic Field}
The $1/R$ magnetic field expected from the current path within the horn is
simulated in  the volume corresponding to the cavity of the horn. The strength of the
field corresponds to a 174 kA current running along the inner conductor (reversed when
simulating anti-neutrino mode). In addition,
the permeation of the magnetic field into the inner conducting cylinder of the
horn from the skin effect (described in Section \ref{sec:systematic})  is included in the simulation. The predicted trajectories for charged particles in the magnetic field in the Geant4 simulation
have been checked in an external study using
the  DRKNYS routine from CERNLIB \cite{cernlib}, an independently implemented numerical method.

 \subsection{Generation of the Primary Proton Beam}
\label{sec:primaryoptics}
 The primary protons are simulated individually, since no correlated effects
 between the protons in a bunch are expected. The properties of the proton
 beam, such as the position and profile, have been simulated using TRANSPORT \cite{transport} and
 verified by upstream beam monitors \cite{moorepac2003}. The protons are generated
 1 cm upstream of the target with the transverse ($x$, $y$) positions
 drawn from random Gaussian distributions with 0 mm mean and 1.51 mm and 
 0.75 mm widths, respectively. Likewise,
 the angular deviations of the proton direction from the $z$ direction,
  $\theta_x$ and $\theta_y$, are drawn from Gaussian distributions with
 0 mrad mean and $0.66$ mrad and $0.40$ mrad width, respectively. 
The number of protons
 that  undergo inelastic interactions in the target (as opposed to scattering out)
 is studied in Section \ref{sec:primarysys}. In particular,
while the default configuration describes a diverging beam, the TRANSPORT simulations indicate
that the protons are expected to be convergent on the target, with a ``waist'' of zero
divergence at the center of the target. 
The simulated beam configuration is such that 99.8\% of the protons are on a trajectory to intersect
 the target. The studies in Section \ref{sec:primarysys} indicate that
reasonable perturbations to the model, including the expected focusing configuration, do 
not affect the predicted neutrino flux by more than $1\%$.
 
\subsection{Particle Tracking and Propagation}
\label{sec:tracking}
Particles, whether they are primary protons or particles produced in the simulation, 
are tracked and propagated within the Geant4 framework with full accounting
for electromagnetic and hadronic interactions and decays. Within each medium,
the Coulomb scattering and energy loss are calculated in each step of the tracking,
and the particle trajectory and energy are updated accordingly. The energy loss and deflection
angles predicted by the framework have been checked in a comparison with the
Bethe-Bloch formalism  and the Highland formula \cite{michel}. The rate of hadronic interactions
for protons, neutrons and charged pions on beryllium and aluminum are
governed by customized cross section tables (see Section \ref{sec:hadronic})
that determine the rate of elastic and inelastic scattering within the target
and horn. The outgoing final state configurations of these interactions are handled by the 
default Geant4 elastic and inelastic scattering algorithms, with the exception of the
primary $p$-Be interactions.

For other particle/nucleus combinations, the default cross section tables in Geant4 are used. Extensive
checks have been performed to ensure that the rate of hadronic
interactions, both elastic and inelastic, are consistent with cross sections
assigned to these processes. The final state configurations of neutrinos produced in decays  
are handled outside of the Geant4 framework as described in Section \ref{sec:statenhance}.

 \subsection{Primary Proton Interactions}
\label{sec:pbeint}
For the vast majority of primary protons, the first beamline component 
encountered is the target. Since $\mbox{p}$-Be interactions are the primary 
source of secondary mesons, a dedicated model (described in Section 
\ref{sec:secondary}) tuned to external data is used to describe the particle 
production in proton interactions in the target slugs, fins and sleeve. Elastic
scattering is handled by existing Geant4 models, while nucleon and pion
quasi-elastic scattering use a dedicated model based on free hadron-nucleus
scattering data. Due to the divergence of the primary beam and scattering, it is possible for a primary proton
to interact outside of the target (usually the aluminum of the horn or the concrete of
the decay region). For these cases, the particle production is handled
by the default Geant4 hadronic model.

For the primary $\mbox{p}$-Be interactions, secondary particles of seven types ($\pipm$, $\Kpm$, $\Kz$, 
$\mbox{p}$, $\mbox{n}$) are generated according to custom production tables describing
the double differential cross sections for the production of each secondary species as
a function of $p_z$ and $p_T$, the components of momentum along and transverse to
the primary proton direction, respectively. 
The total production cross section for a given species, obtained by integrating the double differential cross section,
determines the average multiplicity of the species in each primary $\mbox{p}$-Be reaction
(hadronic interactions excluding elastic and quasi-elastic scattering)
when divided by the reaction cross section. In each such reaction, the multiplicity for
each species in drawn from a Poisson distribution based on the average multiplicity, and the kinematics
drawn from the table of double differential cross sections. The reaction cross sections are 
described in Section \ref{sec:hadronic}, while the specification of the
double differential cross section tables is described in Section \ref{sec:secondary}.

\subsection{\label{sec:decays}Particle decays}

Neutrinos reaching the MiniBooNE detector are produced in the decays of
 charged pions, charged and neutral kaons, and muons. The particle lifetimes, decay
 modes and associated branching ratios, and kinematic distributions
 of the neutrinos produced in the decays assumed in the simulation 
 affect neutrino flux predictions, and are discussed here. 
 The neutrino parent lifetimes and branching ratios
 used in the simulation are given in Table. \ref{tab:decay},
 for $\pi^+$, $K^+$, $K^0_L$, and $\mu^+$ neutrino parents.
 The corresponding decays of negatively-charged particles
 are also simulated. \\
\begin{table}[tb]
\begin{center}
\begin{tabular}{c|rrr} \hline\hline
Particle & Lifetime & Decay mode & Branching ratio \\ 
  &  (ns) &  & (\%)   \\ \hline
$\pi^+$ & 26.03 & $\mup + \num$ & 99.9877 \\
        &       & $e^ + +\nue$     &  0.0123 \\ \hline
$K^+$ & 12.385 & $\mup + \num$      & 63.44 \\
      &        & $\piz+ e^+ + \nue$       &  4.98 \\
      &        & $\piz + \mup + \num$ &  3.32 \\ \hline
$K_L^0$ & 51.6 & $\pim + e^+ + \nue$             & 20.333 \\
        &      & $\pip + e^- + \nueb$       & 20.197 \\
        &      & $\pim + \mup + \num$       & 13.551 \\
        &      & $\pip + \mu^- + \numb$ & 13.469 \\ \hline
$\mup$ & 2197.03 & $e^+ + \nue + \numb$ & 100.0 \\ \hline
\end{tabular}
\end{center}
\caption{\label{tab:decay}{
Particle lifetimes, and neutrino-producing decay modes and branching ratios 
considered in the simulation.}}
\end{table} 
\indent In the two-body decays of charged pseudo-scalar mesons 
$M^+\to l^+ + \nu_l$, where $M=\pi ,K$ and $l=e,\mu$,
neutrinos are produced in the meson rest frame with fixed energy 
$E_{\nu}=(m_{M}^2-m_{l}^2)/(2m_{M})$
and isotropic angular distribution. \\
\indent For kaon semileptonic decays, $K\to\pi + l + \nu_l$ (``$K_{l3}$''),
neutrinos are produced with isotropic angular distribution in
the kaon rest frame. For the neutrino energy distributions,
different parametrizations are used for the $K^{\pm}_{l3}$ and
$K^0_{l3}$ form factors depending on whether electron or
muon neutrinos are produced in the decay. In both cases,
we assume that only vector currents contribute, and that
time-reversal invariance holds. \\
\indent For $K^{\pm}_{e3}$ and $K^0_{e3}$ decays, the electron neutrino
energy distribution in the kaon rest frame is given by \cite{Chounet:1971yy}:
\begin{equation}
\label{eq:ke3decay}
\frac{dN}{dE_{\nu}}\propto
\int_{E_{e,-}}^{E_{e,+}} dE_e (2E_{e}E_{\nu}-m_kE_{\pi}')|f^e_+(t)|^2
\end{equation}
\noindent where $E_e$ is the
 electron energy, $f_+$ is a form
 factor depending on the square of the four-momentum
 transfer to the pion, $t=(p_K-p_{\pi})^2=m_K^2+m_{\pi}^2-2m_K E_{\pi}$,
$E_{\pi}'$ is given by:
\begin{equation}
\label{eq:epiprime}
E_{\pi}'\equiv \frac{m_K^2+m_{\pi}^2-m_e^2}{2m_K}-E_{\pi}
\end{equation}
\noindent and $E_{e,\pm}$ are integration limits on the electron energy:
\begin{equation}
\label{eq:elenergyintlimits}
\left\{
\begin{array}{l}
E_{e,-}=\frac{m_K^2-m_{\pi}^2}{2m_k}-E_{\nu} \\
E_{e,+}=\frac{1}{2}(m_k-\frac{m_{\pi}^2}{m_k-2E_{\nu}}) \\
\end{array}
\right .
\end{equation}
\noindent We assume a linear dependence of the form factor $f^e_+$ on $t$:
\begin{equation}
\label{eq:ke3formfactor}
f^e_+(t)= f^e_+(0)(1+\lambda^e_+ t/m_{\pi}^2)
\end{equation}
\indent For both $K^+_{e3}$ and $K^0_{e3}$ decays, the coefficient 
$\lambda^e_+$ for the linear expansion of the form factor used is 
$2.82\cdot 10^{-2}$ \cite{Hagiwara:2002fs}. \\
%
%%%%%%%%%%%%%%%%%%%%%%%%%%%%%%%%%%%%%%%%%%%%%%%%%%%%%5
%
\indent For $K^{\pm}_{\mu 3}$ and $K^0_{\mu 3}$ decays, the muon neutrino
energy distribution in the kaon rest frame is given by \cite{Chounet:1971yy}:
\begin{equation}
\label{eq:kmu3decay}
\frac{dN}{dE_{\nu}}\propto
\int_{E_{\mu ,-}}^{E_{\mu ,+}} dE_{\mu} |f^{\mu}_+(t)|^2
[A+B\xi(t) + C\xi(t)^2]
\end{equation}
\noindent where:
\begin{equation}
\label{eq:kmu3abc}
\left\{
\begin{array}{l}
A \equiv m_K(2E_{\mu}E_{\nu}-m_K E_{\pi}')+m_{\mu}^2(E_{\pi}'/4-E_{\nu}) \\
B \equiv m_{\mu}^2(E_{\nu}-E_{\pi}'/2) \\
C \equiv m_{\mu}^2E_{\pi}'/4 \\
\end{array}
\right .
\end{equation}
\indent The quantities $E_{\pi}'$, $E_{\mu ,-}$, $E_{\mu ,+}$, $f^{\mu}_+(t)$
appearing in Equations~\ref{eq:kmu3decay} and \ref{eq:kmu3abc} are defined as 
in the $K_{e3}$ case
(see Equations~\ref{eq:epiprime}, \ref{eq:elenergyintlimits},
\ref{eq:ke3formfactor})
substituting $e\to\mu$. In the simulation, we take
$\xi (t)\simeq \xi (0) = -0.19$ \cite{Hagiwara:2002fs}. \\
%
%%%%%%%%%%%%%%%%%%%%%%%%%%%%%%%%%%%%%%%%%%%%%%%%%%%%%%%%%%%%%
%
\indent Concerning electron (anti-)neutrinos from $\mu^{\pm}$ 
decays, we neglect terms proportional
to the electron mass and assume the following neutrino energy and
angular distribution \cite{gaisser}:
\begin{equation}
\label{eq:nuemudecay}
\frac{dN}{dxd\Omega_{\nu}}= 
 \frac{12x^2}{4\pi}(1-x)[1\mp P_z\cos (\theta_{\nu})]
\end{equation} 
\noindent where $\cos (\theta_{\nu})$ is the neutrino emission
 angle with respect to the beam direction $z$, $P_z$ is the projection
 along $z$ of the muon polarization
 vector in the muon rest frame, and $x=2E_{\nu}/m_{\mu}$, with $0<x<1$.
 The muon polarization vector is estimated on an event-by-event basis.
 For $\pi^+\to\mu^+\to\nu_e$ decays, the muon polarization in the muon
 rest frame is calculated from the known muon polarization
 in the pion rest frame, and boosting the polarization vector into the
 muon rest frame. The muon polarization for muons proceeding from $K^{\pm}$
decays is computed in the same way, with the simplifying assumption that
all $K^{\pm}$ decays proceed via the $K^{\pm}_{\mu 2}$ decay mode.

\subsection{Statistical Enhancements}
\label{sec:statenhance}
Running the Geant4 beamline simulation and recording the
outgoing neutrinos proton-by-proton would not provide enough neutrinos at the MiniBooNE
detector to allow for a precise determination of the flux across the entire
phase space of interest. As a result, several modifications are made to 
enhance the beam Monte Carlo simulation statistics.

A large statistical enhancement is gained by ``redecaying'' the parent
particle of the neutrino.  For each neutrino produced in the beam
Monte Carlo, the particle decay which produced the neutrino is
performed 1000 times.  Each redecay is performed at the same
location, but the kinematics of the decay are randomly redrawn each
time from the decay distributions, resulting in different momenta
for the daughter particles in each draw.

A similar technique is used to boost the statistics of neutrinos from
muon decay.  Most muons produced in the secondary beam do not decay
before stopping in the beam stop or the walls of the decay region due 
to their long lifetime.  To better
estimate the muon decay-in-flight component of the neutrino flux, each
time a muon is produced in the simulation, 19 identical copies are created
and independently propagated through the simulation.  To
account for the resulting overproduction of neutrinos, the weight for
each muon-decay neutrino is correspondingly reduced by a factor of 20.

Another weighting technique is used to determine the high energy
neutrino flux.  The MiniBooNE neutrino energy spectrum peaks between
500 and 600 $\mev$ with a long high energy tail extending past 6
$\gev$.  Since fewer neutrinos are produced at these higher energies,
statistical fluctuations are much larger, increasing the
uncertainty in the shape of the high energy tail.  This problem is
made worse by the redecay procedure described above since high energy
parent particles tend to decay to high-energy, forward-going
neutrinos, which resulting in a significant fraction of the 1000
redecays producing neutrinos pointed at the detector, all with similar
energies.  To reduce the statistical uncertainty in the prediction of the
high-energy tail, the meson production cross sections for proton-beryllium
interactions are multiplied by an exponential function in longitudinal
meson momentum.  Each event is de-weighted by its corresponding cross
section enhancement to preserve the correct neutrino/proton ratio.
This provides an artificially large production of neutrinos at high
energies with small event weights, thus reducing the statistical
uncertainty in the high energy tail of the predicted neutrino flux.

The neutrino flux at the MiniBooNE detector is determined by projecting
the path of the neutrino to the plane containing the center of the detector,
541 meters from the face of the target. Neutrinos which are on a path to
pass through the detector (within 610.6 cm of the center of the detector at this plane,
accounting for the vertical displacement) are recorded in the flux distributions used
for Monte Carlo event simulation in the detector. For the simulation of neutrino
interactions outside the detector in the concrete walls of the detector hall or in the 
dirt beyond, a larger radius of 1400 cm is used to determine the flux distributions.

\section{Hadronic Interactions in the Beamline}
\label{sec:hadronic}
Hadronic cross sections  play an important role in determining the
properties of the neutrinos produced in the BNB. Most notably,
hadronic cross sections on beryllium and aluminum, the 
materials composing the target and horn, respectively, govern
the rate of primary proton interactions in the target, as well as the
rate of absorption of pions in the target and the horn. The breakdown
of the proton cross sections between elastic, quasi-elastic and other forms
of interactions govern the fraction of protons that scatter out of the
target before interacting. The analogous breakdown of the cross section
for pions is particularly important at high momentum, where forward-going
pions may intersect a considerable amount of material in the target or the
horn before entering the decay region.

The cross sections fall into three categories: elastic (coherent)
scattering, inelastic scattering, and quasi-elastic scattering. 
In the first, the incident hadron scatters coherently from the
nucleus as a whole. The rest of the total hadronic cross section
is due to inelastic processes. A subset of these processes involve hadron
scattering with the nucleons within the nucleus in a manner analogous
to the elastic scattering of hadrons off free nucleons; this is referred
to as quasi-elastic scattering. 
The remainder of the inelastic
cross section includes the particle production processes  discussed in Section \ref{sec:secondary}.
The relevant momentum range in the flux prediction are at and below the primary proton momentum ($8.89\gevc$)
for nucleons. The corresponding momentum range for the pions produced by these protons is
$0-6\gevc$

Wherever possible, measured cross sections have been used in the simulation. In some
cases, the measured and calculated cross sections are extrapolated
to cross sections for other particles that are related by
isospin. Measurements exist primarily for total hadronic cross
sections and inelastic cross sections, from which the elastic cross
section can be inferred. Theoretical guidance is needed primarily
for the total hadronic cross sections for pion-nucleus scattering
and quasi-elastic scattering. Table \ref{tab:bexsecsources} summarizes
the source of nucleon and pion cross sections. Details of the parametrizations
used to describe the momentum dependence of each cross section are given
in Section \ref{sec:hadpar}.

\begin{table}
\begin{center}
\begin{tabular}{l|ll|l}\hline\hline
	& p-(Be/Al)	& n-(Be/Al)			& $\pipm$-(Be/Al)           \\ \hline
$\stot$	& Glauber 	& Glauber		        & Data ($p<0.6/0.8\gevc$)   \\
	&              	& (checked with data) 		& Glauber ($p>0.6/0.8\gevc)$\\
$\sine$	& Data		& (same as $\mbox{p}$-Be/Al)  	& Data	   		    \\
$\sqel$	& Shadow	& Shadow	 	        & Data ($p <0.5\gevc$ )   \\
	&		& 		 		& Shadow $(p>0.5\gevc)$ \\ \hline
\end{tabular}
\vskip 0.25 cm
\caption{\label{tab:bexsecsources} Origin of hadron-beryllium cross sections used in 
the Geant4 simulation. ``Glauber'' indicates that the Glauber model calculations described
in Section \ref{sec:sela} are used. These have been cross checked by $\mbox{n}$-Be $\stot$ data. ``Data'' indicates
that existing measurements are directly parametrized. ``Shadow'' refers to the calculation of
$\sqel$ using the shadowed multiple scattering model described in Section \ref{sec:ine}.}
\end{center}
\end{table}

\begin{figure}[t]
\begin{center}
\includegraphics[width=14.0 cm]{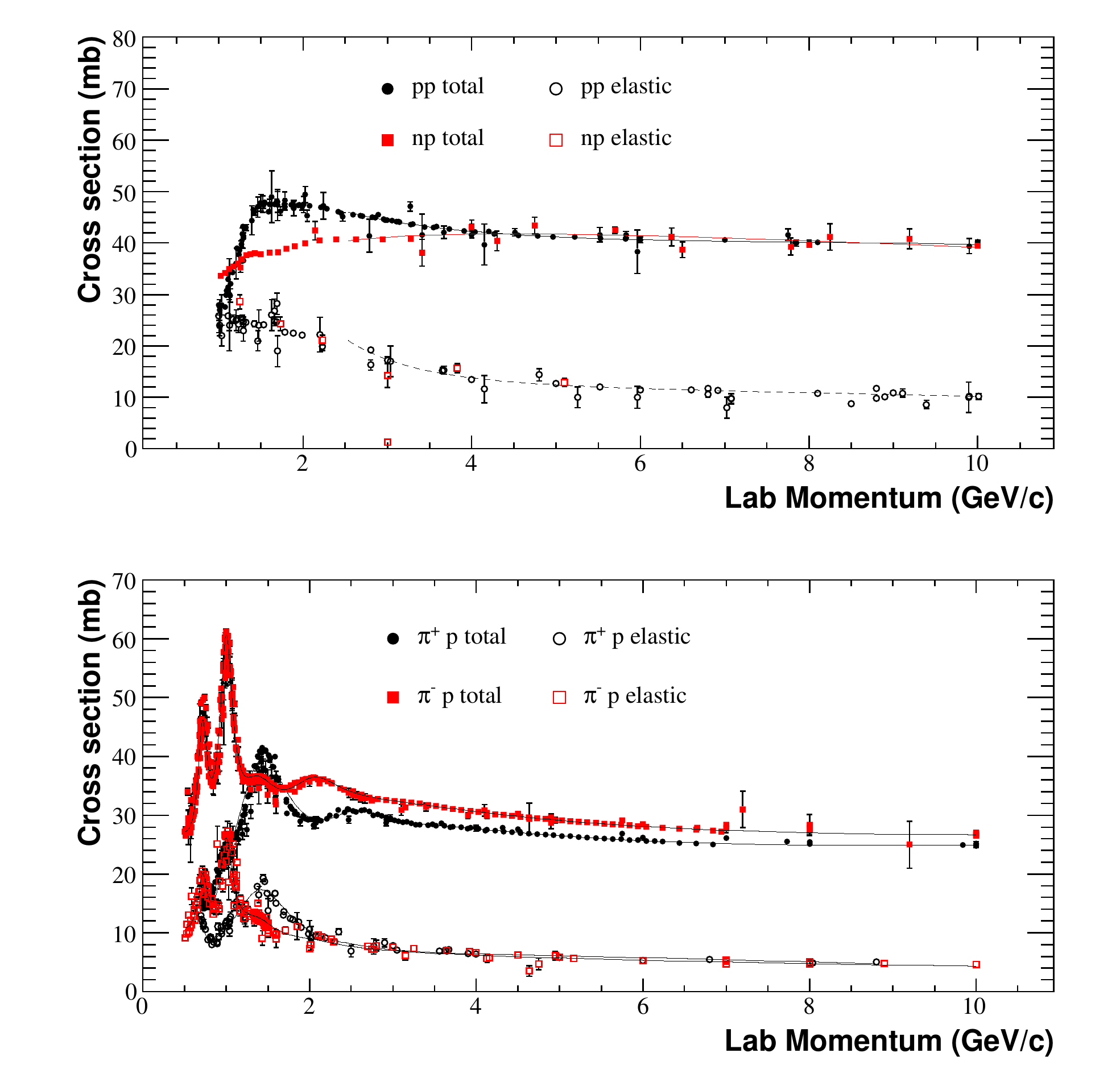}
\caption{\label{fig:fn_cs} Total and elastic hadron-proton
cross sections (Top: proton/neutron, Bottom: $\pip$/$\pim$) compiled
by the Particle Data Group and the COMPAS collaboration\cite{pdg}, with the 
parametrizations used in the Glauber model calculations.}
\end{center}
\end{figure}

\begin{figure}[t]
\begin{center}
\includegraphics[width=14.0 cm]{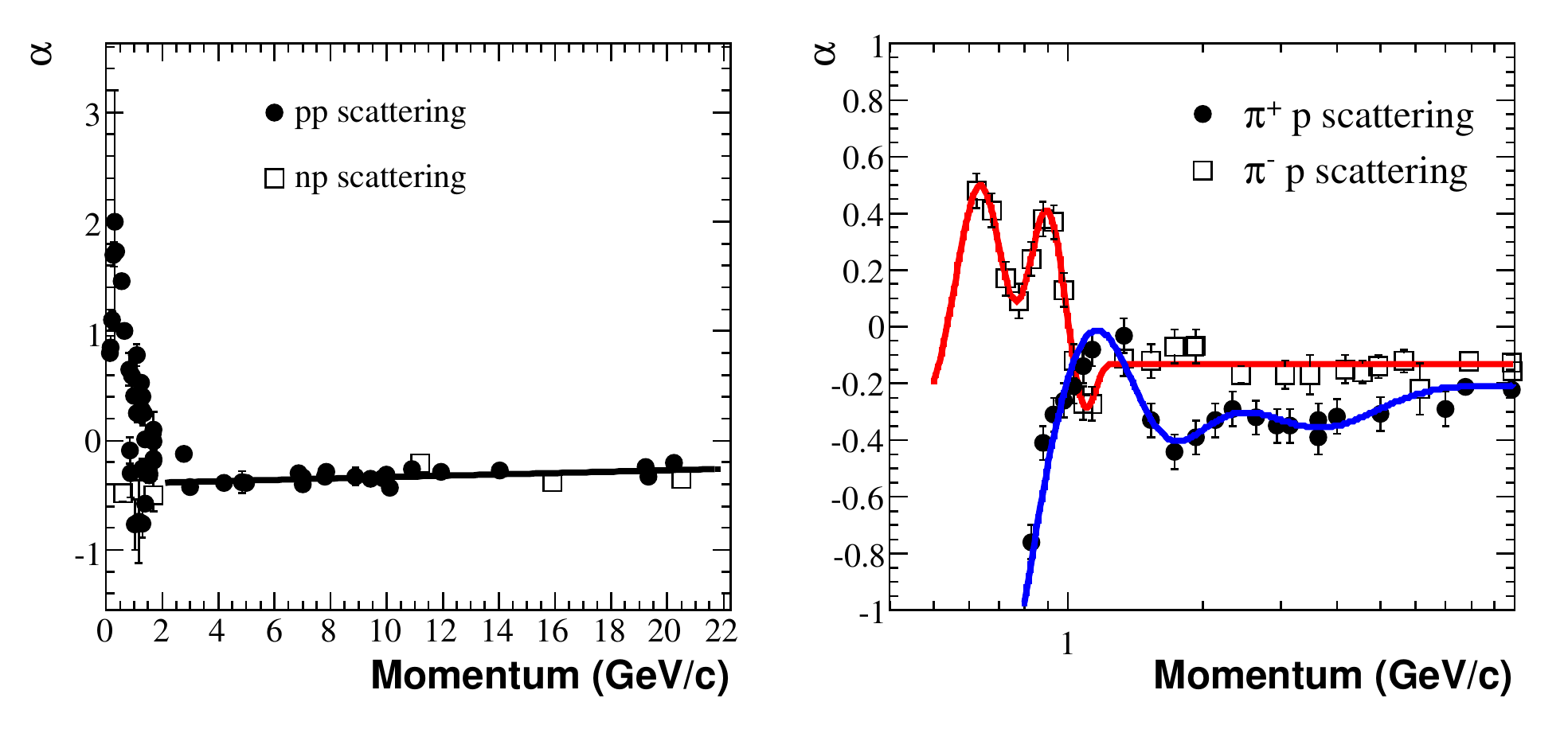}
\caption{\label{fig:alpha_nuc_pion} Measured values of $\alpha$ the
real-to-imaginary ratio of the forward scattering amplitude for $\mbox{p}-\mbox{p}$ and
$\mbox{n}-\mbox{p}$ scattering (left) and $\pip-\mbox{p}$ and $\pim-\mbox{p}$ scattering (right) with parametrizations.
The parametrizations used in the Glauber model calculation are shown.}
\end{center}
\end{figure}

\begin{figure}[t]
\begin{center}
\includegraphics[width=6.7 cm]{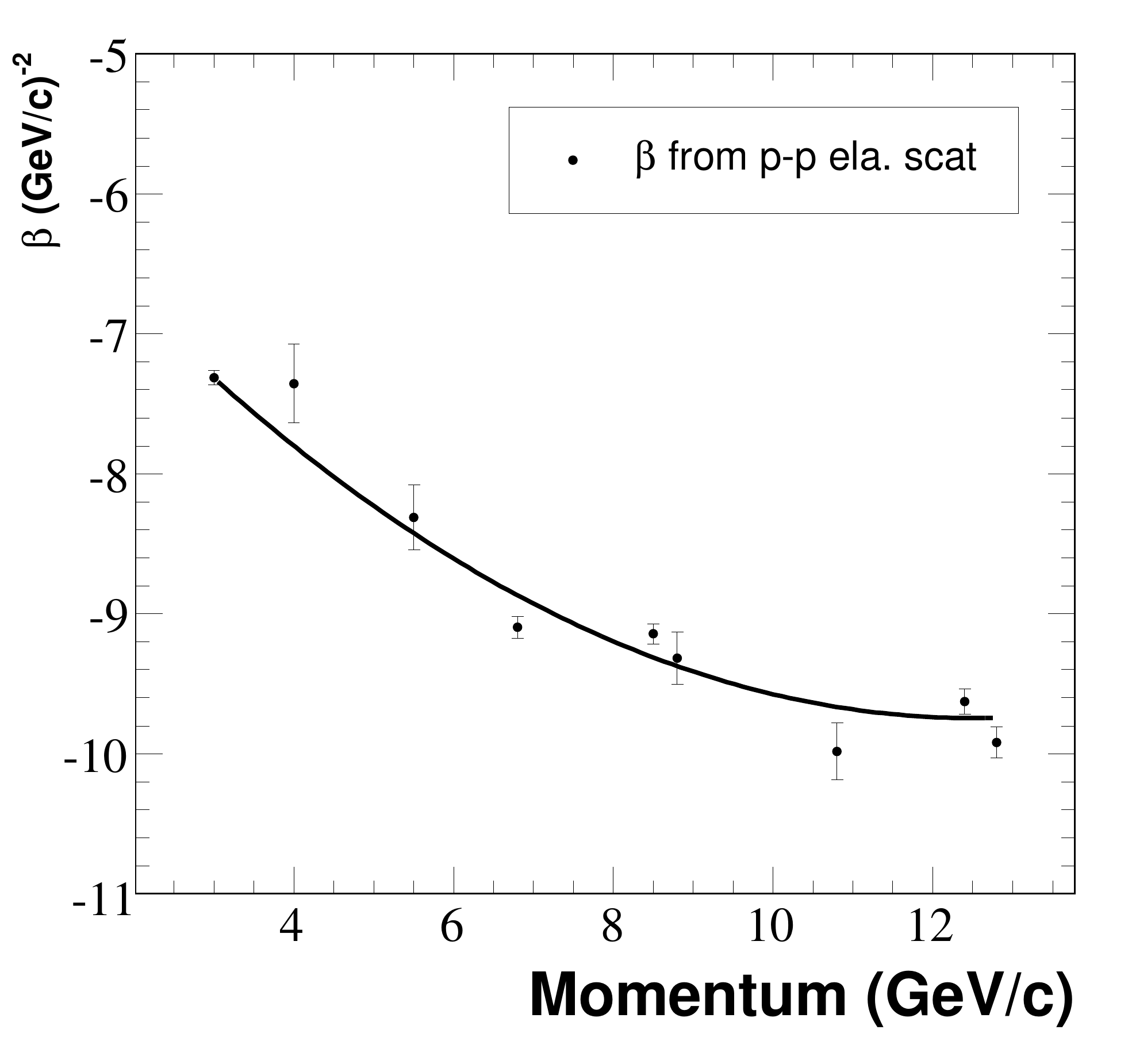}
\includegraphics[width=6.7 cm]{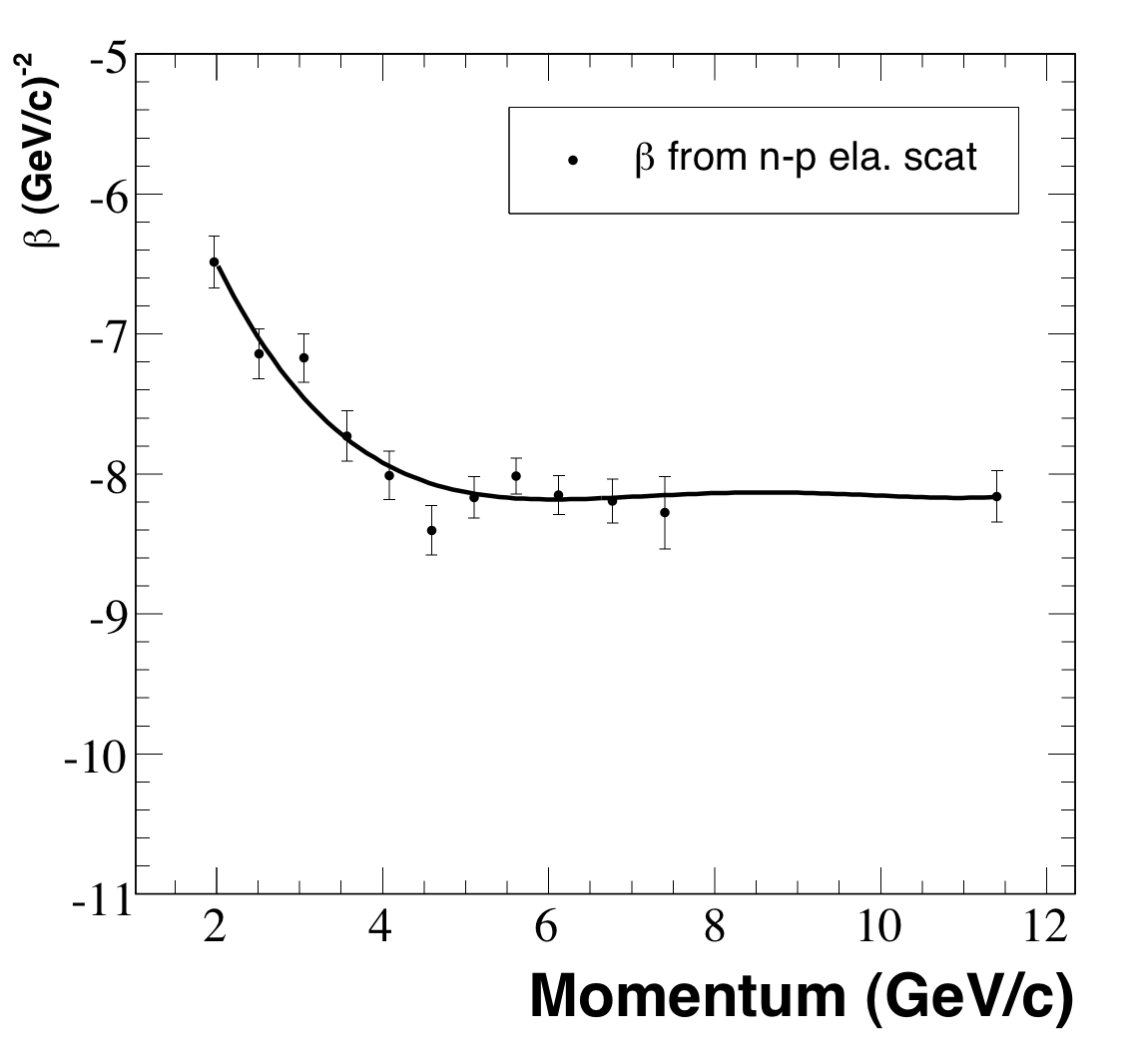}
\caption{\label{fig:betafits_gamfix} $\beta$ parameters  for $\mbox{p}-\mbox{p}$ (left) and $\mbox{n}
-\mbox{p}$ (right) scattering obtained from fits to the data with the parametrizations used in the
Glauber model calculation.}
\end{center}
\end{figure}

\begin{figure}[t]
\begin{center}
\includegraphics[width=14.0 cm]{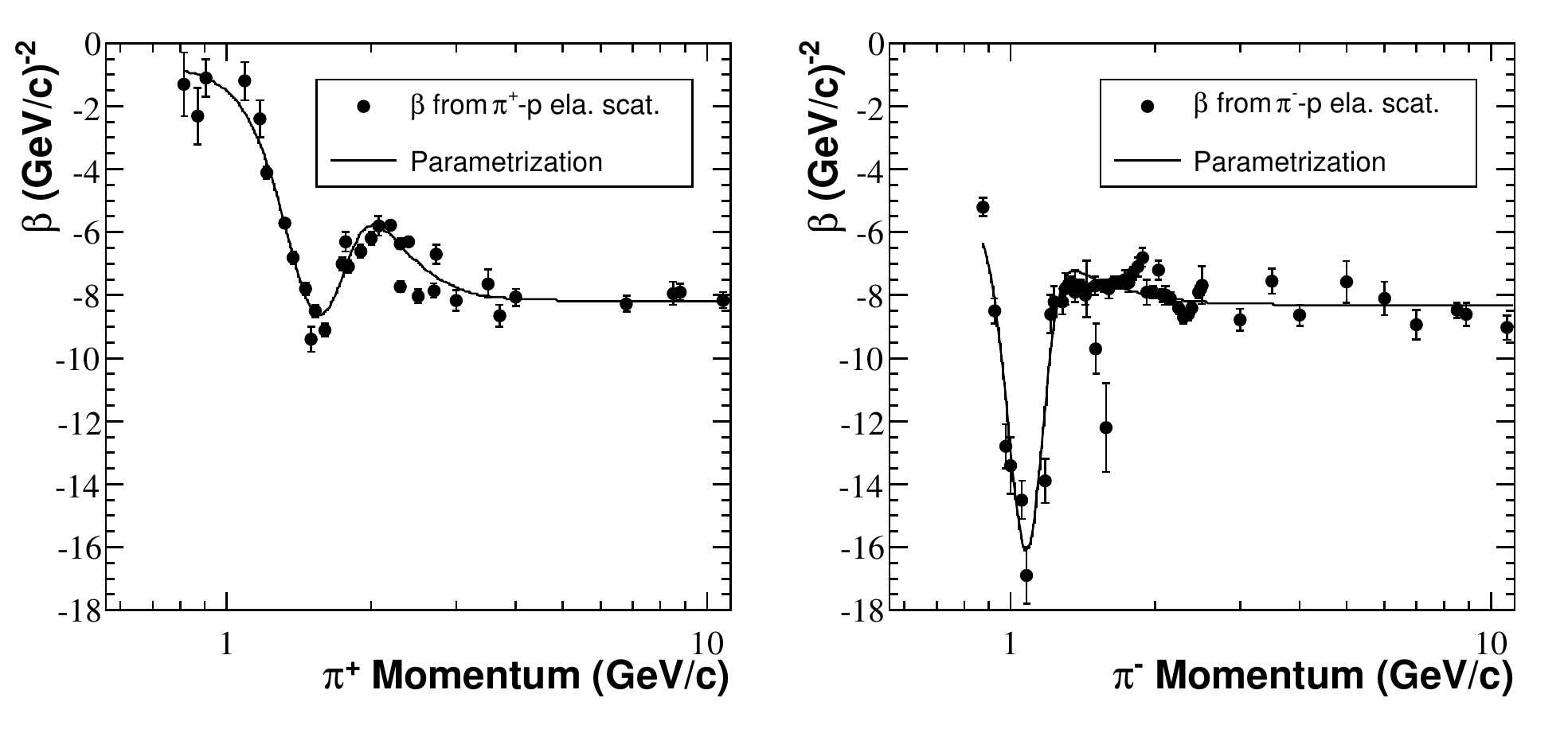}
\caption{\label{fig:beta_pippim_all} Compilation of measured $\beta$ parameters 
for $\pip-\mbox{p}$ elastic scattering (left) and $\pim-\mbox{p}$ elastic scattering (right) versus incident
pion momentum with the parametrizations used in the Glauber model calculation. The measured values of $\beta$ 
include the Lasinski compilation \cite{lasinski}
as well as our own fits to the $q^2$ distributions measured in References \cite{coffin,lindenbaum,cern}.}
\end{center}
\end{figure}

\subsection{Total and Elastic Scattering}
\label{sec:sela}
The elastic scattering cross sections for protons, neutrons and charged
pions have been obtained by calculating the total hadronic cross section
$\stot$ using the Glauber model \cite{glauber} and subtracting the measured inelastic cross
sections described in Section \ref{sec:ine} assuming $\stot=\sine+\sela$. Direct measurements of $\stot$
for hadron-nucleus interactions in the relevant energy range 
exist only for neutrons and for pions in the $\Delta(1232)$ resonance region.  Wherever possible,
we compare the calculated results with the existing measurements to check their validity.
No direct measurements of the total elastic cross section ($\sela$) exist in this momentum range.

The calculation of $\stot$ follows the work described in Reference \cite{franco}, where hadron-{\bf nucleus}
elastic scattering is modeled as the coherent sum of scattering amplitudes from hadron-{\bf nucleon}
scattering. The amplitude for forward elastic scattering is calculated allowing $\stot$ to be obtained
via the optical theorem.
 At each incident hadron energy, these amplitudes are summarized by three parameters, namely the total cross section for hadron-nucleon scattering ($\sigma_n$), the ratio of the real and imaginary parts of the forward scattering amplitude ($\alpha$), and the differential cross section
in $t=|q^2|$, the square of the 4-momentum transfer. The latter is parametrized as
an exponential distribution in $t$. All together, the hadron-nucleon scattering amplitude
can be expressed as:
\begin{equation}
f(q) = \frac{(i+\alpha) k \sigma_n}{4\pi}e^{\beta t/2}
\end{equation}
where $k$ is the wave number of the incident hadron. This form identically satisfies the optical
theorem. 

The Glauber model for elastic scattering represents the nucleus as a collection
of nucleons distributed in a spherically symmetric state with radial distributions given
by the independent harmonic oscillator form (for beryllium) or the Woods-Saxon form (for aluminum) \cite{woodssaxon}. The
scattering amplitude for a given configuration of nucleons is obtained by considering the
phase shift due to the individual hadron-nucleon scattering amplitudes. The total scattering
amplitude for the nucleus is calculated by averaging over all nucleon configurations weighted
by the nucleon density distribution. The total cross section $\stot$ is obtained
by applying the optical theorem to the resulting forward scattering amplitude. As mentioned
above, $\sela$ at a particular incident hadron momentum is calculated via the relation
$\stot = \sela+\sine$ using the values of $\sine$ described below. While it is in principle
possible to obtain $\sela$ from the Glauber model by obtaining the elastic cross section
as a function of $q^2$ and integrating, the assumptions of the model are most valid in the forward
direction. An extraction of $\sela$ using the model requires integrating the differential cross section
outside of this region.

The hadron-nucleon scattering parameters $\sigma_n$, $\alpha$ and $\beta$ are obtained
from the literature. In particular, $\sigma_n$ and $\alpha$ for $\mbox{p}-\mbox{p}$, $\mbox{p}-\mbox{n}$, $\pip-\mbox{p}$ and $\pim-\mbox{p}$ elastic
scattering have been compiled by the Particle Data Group (PDG) \cite{pdg}. 
The compiled data on $\sigma_n$ and $\alpha$ and our parametrization of  their momentum dependences are shown in Figure \ref{fig:fn_cs} and \ref{fig:alpha_nuc_pion}. In addition
to the PDG compilation, a compilation of $\alpha$ measurements by CERN\cite{cern}, as well
as the measurements of Foley {\em et al.} \cite{foley}, have been included in the parametrization.
This latter data are at momenta above the region of interest ($[7-10]\gevc$) but are nonetheless
useful in determining the momentum evolution of $\alpha$ for $\pipm-\mbox{p}$ scattering. The
measurements at momenta less than $3.5\gevc$ come entirely from the CERN compilation.
The PDG compilation of $\alpha$ in  nucleon-nucleon scattering adequately covers the
range of interest for the flux prediction.

Unfortunately, the $\beta$ measurements for hadron-nucleon elastic scattering have not been
compiled by the PDG. We have taken data from a number of experiments
(for $\mbox{p}-\mbox{p}$\cite{clyde,colleti,alexander,lindenbaum}, for $\mbox{n}-\mbox{p}$ \cite{perl, gibbard}, and for $\pipm-\mbox{p}$
 \cite{coffin,lindenbaum,cern}) and used the reported $t$ distributions to extract $\beta$.
Further, a compilation by Lasinski {\em et al.} \cite{lasinski} is used to supplement our
own compilation for $\pipm-\mbox{p}$ scattering at low momentum. The compiled $\beta$
values and the parametrized momentum dependences are shown in Figure \ref{fig:betafits_gamfix}
for $\mbox{p}-\mbox{p}$ and $\mbox{n}-\mbox{p}$ scattering, and Figure \ref{fig:beta_pippim_all} for $\pipm-\mbox{p}$ scattering.

\begin{figure}[t]
\begin{center}
\includegraphics[width=14.0 cm]{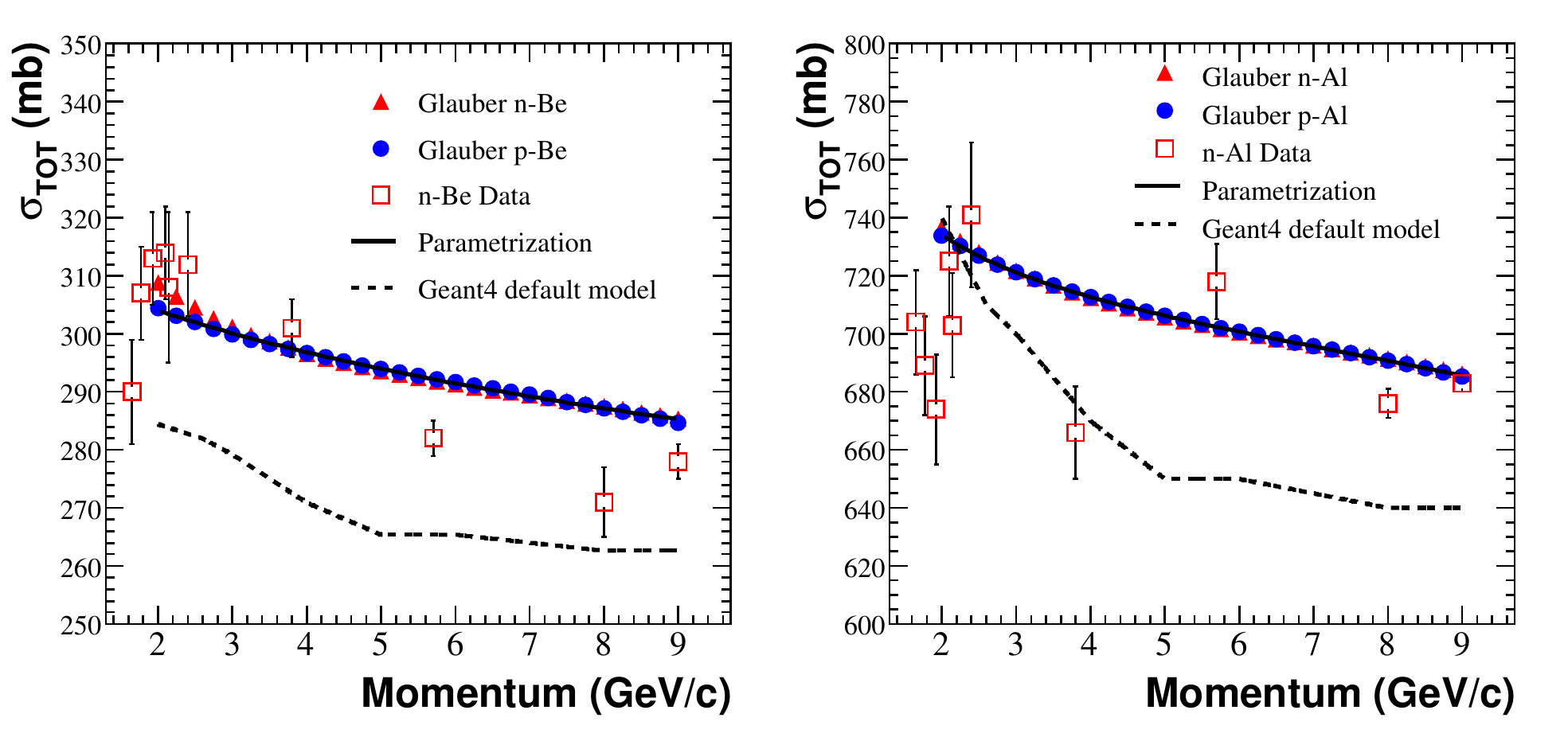}
\caption{\label{fig:nuctotxsec} Total hadronic cross sections calculated using the
Glauber model and the optical theorem for beryllium (left) and aluminum (right)
targets. The calculated results for neutrons(protons) are shown as triangles(circles).
The parameterization used in the flux prediction is shown as the solid line, while the default Geant4 
parametrization is shown as a dashed line. The measured
 values of $\stot$ for $\mbox{n}$-Be/Al from References \cite{neutnuc1,neutnuc2,neutnuc3,neutnuc4,neutnuc5} are shown as squares.}
\end{center}
\end{figure}

\begin{figure}[t]
\begin{center}
\includegraphics[width=6.8 cm]{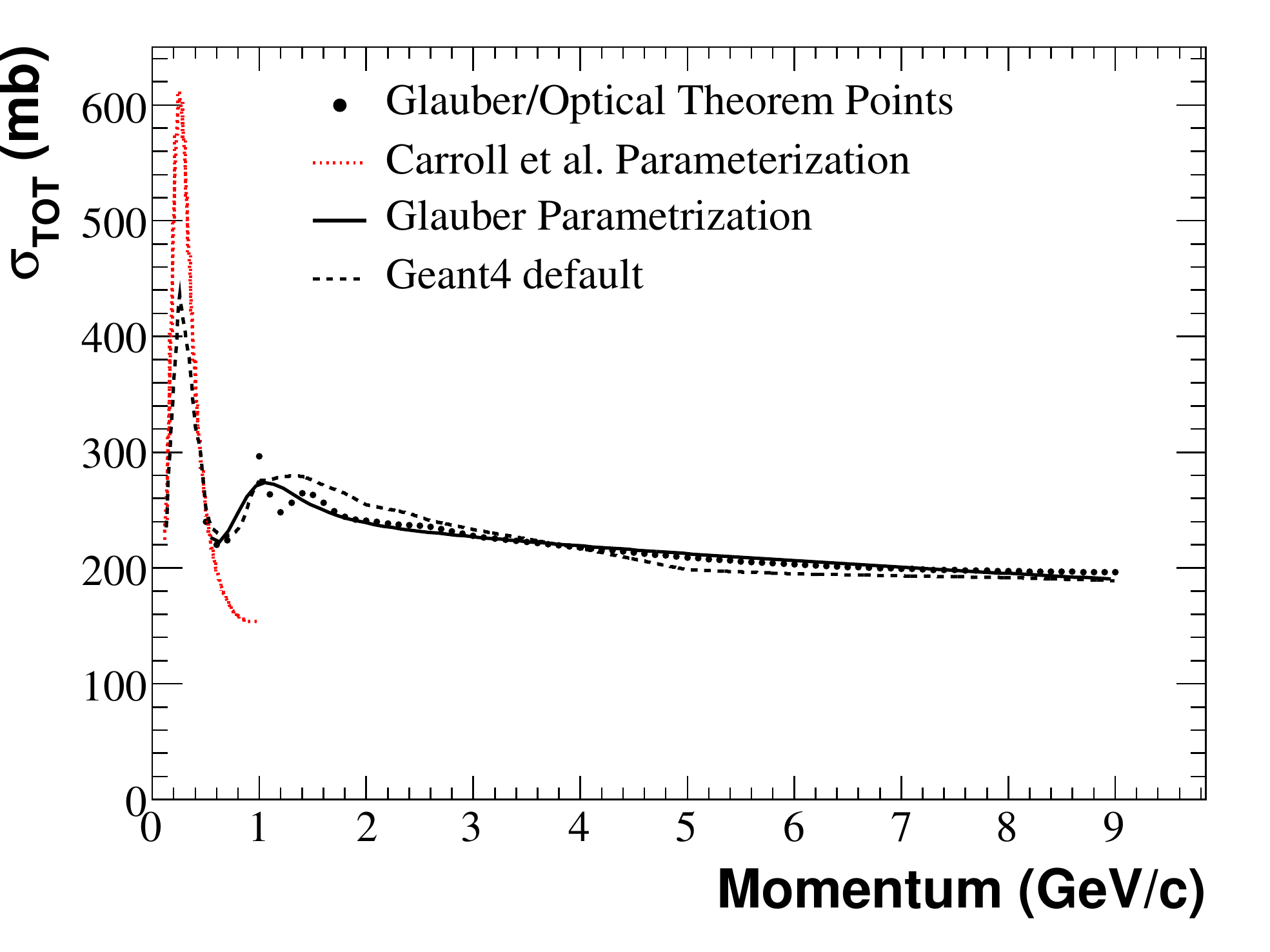}
\includegraphics[width=6.8 cm]{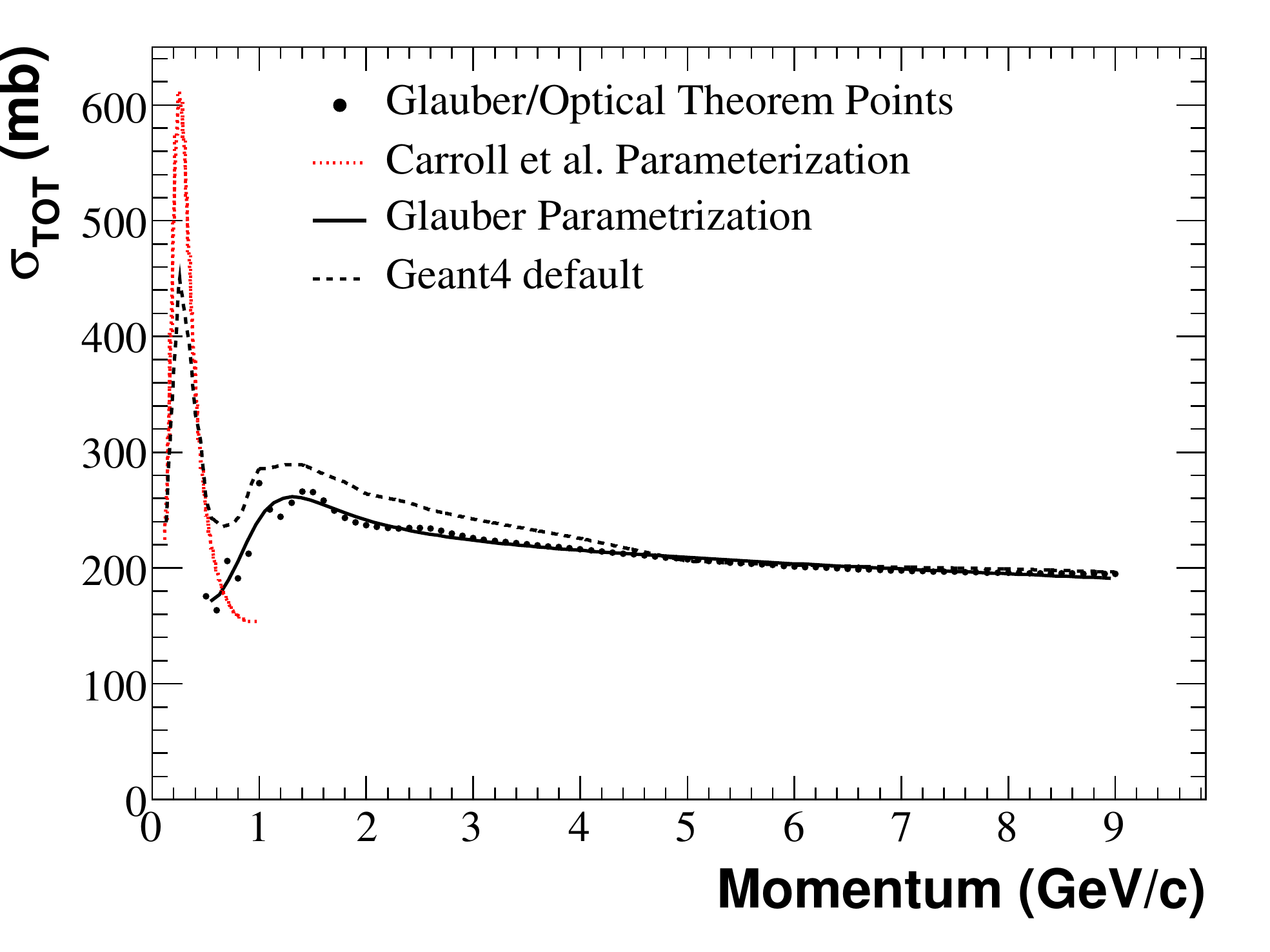}
\caption{\label{fig:pibetotxsec} Total hadronic cross sections for $\pipm$-Be
calculated using  the
Glauber model  (points) for $\pip$ (left) and $\pim$ (right). The 
Breit-Wigner parametrization based on the Carroll data \cite{carroll}
on the $\Delta(1232)$ resonance is shown as a dotted line, while the parametrization
of the Glauber model points is shown as a solid line. The Geant4 default model
is shown as a dashed line.}
\vskip 1.0 cm
\includegraphics[width=6.8 cm]{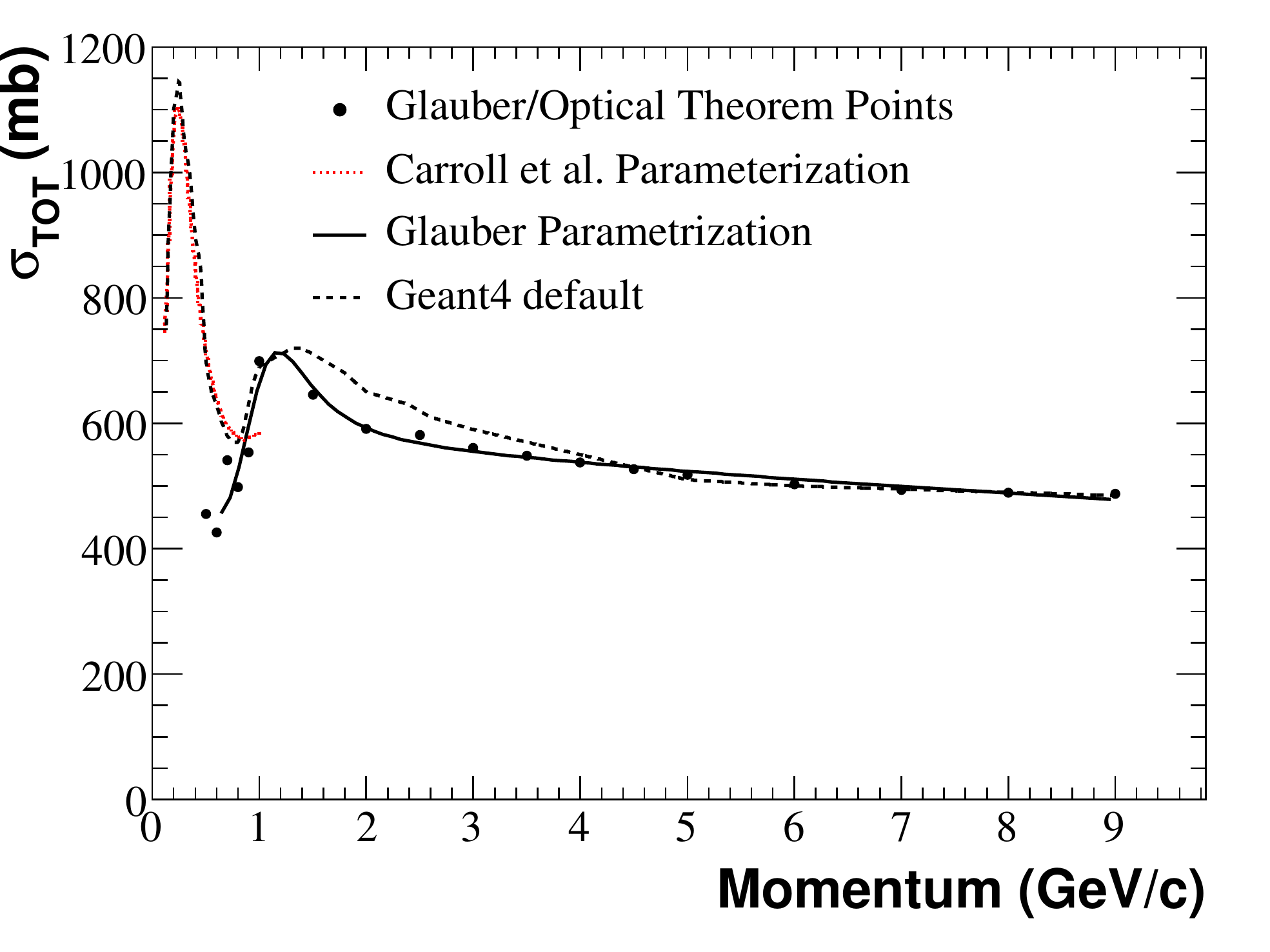}
\includegraphics[width=6.8 cm]{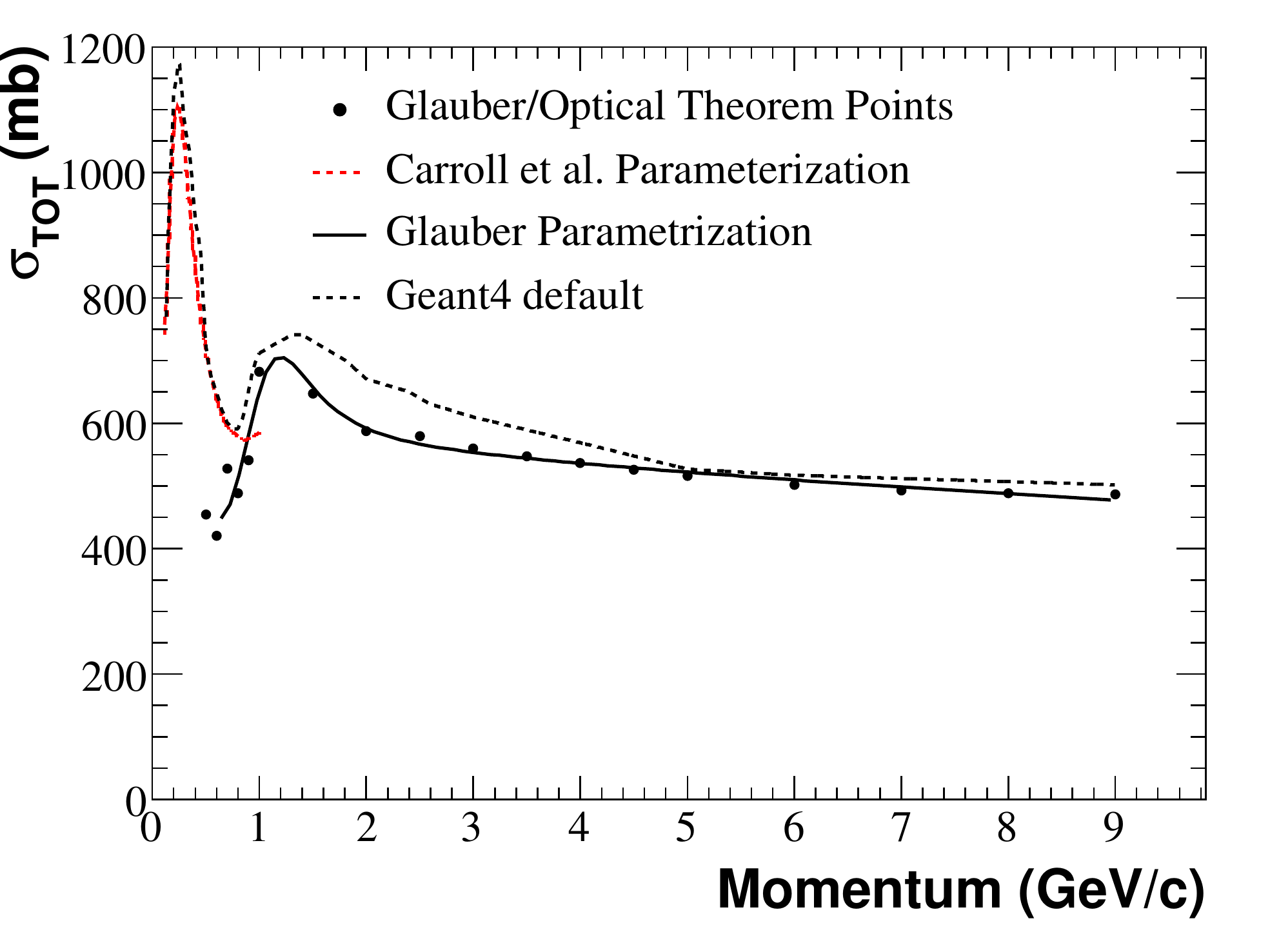}
\caption{\label{fig:pialtotxsec} Total hadronic cross sections for $\pipm$-Al
calculated using the Glauber model (points) for $\pip$ (left) and $\pim$ (right). 
The Breit-Wigner parametrization based on the Carroll data \cite{carroll}
on the $\Delta(1232)$ resonance is shown as a dotted line, while the parametrization
of the Glauber model points is shown as a solid line. The Geant4 default model
is shown as a dashed line.}
\end{center}
\end{figure}

The resulting total cross sections for nucleon-nucleus scattering (beryllium and aluminum) are shown
in Figure \ref{fig:nuctotxsec}. The calculated values of $\stot$ are compared with 
measurements of $\stot$ for $\mbox{n}$-Be data. The model predictions agree with the data
to within several percent, and indicate that $\stot$ for proton-nucleus and neutron-nucleus
interactions are very similar except at the lowest energies, as expected from isospin symmetry.
The success of the  model in reproducing $\stot$ neutron-nucleus is taken as an indication
that the model can be used for proton-nucleus and pion-nucleus interactions, where such
a check with data is not possible. The spread in values between the data and the model is
considered a source of systematic uncertainty.

The $\stot$ values obtained for pion-nucleus interactions are shown in Figure \ref{fig:pibetotxsec}
for $\pipm$-Be interactions and Figure \ref{fig:pialtotxsec} for $\pipm$-Al interactions. The calculated
points are parametrized by the black curve. At low momentum ($<600\mevc$ for beryllium, $<800\mevc$ for aluminum), 
where the $\Delta$ resonance dominates the cross section, parametrizations based on $\stot$ measurements 
by Carroll {\em et al.} are used \cite{carroll}. 
While not used in the flux prediction, the $\stot$ values used in the GHEISHA model (the Geant4
default) are shown as a dashed black line for comparison.

\begin{figure}[t]
\begin{center}
\includegraphics[width=14 cm]{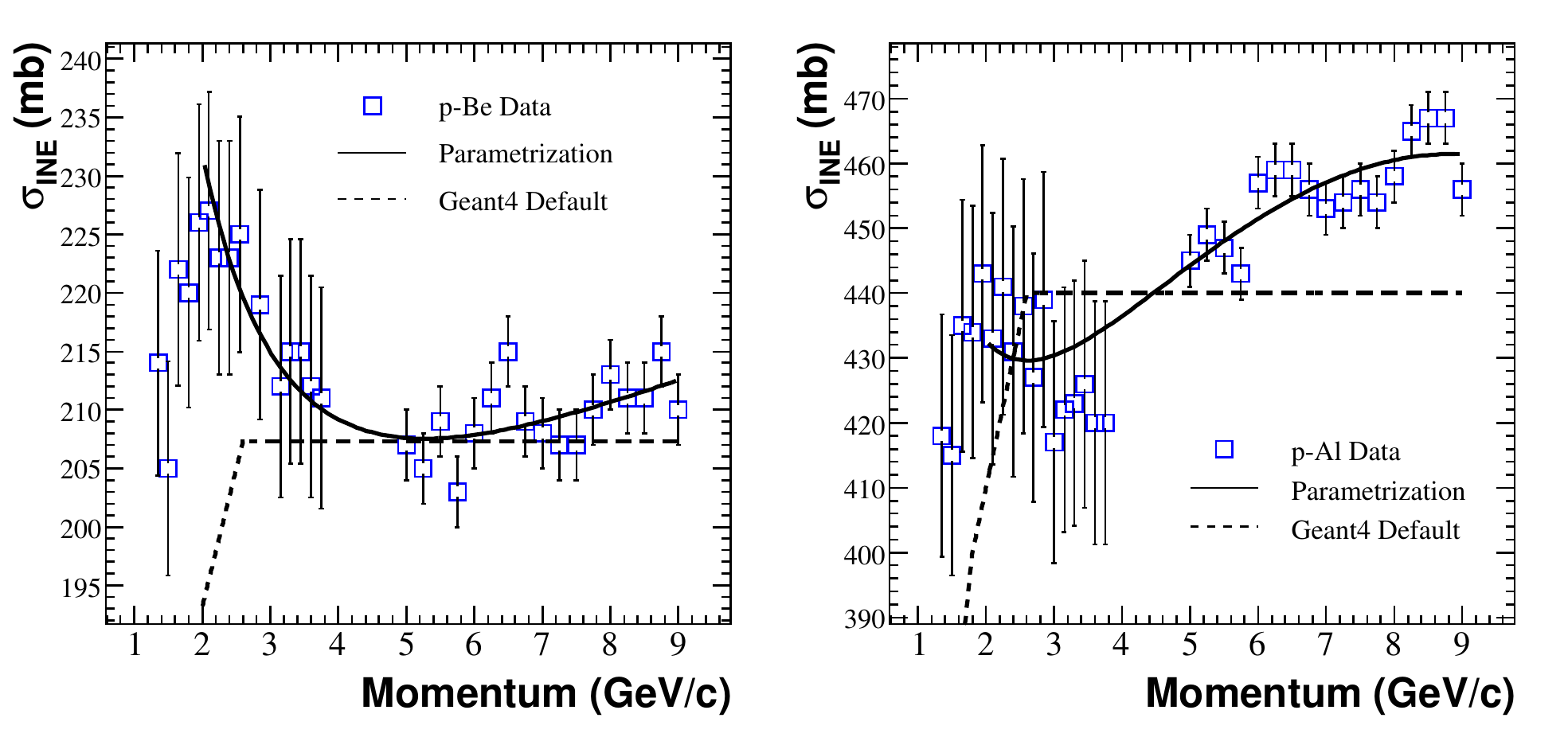}
\caption{\label{fig:nucleoninelastic} $\mbox{p}$-Be (left) and $\mbox{p}$-Al (right) inelastic
 cross sections measured from Gachurin {\em et al.} \cite{gachurin} $(1-4\gevc)$ and Bobchenko {\em et al.}\cite{bobchenko} $(5-9\gevc)$. 
The solid line is the parametrization used in the flux prediction, while the dashed line shows the Geant4 default parametrization. }
\end{center}
\end{figure}

 \begin{figure}[h]
\begin{center}
\includegraphics[width=6.8 cm]{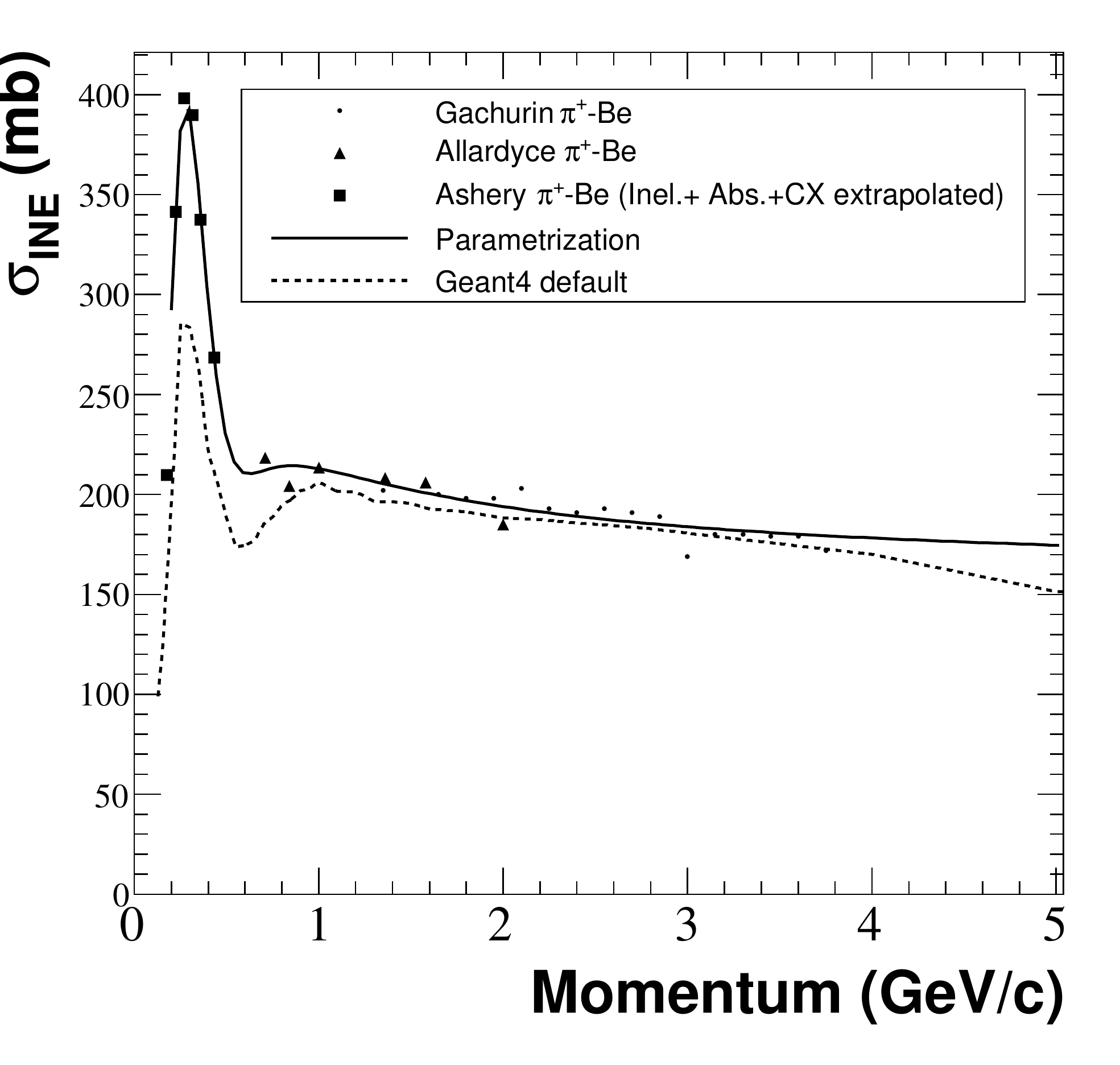}
\includegraphics[width=6.8cm]{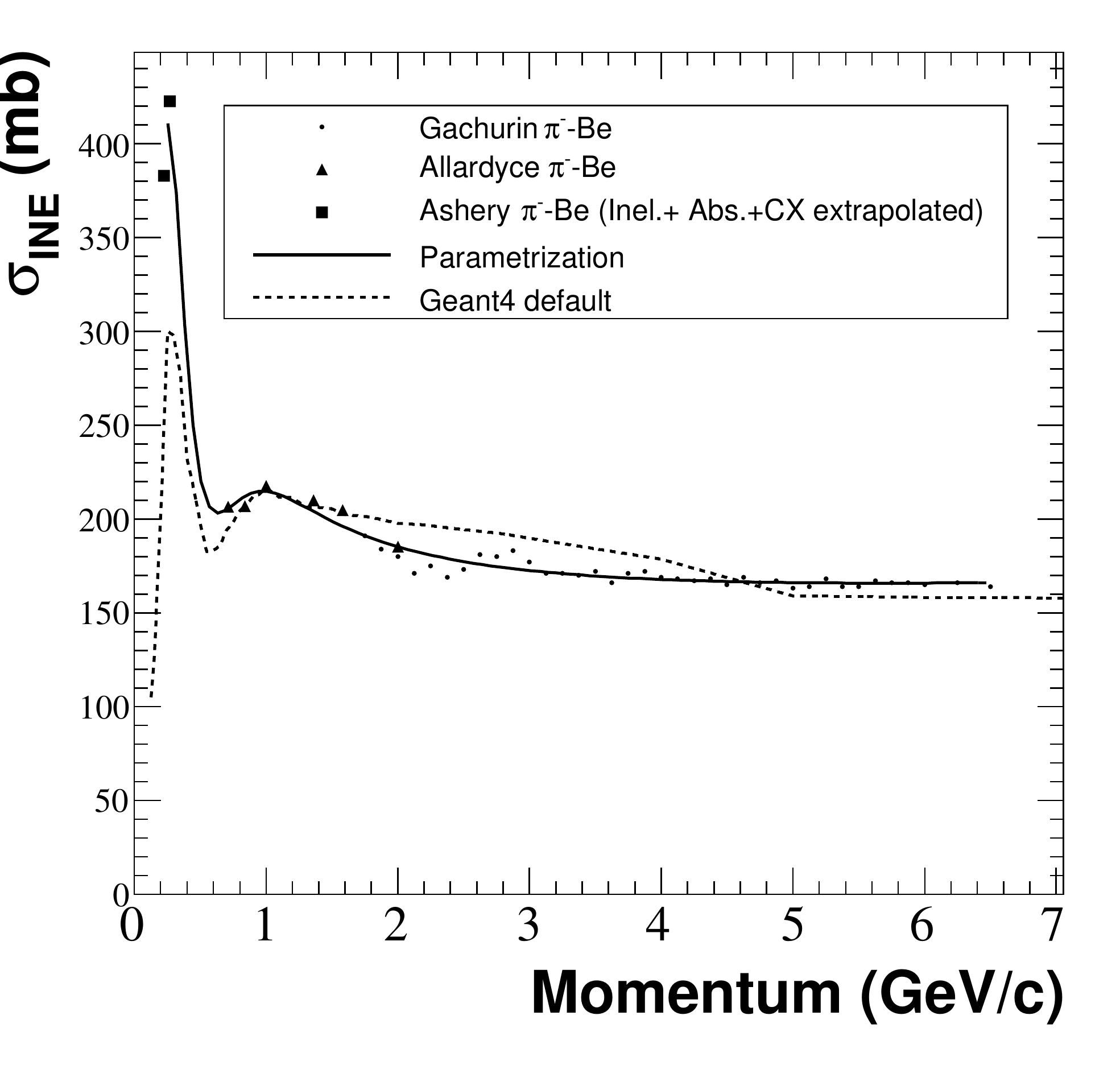} \\
\includegraphics[width=6.8 cm]{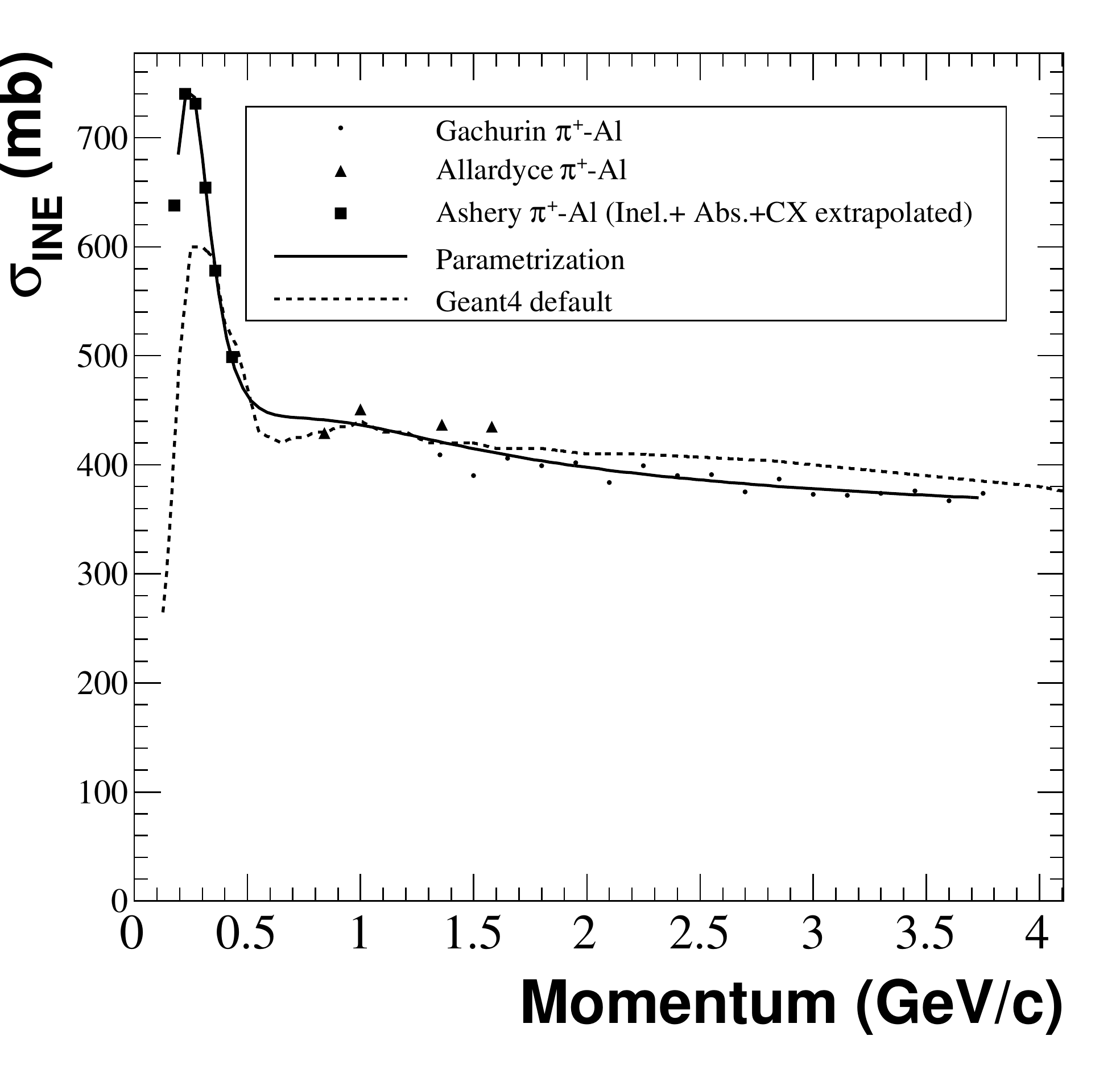}
\includegraphics[width=6.8 cm]{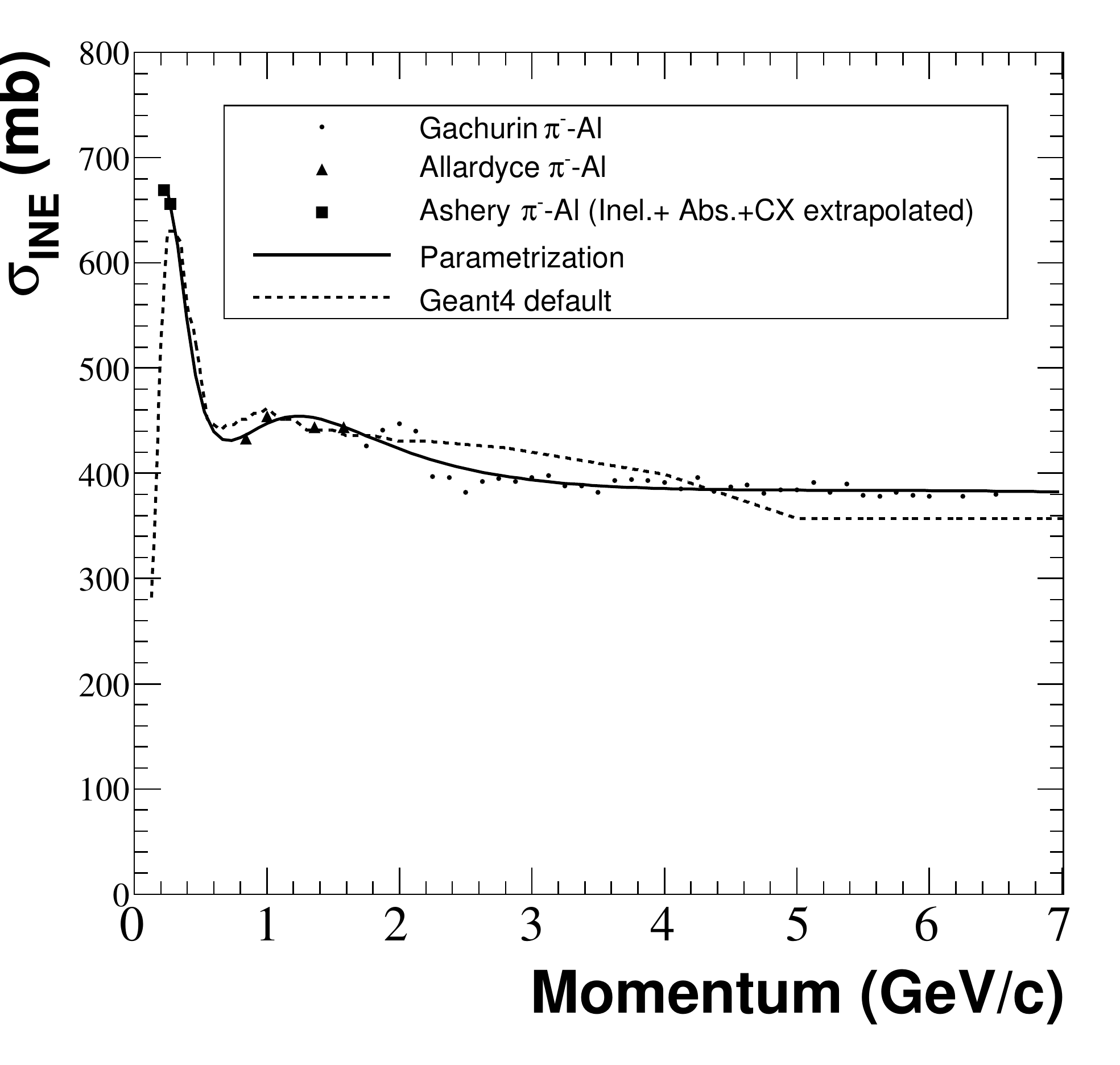}
\caption{\label{fig:pipminelastic}Inelastic cross sections for $\pip$-Be (top left), $\pim$-Be (top right),
$\pip$-Al (bottom left) and  $\pim$-Al (bottom right) as
measured in References \cite{ashery} (squares), \cite{allardyce} (triangles) and \cite{bobchenko} (circles).  
The solid lines are the parametrizations used in the flux prediction, while the dashed lines  
are the default Geant4 parameterizations.}
\end{center}
\end{figure}

\begin{figure}[t]
\begin{center}
\includegraphics[width=14 cm]{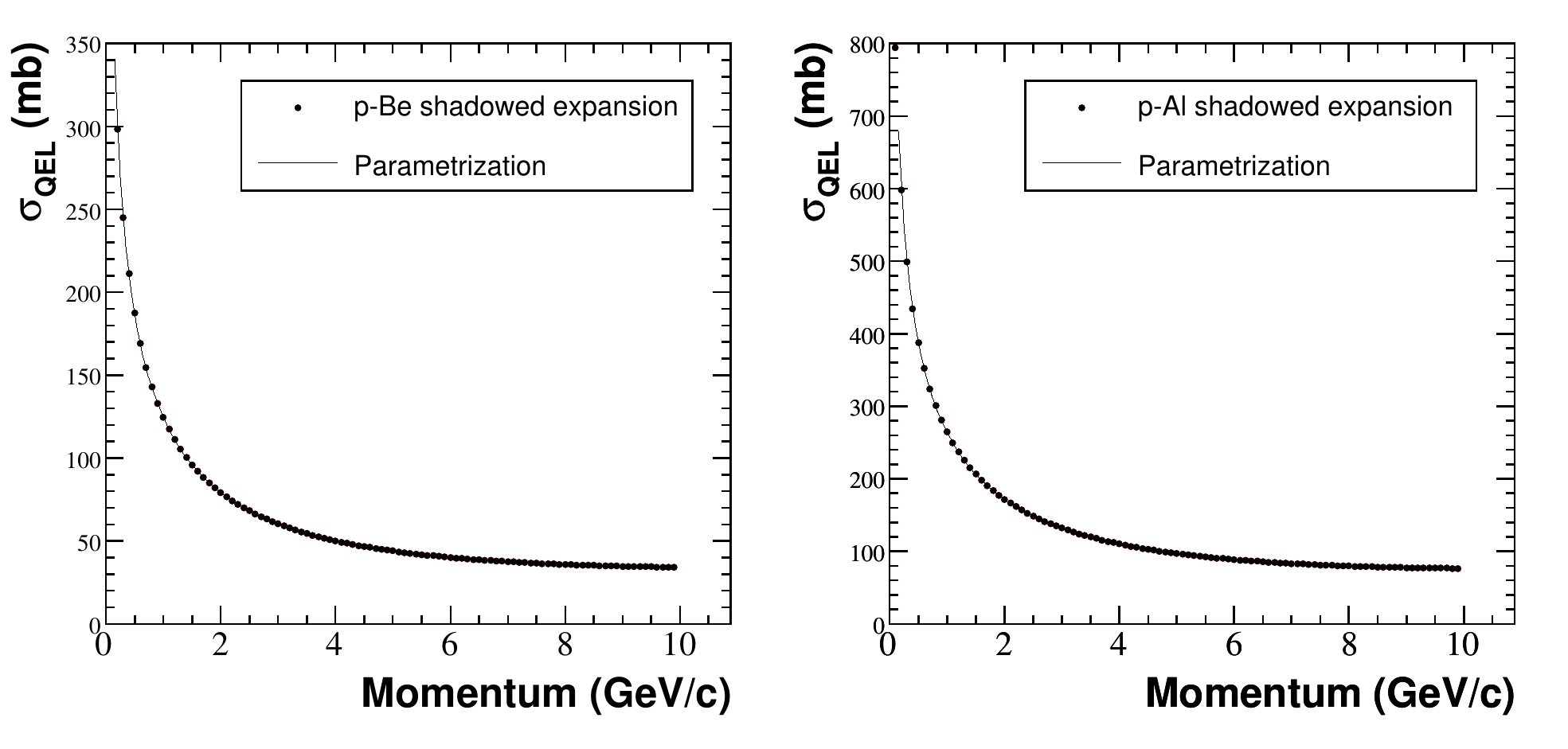}
\caption{\label{fig:nucleonquasielastic} Calculated values of $\sqel$ for $\mbox{p}$-Be (left) and $\mbox{p}$-Al (right) interactions along with
the parametrization used in the flux prediction.}
\end{center}
\end{figure}

 \begin{figure}[h]
\begin{center}
\includegraphics[width=6.8 cm]{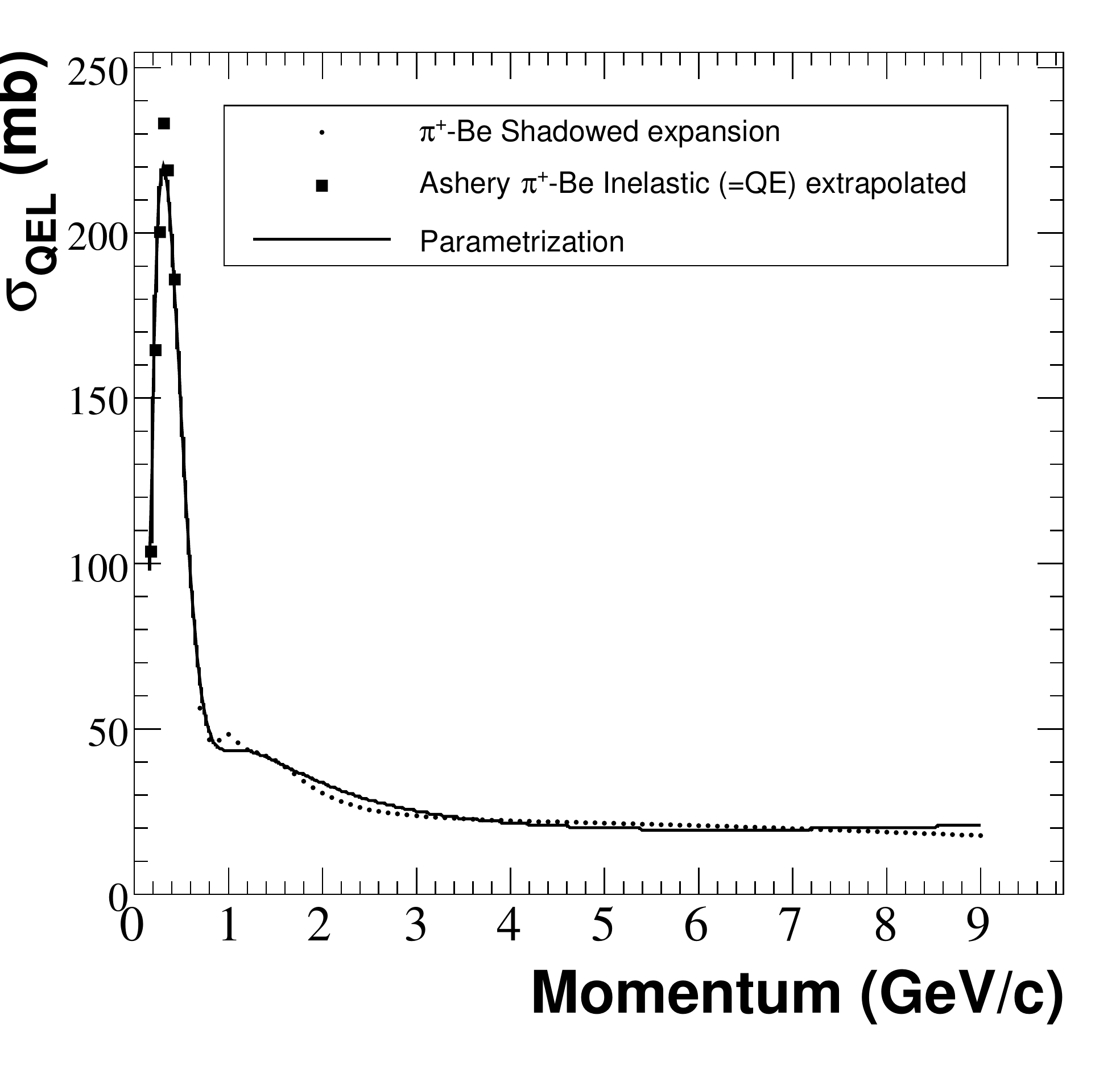}
\includegraphics[width=6.8 cm]{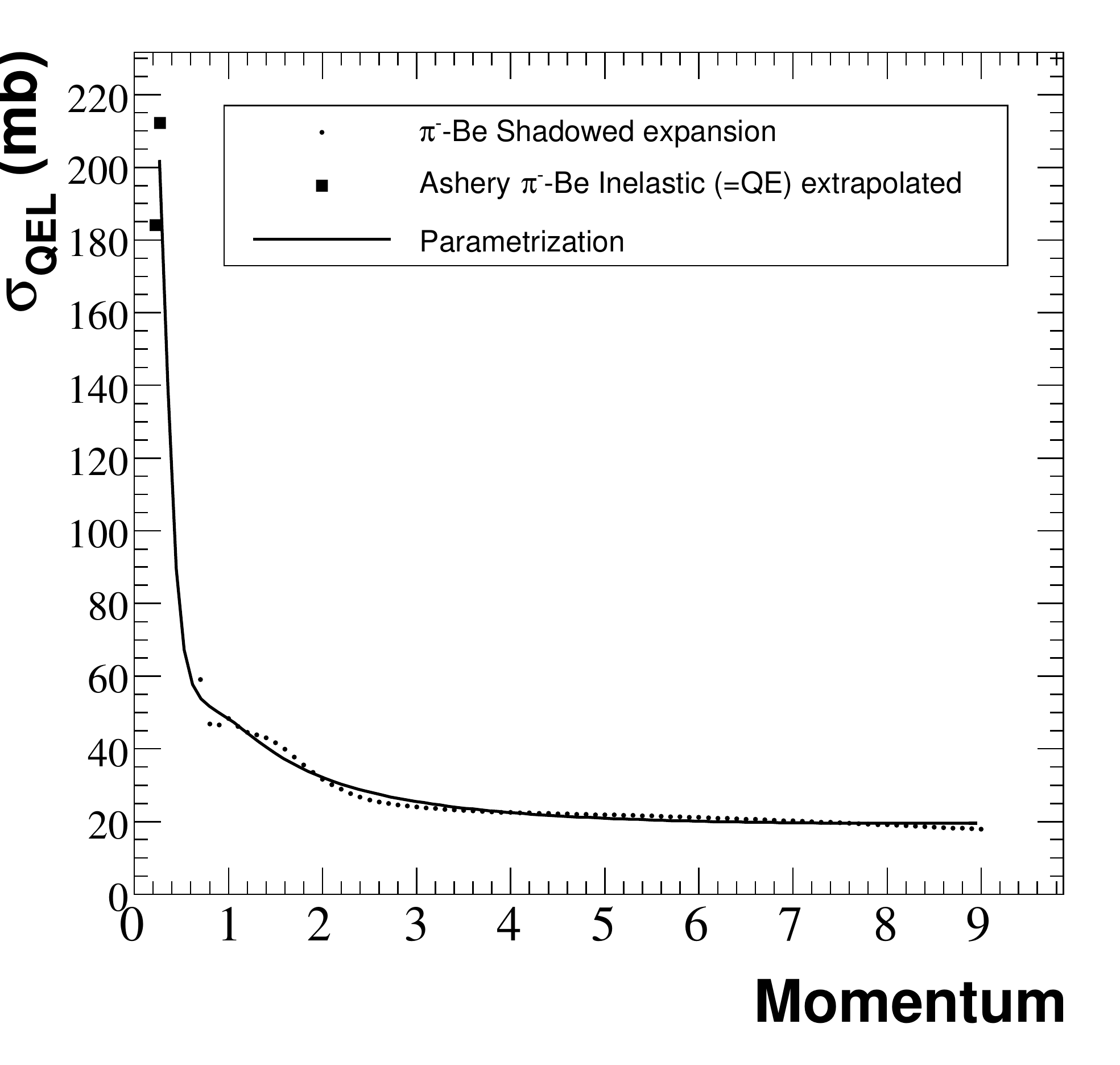} \\
\includegraphics[width=6.8 cm]{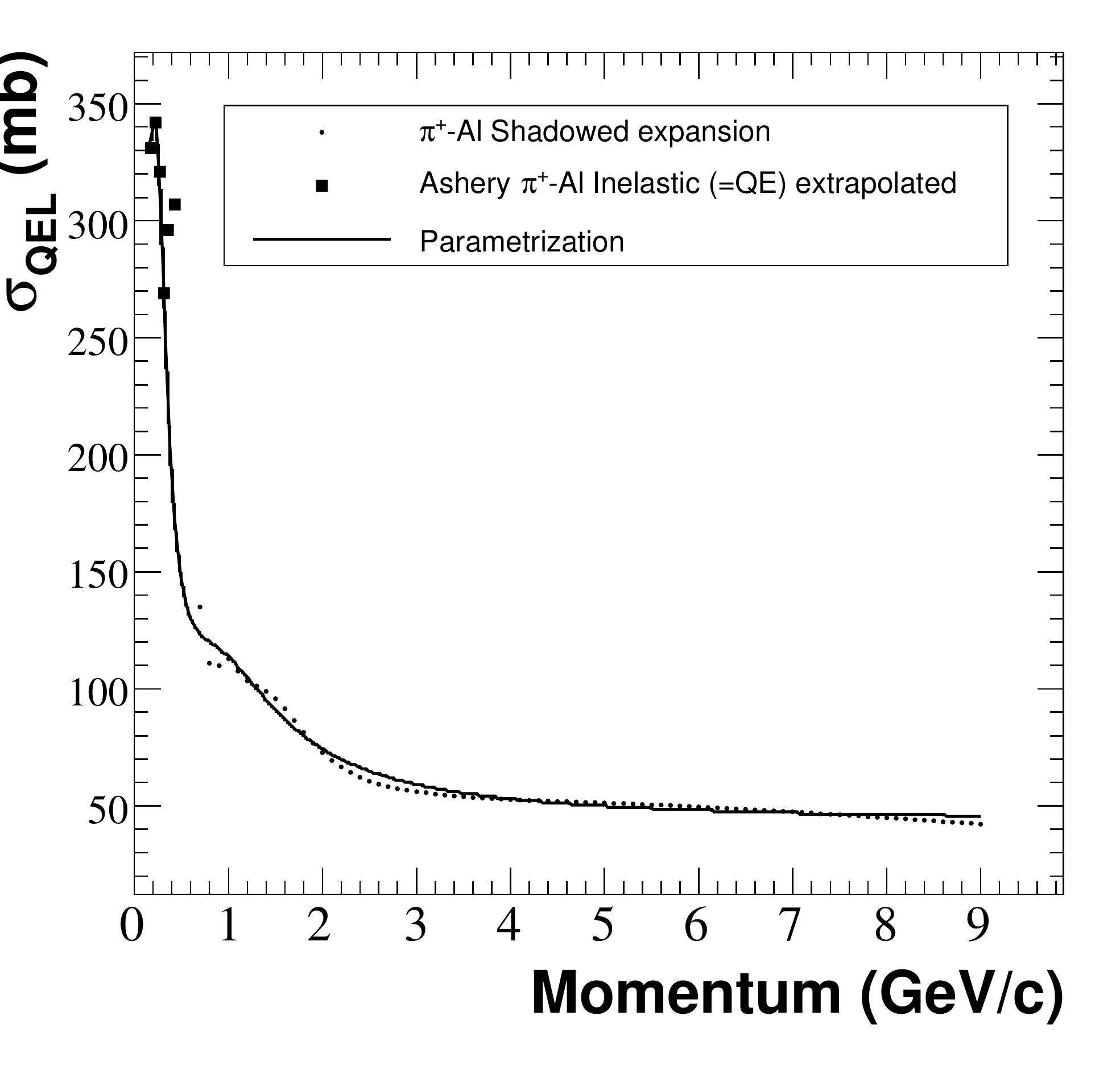}
\includegraphics[width=6.8 cm]{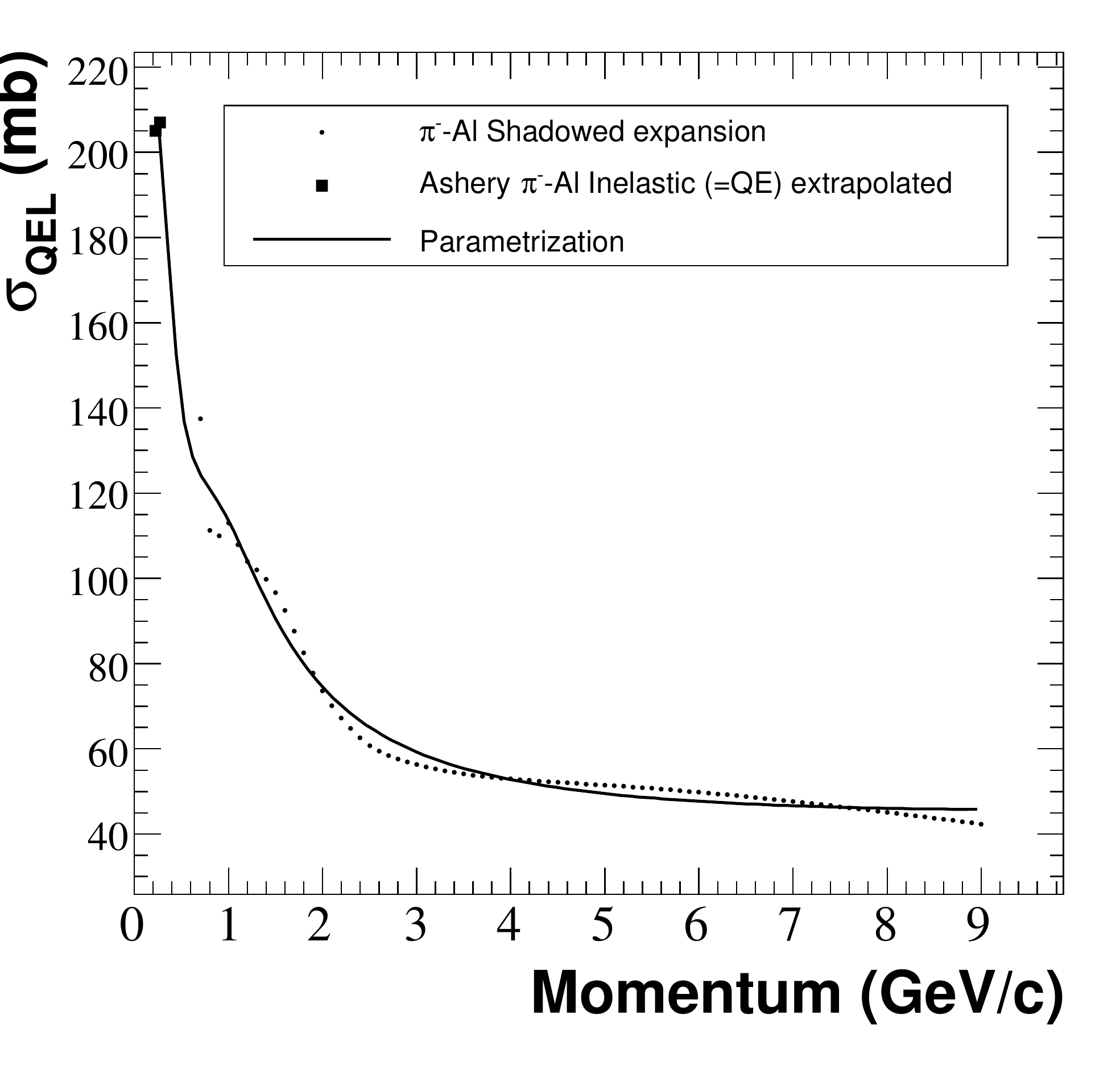}
\caption{\label{fig:pipmquasielastic}Quasi-elastic cross sections for $\pip$-Be (top left), $\pim$-Be (top right),
$\pip$-Al (bottom left) and  $\pim$-Al (bottom right) as
measured in References \cite{ashery} (squares) and calculated using the shadowed scattering model
(circles). The solid black lines are the parametrizations used in the flux prediction.}
\end{center}
\end{figure}

\subsection{Inelastic and Quasi-elastic Processes}
\label{sec:ine}
In the case of inelastic scattering ($\sine$), a much larger set of cross section measurements  exists 
eliminating the need for theoretical models. The entire momentum range
of interest for nucleon-nucleus inelastic scattering and a large subset of the momentum 
range for pion-nucleus inelastic scattering has  been measured. 

The available measurements of $\sine$ for p-Be and p-Al interactions
in the relevant momentum range are shown in Figure \ref{fig:nucleoninelastic}. The Gachurin {\em et al.} data \cite{gachurin} spans the low
momentum region, while the Bobchenko {\em et al.}\cite{bobchenko} data covers the high momentum region
up to $9\gevc$. Together, they cover the entire momentum range of interest for MiniBooNE. The 
parameterization used to model the 
momentum dependence is shown as a solid black line. Likewise,  $\sine$ for
$\pipm$-nucleus interactions is shown in Figure \ref{fig:pipminelastic}. The Ashery {\em et al.}
\cite{ashery} measurements are used around the $\Delta$ resonance, while the Allardyce {\em et al.}
\cite{allardyce},  Gachurin {\em et al.} \cite{gachurin}, and Bobchenko {\em et al.} \cite{bobchenko} data are used at higher momentum. The low momentum data do not include beryllium; for these points,
the cross sections are extrapolated using the cross sections measured on different elements at
the same momentum. The measured cross sections are parametrized as $A^n$ based on the $A$-dependence
of the cross sections, with typical values of $n$ ranging from $0.6$ to $0.8$. The 
resulting function is used to infer the cross section at $A=9$.

A subset of the inelastic interactions results from quasi-elastic scattering, where hadrons scatter
off the individual nucleons in the nucleus in a manner analogous to hadron elastic scattering off
free nucleons.
The rate of this process relative to other forms of inelastic scattering is important since it  allows
the incoming hadron to emerge from inelastic scattering with its initial momentum largely intact, 
whereas it would otherwise have been effectively absorbed or significantly reduced in momentum. 

Unfortunately, the available measurements of $\sqel$ are sparse, with only a few measurements for pions at low momentum. As a result, we must appeal
to a theoretical calculation for this part of the inelastic cross section. This can be effected via
the shadowed multiple scattering expansion, in which  $\sqel$ is calculated as the incoherent sum
of the cross section for hadrons to scatter off the nucleons in the nucleus, accounting for the
attenuation of the hadron wavefunction as it traverses through the nucleus \cite{glauber}. The cross section for
multiple scattering of the hadron within the same nucleus can also be calculated in this formalism.
This is found to be a small fraction of the single-scatter cross section in our case. 

The calculated values of $\sqel$ for nucleon-nucleus quasi-elastic scattering 
are shown in 
Figure \ref{fig:nucleonquasielastic}, while the values for pion-nucleus 
scattering from Reference \cite{ashery} are shown in Figure 
\ref{fig:pipmquasielastic}. The latter figure includes measurements of 
$\sqel$ for $\pipm$-nucleus interactions around the $\Delta$ resonance. The 
calculated values, along with the measurements, have been incorporated into 
the parametrizations of the momentum dependence of $\sqel$ for each of the 
hadron/nucleus combinations. As before, these measurements do not exist for 
beryllium and  have been extrapolated assuming an $A^n$ dependence, where $n$
 has been  determined from the $A$-dependence of the measured cross section at
each momentum from Reference \cite{ashery}. The resulting values of $n$ range 
from 0.3 to 1.0, varying with incident pion momentum. 
% For $\pipm$-(Be/Al) interactions, the only measurement of quasi-elastic is from the Ashery {\em et al.} \cite{ashery}
%on the $\Delta(1232)$ resonance as shown in Figure \ref{fig:pipminelastic}. Once again, the
%beryllium cross sections have been extrapolated from the measurements on other elements.

\begin{table}[ht]
\begin{center}
\begin{tabular}{c|ccccc} \hline\hline
		& $A$	&	$B$	 & $n$	   & $C$            & $D$          \\ \hline
$(\mbox{p/n})$-Be&  		&		&		&			&       \\
$\stot$  	& 307.8 	&  0.897	& 0.003  &-2.598  &-4.973     \\
$\sine$  	& 186.7 	&  104.3 	&-1.039 	   &10.38   &-15.83   \\
$\sqel$   	& 164.8 	& -40.09   & 0.408    & 21.40  	& -61.45   \\ \hline
$(p/n)$-Al	&		&		&		&		&		\\
 $\stot$     	& 760.3	&-0.056   	& 2.485    	& 6.173	& -41.60 \\
 $\sine$   	& 470.9 	&-0.259    & 2.429	& 48.86  	& -87.19 \\
 $\sqel$	& 255.7 	& 8.792    & 0.0024   & 32.24  	&-155.9 \\ \hline
\end{tabular}
\vskip 0.2 cm
\caption{\label{tab:nucleonpar} Parameter values for proton and neutron on beryllium and aluminum  
cross sections using Equation \ref{eq:regge}}
\end{center}
\end{table}

\begin{table}[gt]
\begin{center}
\begin{tabular}{l|ccccccccc} \hline\hline
            	& $\theta_0$ & $\theta_s$ & $A$     &   $B$   & $n$     & $C$    &  $N_R$ & $M_R$ & $\Gamma_R$ \\ \hline
$\pip$-Be &		&		&		&		&	       	&              &              &                &                            \\
$\stot$	& 0.814	& 3.418 	& 237.6	& 111.3	&-4.186 & -9.792  &     --       &  --   &   --       \\ 
$\sine$ 	& 0.400     & 5.142   	& 162.3 	& -99.79 	& -2.407  	& -0.423& 850.3 & 1.201& 0.375     \\
$\sqel$ 	& 0.635     & 3.784    & -2.38   	& -81.84 	& -2.702  	& 3.173 & 379.9 & 1.201& 0.558     \\ \hline
$\pip$-Al 	&		&		&		&		&		&               &               &              &              \\
$\stot$ 	& 0.931	& 3.186 	&569.1  	& 511.3 	& -3.79   & -18.50&     -- &  --   &   --       \\
$\sine$ 	& 0.295  	& 2.307	&1537.4  	&-1109.4 & 0.057   & 14.40 & 510.7 & 1.189& 0.185     \\
$\sqel$ 	& 0.698	& 2.134	& 40.38   	& 89.20  	& -1.575  & 0.335 & 229.4  & 1.189& 0.187      \\ \hline
\end{tabular}

\vskip 0.5 cm
\begin{tabular}{l|ccccccccc} \hline\hline
                    & $\theta_0$ & $\theta_s$ & $A$     &   $B$   & $n$     & $C$  &  $N_R$ & $M_R$ & $\Gamma_R$ \\ \hline
$\pim$-Be &		&		&		&		&	       	&              &              &                &                            \\ 
$\stot$ 	& 0.814	& 3.418     & 237.6 	& 111.3 	& -4.186 & -9.792    &   --   &   --  &    --      \\
$\sine$ 	& 0.600   	& 2.874    	& 92.66 	& 112.2 	&-0.486  & 7.500     &371.5  & 1.201& 0.233      \\
$\sqel$ 	& 0.626   	& 2.504    	& -1.559  	& 46.41  	&-0.633  & 1.874     &189.0  & 1.201& 0.185      \\ \hline
$\pim$-Al &		&		&		&		&	       	&              &              &                &                            \\
$\stot$ 	& 0.931     & 3.186   	& 569.1 	& 511.3 	&-3.79    & -18.50   &   --   &   --  &    --       \\
$\sine$ 	& 0.706     & 1.685   	& 997.8 	&-457.8 	&0.611   & 233.4    & 446.8 & 1.189& 0.305      \\
$\sqel$ 	& 0.633     & 2.199   	& 32.52  	& 85.15  	&-1.225   & 1.383  &129.1  & 1.189& 0.305      \\ \hline
\end{tabular}
\vskip 0.2 cm
\caption{\label{tab:pionpar} Parameter values for $\pipm$-(Be/Al) hadronic cross sections. For $\sine$ and $\sqel$,
Equation \ref{eq:reggebw} is used. For $\stot$, the parametrization of Carroll {\em et al.}\cite{carroll} is used at low momentum,
while Equation \ref{eq:regge} with a threshold term is used at high momentum.}
\end{center}
\end{table}

\subsection{Explicit forms for the Cross Section Parametrizations:}
\label{sec:hadpar}
In summarizing the momentum dependence of the nucleon and pion cross sections,
we have made use of the following form:
\begin{equation}
\label{eq:regge}
\sigma =  A + B \times  p^n  + C \times \ln^2 p + D \times \ln p 
\end{equation}
where $p$ is the momentum of the incident particle in $\gevc$.
While this form is inspired by Regge theory\cite{pdg1996}, it is used as a purely
empirical description of the cross section. No physical significance is attributed
to the parameters apart from the ability of the parametrization to describe
the measured or calculated cross sections with the appropriate parameters. The
parameters used in the flux prediction for proton and neutron hadronic cross sections
on beryllium and aluminum are given in Table \ref{tab:nucleonpar}.

For $\sine$ and $\sqel$ in pion-beryllium/aluminum scattering, a more complicated form 
is needed in order to describe the peak in the cross section near the $\Delta$ resonances: 
\begin{equation}
\begin{split}
\label{eq:reggebw}
\sigma ~ = ~&   N_{R} \left| \frac{-m(p) \Gamma_{R}}{M^2_{R} - m(p)^2 + im(p) \Gamma_{R}} \right|^2   \\
	    &~ +~  \left[\;1+ \tanh(\theta_s (p-\theta_0))\;\right]\times \left(A + Bp^n + C \ln^2 p \right )\\  
\end{split}
\end{equation}
where $p$ is the momentum of the incident particle in $\gevc$. 
The first term describes a Breit-Wigner resonance, where $m(p)$ is the invariant
mass of the pion/target nucleon system in $\gevcc$ assuming that the target is a nucleon 
at rest. The second term is 
a simplified version of Equation \ref{eq:regge} with a threshold behavior described by
a hyperbolic tangent function. The threshold function allows the second term
to dominate at pion momenta above the $\Delta(1232)$ resonance. Here also, the approach
is purely empirical; the parameters, including the resonance terms are extracted in such
a way to reproduce as closely as possible the measurements, without assigning any physical
significance to any of the parameters. In particular,
the various $\Delta$ resonances are not modeled individually. The parameters used in the
flux simulation using Equation \ref{eq:reggebw} are given in Table \ref{tab:pionpar}.

As mentioned in Section \ref{sec:sela}, the parameterization of Carroll {\em et al.}\cite{carroll}
is used
for the total cross sections on $\pipm$  scattering on beryllium and aluminum for momenta up
to $600\mevc$ in the former case and $800\mevc$ in the latter. At higher momentum, the second
term of Equation \ref{eq:reggebw} is used with the parameters shown in Table \ref{tab:pionpar}

\newpage

\section{Secondary Particle Production Cross Sections}
\label{sec:secondary}
The primary source of the neutrino flux at MiniBooNE is the decay of secondary particles produced
in $\mbox{p}$-Be interactions. The knowledge of the neutrino flux thus critically depends on
the understanding of the meson production in $\mbox{p}$-Be interactions. Most of the
$\num$ flux at the MiniBooNE detector comes from  $\pip\rightarrow \mup + \num$ decays,
while the $\nue$ flux is dominated by three-body decays of kaons ($\Kp$ and $\KL$)
and the decay of muons (primarily produced in the decay of pions). The tables in the
Monte Carlo simulation describing the double differential cross sections which specify the multiplicity
and kinematic properties of the protons, neutrons, $\pipm$, $\Kpm$ and $\KL$ produced in
$\mbox{p}$-Be interactions at $8.89\gevc$  are based
on hadron production measurements with similar kinematic configurations wherever possible.
In the case of $\pipm$, $\Kp$ and $\Kz$ production, the double differential cross sections
are summarized as parametrizations. The parametrizations are evaluated at each point
within the table to determine the corresponding cross sections. For protons, neutrons,
and $\Km$, the cross sections are based on a model of hadronic interactions.

\begin{figure}
\begin{center}
 \includegraphics[width=140mm]{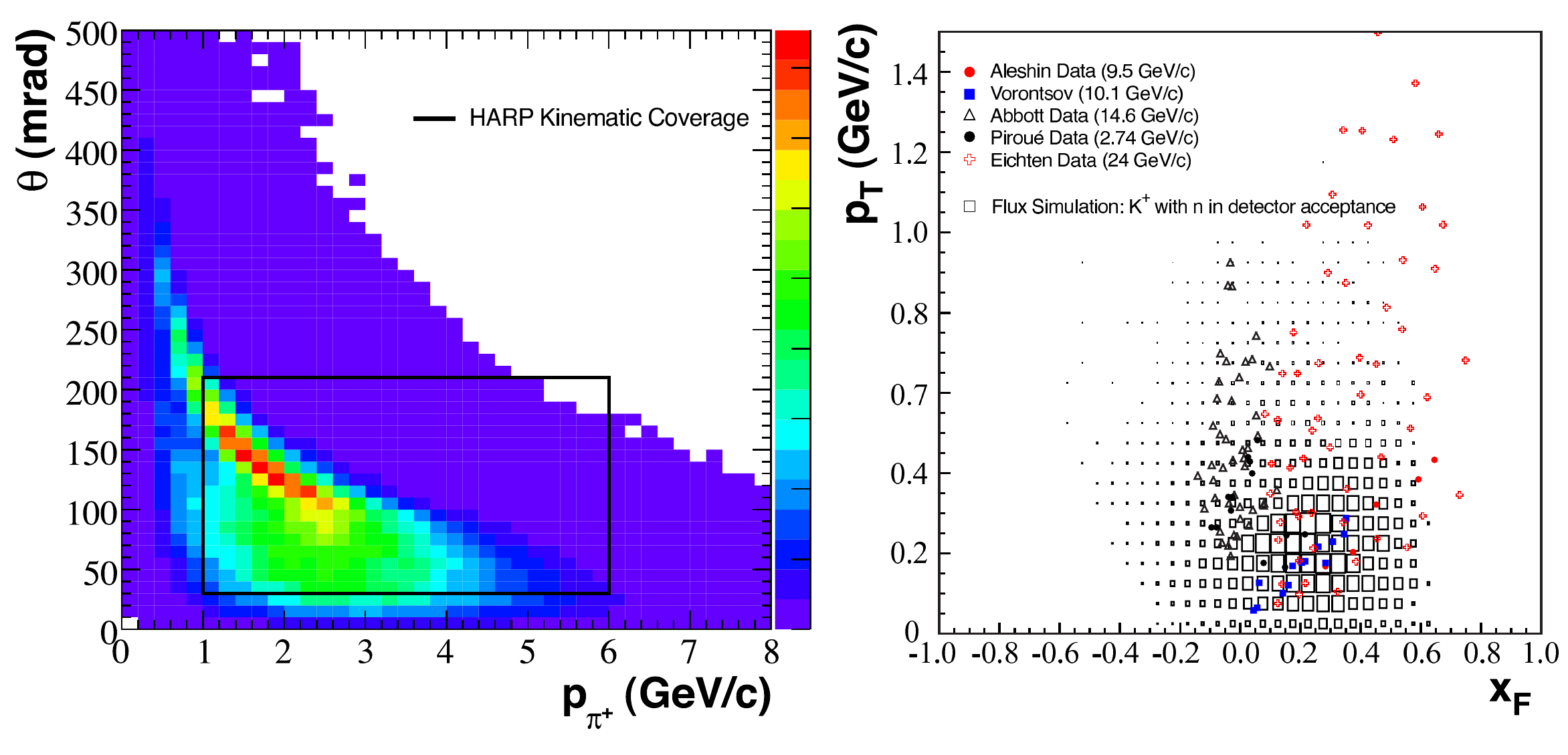}
     \caption{
 \label{fig:prodcoverage} Left: Production angle $\theta$ vs. momentum $p$
for the $\pip$ in the flux simulation that contribute to the $\num$ flux at the
MiniBooNE detector. The color scale indicates the relative cross section for $\pip$
production in each bin of angle and momentum. The black box marks the kinematic range covered by
the HARP measurements \cite{harp}. Right: Transverse momentum $p_T$ vs. the Feynman-scaling
variable $x_F$ for the $\Kp$ in the flux simulation that contribute to the neutrino
flux at the MiniBooNE detector (black squares). The colored points indicate
the kinematic regions measured by $\mbox{p}$-Be $\Kp$ production measurements \cite{jocelyn}. 
}
\end{center}
\end{figure}

\subsection{Pion Production Measurements}
The cross section tables for $\pipm$ production in p-Be interactions are based on parameterizations
of measurements taken by the HARP \cite{harp} and BNL E910 \cite{e910} experiments.
The HARP experiment measured the \pipm differential production cross section for 
$\mbox{p}$-Be interactions using replicas of the MiniBooNE beryllium target at the same 
incident proton momentum of 8.89 \gevc. However, since the analysis of the data from the 
replica targets is not complete, the data used in modeling the pion production
is from the thin target run, where  a $5\%$ interaction length beryllium 
target was measured. The pion tracks are binned in total pion momentum $p_\pi$ ranging from 
0.75 to 6.5 \gevc and angle $\theta_\pi$ with respect to the incident proton direction from 
30 to 210 \mrad. The measurements from the experiment represent the average 
differential cross section over the bin. A complete covariance matrix is also reported to  
account for bin-to-bin correlations in the uncertainties.  The quoted normalization 
uncertainty is $\sigma_{HARP}=2\%$. 

The left plot of Figure \ref{fig:prodcoverage} shows
the kinematic distribution (in terms of $\theta_\pi$ and $p_\pi$) from the Monte Carlo simulation
of pions produced in the target that decay to produce  neutrinos in the MiniBooNE detector. The black square indicates the kinematic range covered by the HARP measurements.

The BNL E910 experiment measured the \pipm differential cross section for 
$\mbox{p}$-Be interactions at three different energies of the incident protons (6.4, 12.3,
17.5 and 17.6 \gevc). 
Data is binned in pion momentum $p_\pi$ ranging from 0.4 to 5.6 \gevc
 and angle $\theta_\pi$ from 18 to 400 \mrad. The extended coverage of the E910 measurements
in the forward angular region provides further constraints of the pion production in this
kinematic region.
A covariance matrix was not reported for these measurements, hence we use  
a diagonal bin-to-bin covariance matrix. The quoted normalization error is $\sigma_{E910} = 5\%$.

The $\pip$ production cross sections (momentum distribution of pion production
in bins of production angle) from the two experiments are shown in Figures \ref{fig:harpfit}
(HARP $8.89\gevc$), \ref{fig:e910fit6} (E910 $6.4\gevc$), and \ref{fig:e910fit12} 
(E910 $12.3\gevc$).
The corresponding $\pim$ production measurements are shown in Figures \ref{fig:harpfitpm},
\ref{fig:e910fit6pm} and \ref{fig:e910fit12pm}. While a significant body of historical 
$\mbox{p}$-Be pion production data exists
(\cite{chopi,vorontsovpi,allabypi,marmerpi,aleshin}), the measurements are removed from the 
primary beam momentum in the BNB, have insufficient kinematic coverage, or have inconsistencies 
that led to the exclusive  use of the latest data from E910 and HARP. The E910 $17.5\gevc$ and $17.6\gevc$ data 
is also not used for the first reason.

%Figure \ref{fig:pipfsdata} shows the invariant cross section $E\frac{d^3 \sigma}{dp^3}$ for $\pip$ 
%production  as a function of $p_\pip$ in angular bins. The E910 measurements are rescaled using 
%Feynman-scaling  to correspond to the relevant $8.89\gevc$ proton momentum. This was achieved by %preserving the value of the Feynman-scaling variable $x_F=\frac{p_\parallel^{CM}}{p_\parallel^
%{max,CM}}$ and $p_t^{CM}$ for each data point.

\subsection{Sanford-Wang Fit to the Pion Production Data}

Following the K2K experiment \cite{k2kprd}, the parametrization of 
Sanford and Wang (SW) \cite{swpar} is used to describe the  $\pipm$ differential production 
cross section across different incident primary beam momenta.  The SW parameterization for the production cross section of a given meson species is given by
\begin{equation}
\frac{d^2 \sigma}{dp d\Omega}(p,\theta)=c_1 p^{c_2}\Big(1-\frac{p}{p_B - c_9}\Big) \exp\Big( -c_3 \frac{p^{c_4}}{p_B^{c_5}}-c_6\theta(p-c_7p_B \cos^{c_8}\theta) \Big)
\label{eq:sw}
\end{equation}
where $\frac{d^2 \sigma}{dp d\Omega}$ is the double differential cross section, 
 $p$ is the total momentum of the meson in $\gevc$, $\theta$
is the angle of the meson with respect to the incident proton in radians, $p_B$ is the 
momentum of the incident proton in $\gevc$, and $c_1, .., c_9$ are parameters to be determined in the
fit to the production data. For the fits to the pion production data, $c_9$ is set to unity; for the
kaon production fits (see below), it is a free parameter. 
The parametrization allows one to relate production data at different incident proton
energies, to smoothly interpolate the behavior of the cross section between
measured points, and to extrapolate into regions where production data do not exist.
Due to the strong correlation between the $c_3$, $c_4$,  and $c_5$ parameters, the value of  $c_3$
is fixed to unity for the $\pip$ production fit. For $\pim$, the $c_3$ parameter is initially floating, but then fixed to its initial best-fit value when the fit is iterated. The correlation results from the limited range of proton momentum covered
by the measurements (6.4, 8.9 and 12.3 $\gevc$). As a result, the data has limited ability
to constrain the cross section dependence on the proton momentum. The predicted pion production properties, however, are not affected by this indeterminacy.

The values of the parameters $c_i$ are determined from a fit to the $\pipm$ production
cross section data by minimizing the following $\chi^2$ function
\begin{equation}
\label{eq:swchi2}
% \chi^2=\sum_{i,j,k} \left(D_i - N_k T_i \right) {\bf V}_{ij}^{\bf -1}  \left(D_j - N_k T_j \right) + 
%  \sum_k \frac{(N_k -1)^2}{\sigma_k^2},
 \chi^2= \sum_k \left[ \left(\sum_{i,j} \left(D_{i,k} - N_k T_i \right) {\bf V}_{ij,k}^{\bf -1}  \left(D_{j,k} - N_k T_j \right)\right) + \frac{(N_k -1)^2}{\sigma_k^2}\right],
\end{equation}
where $D_{i,k}$ is the $i$-th data point for the $k$-th data set, $T_i$ is the value of the SW function for the kinematic
parameters for that data point, ${\bf V}_{ij,k}$ is the bin-to-bin covariance matrix for the
$k$-th data set,
$N_k$ is relative normalization fit parameter for data set $k$ and $\sigma_k$ is the quoted
normalization uncertainty for data set $k$. There are three data sets used in the fit, namely
HARP and E910 $6.4$ and $12.3 \gevc$. 
The normalization uncertainties for data sets at different beam momenta from the same experiment are the same and varied in common across these data sets during the fit procedure.

%where $D_i$ is the $i$-th data point, $T_i$ is the value of the SW function for the kinematic
%parameters of the $i$-th data point, ${\bf V}_{ij}$ is the bin-to-bin covariance matrix,
%$N_k$ is relative normalization fit parameter for data set $k$ and $\sigma_k$ is the quoted
%normalization uncertainty for data set $k$. A data sets correspond to production cross %section
%measurements from a particular experiment at a single primary energy. Typically,
%the normalization uncertainties for data sets at different energies from the same experiment %are the same.
%The MINUIT program \cite{minuit} is used to minimize the $\chi^2$.

%The normalization of measurements within each data set is allowed to vary freely within its normalization uncertainty through the relative %normalization parameters $N_k$  which is achieved by introducing the normalization pull terms  in the $\chi^2$ function. 
%The MINUIT program \cite{minuit} is used to minimize the $\chi^2$.

Since the data represent the average differential
cross section over a range of angle and momentum (``bin''), bin-centering corrections must be applied.
The bin-centering corrections are model-dependent since one must assume how
the production cross section varies within a bin. Here,  the SW parameterization is used to evaluate
the bin-centering correction
\begin{equation}
 \mbox{BCC}_{ij}=\frac{\mbox{SW}(p^c_i,\theta^c_j)(\Delta p_i \Delta \cos\theta_j)}{\int_{p_i}^{p_{i+1}} \int_{\theta_j}^{\theta_{j+1}} \mbox{SW}(p,\theta) \sin \theta dp d\theta},
\end{equation}
where $(p^c_i,\theta^c_j)$ is the center of bin $(i,j)$ in the $(p,\theta)$ space, $\mbox{SW}(p^c_i, \theta^c_j)$ is the double differential cross section returned by Equation \ref{eq:sw},
and $\Delta p_i$ and $\Delta \cos\theta_j$ are the bin widths.
The MINUIT fits are iterated with bin-centering corrections until convergence is achieved. 
Since we are concerned for the most part with pion production at $8.89\gevc$ proton
momentum, the dependence on the proton momentum is not important in predicting neutrino fluxes.

For the fit to the $\pip$ data, the minimized $\chi^2$/degree of freedom (DOF)  
using the reported experimental uncertainties is 1.8.  To obtain parameter uncertainties
the fit is performed with the covariance matrices scaled by this factor to obtain
the parameter uncertainties, resulting in an effective $\chi^2$ of unity.
The fit parameters are shown in the first row of Table \ref{tab:swpippar}.
The error matrix, shown below the parameters in Table \ref{tab:swpippar},  is obtained by varying 
the parameters in such a way that the resulting variations in the SW function cover the spread in the
data points. This corresponds to an envelope of parameter variations in which the resulting $\chi^2$ is within
$8.14$ of the minimum determined by the fit. While this  
corresponds to a 68\% confidence level parameter envelope for 7 parameters, the $\chi^2$ difference
is set by the desire to have the variations cover the deviations of the data points and their uncertainties.
The normalization factors obtained from the two data sets also are compatible within
the systematic uncertainties quoted by the two experiments ($N_{HARP}=0.966$, $N_{E910}=1.048$).

\begin{figure}
\begin{center}
 \includegraphics[width=150mm]{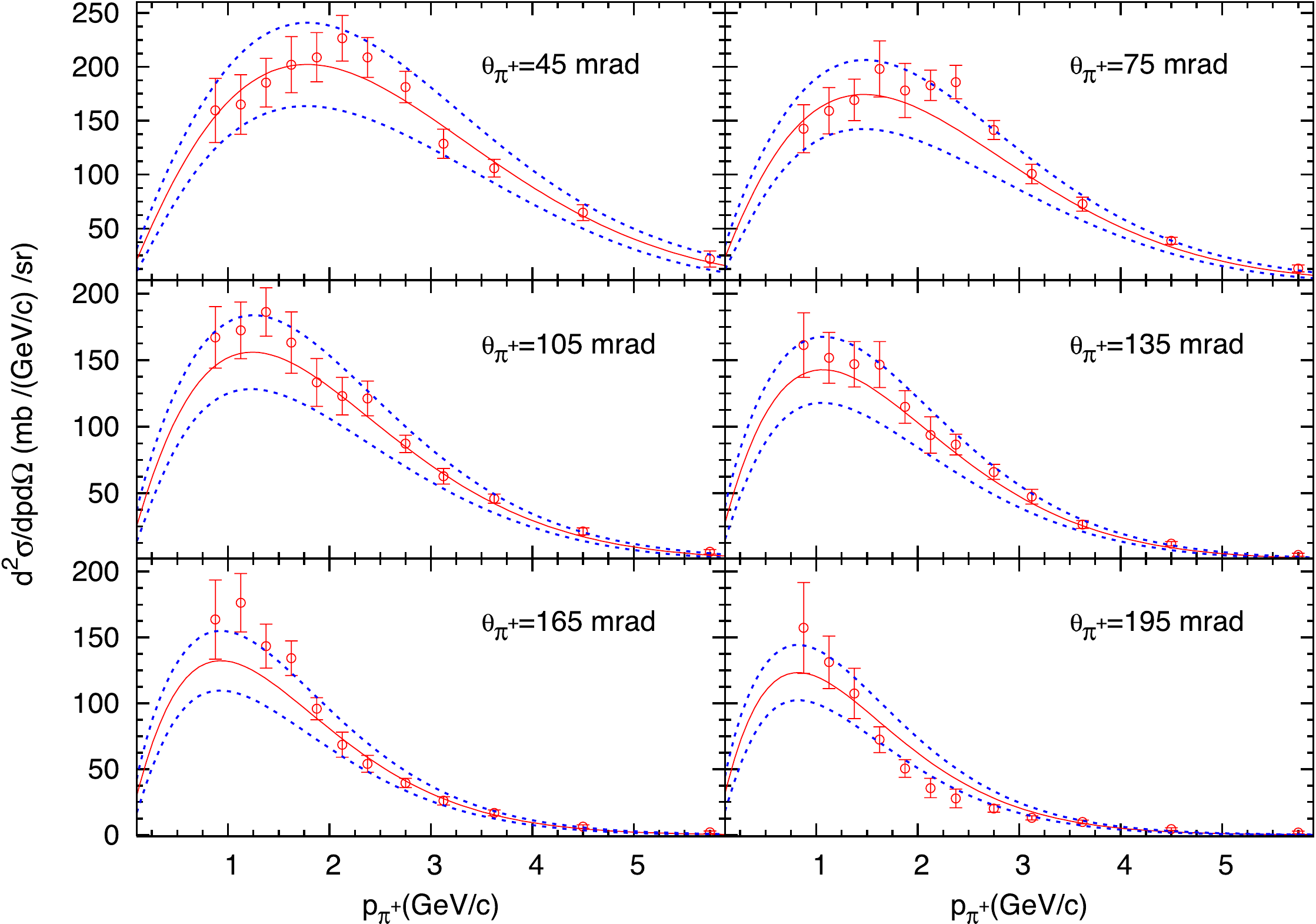}
 % e910_1norm_fracerr.eps: 0x0 pixel, 300dpi, 0.00x0.00 cm, bb=
     \caption{Comparison of HARP \pip production cross section data  \cite{harp}(circles)
		versus $p_\pi$ in bins of $\theta_\pi$ 
              from 8.89 \gevc $\mbox{p}$-Be interactions and best fit SW model (solid lines). 
              The dashed lines represent the uncertainty band resulting from 
              varying the parameters within their correlated uncertainties, as described in the text.}
 \label{fig:harpfit}
\end{center}
\end{figure}

\begin{figure}
\begin{center}
 \includegraphics[width=150mm]{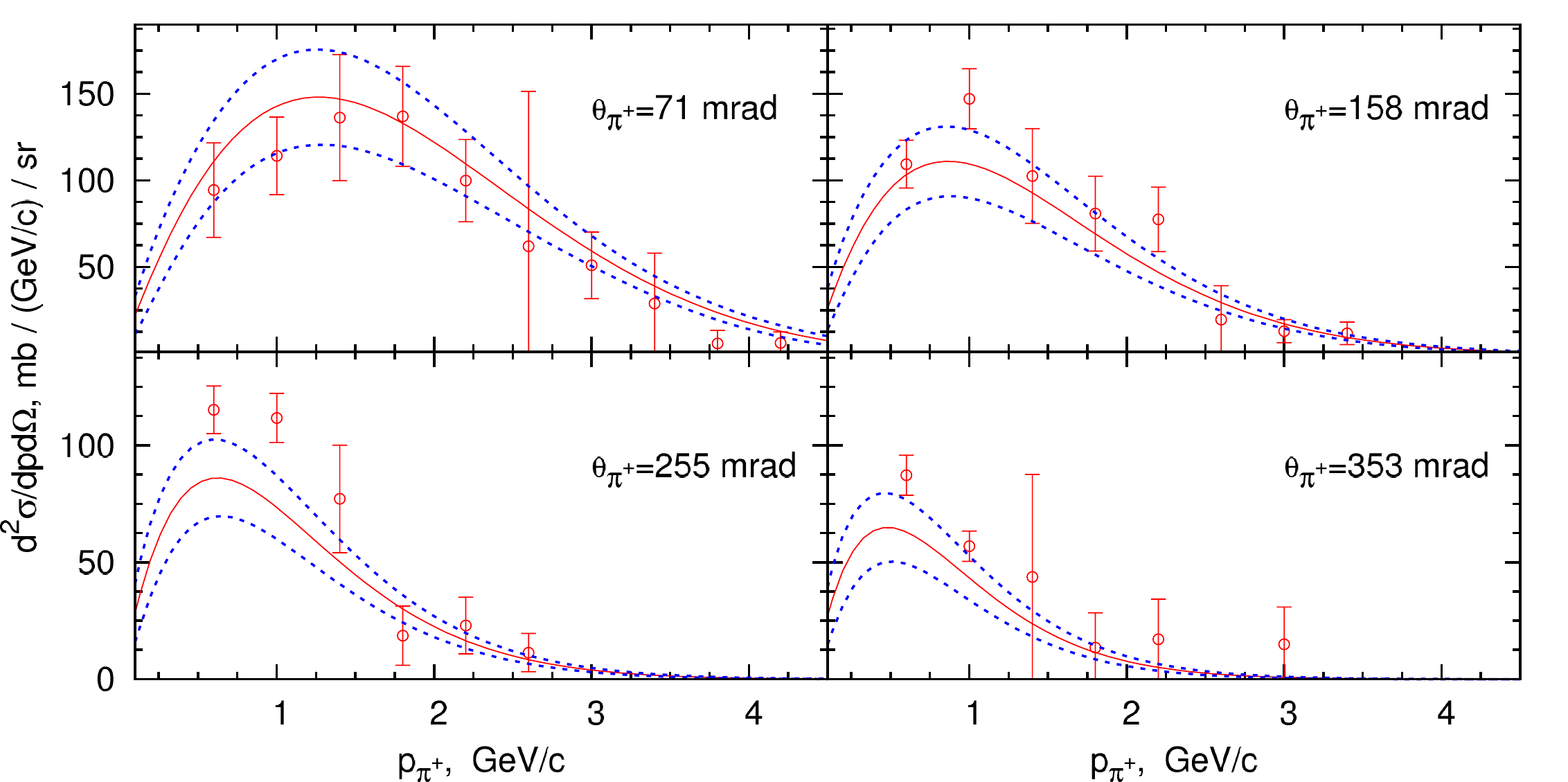}
 % e910_1norm_fracerr.eps: 0x0 pixel, 300dpi, 0.00x0.00 cm, bb=
     \caption{Comparison of E910 \pip production cross section data \cite{e910} (circles) 
		versus $p_\pi$ in bins of $\theta_\pi$ 
              from 6.4 \gevc $\mbox{p}$-Be interactions and best fit SW model (solid lines). 
              The dashed lines represent the uncertainty band for 68\% confidence 
              level for 7 fit parameters.}
 \label{fig:e910fit6}
\end{center}
\end{figure}

\begin{figure}
\begin{center}
 \includegraphics[width=150mm]{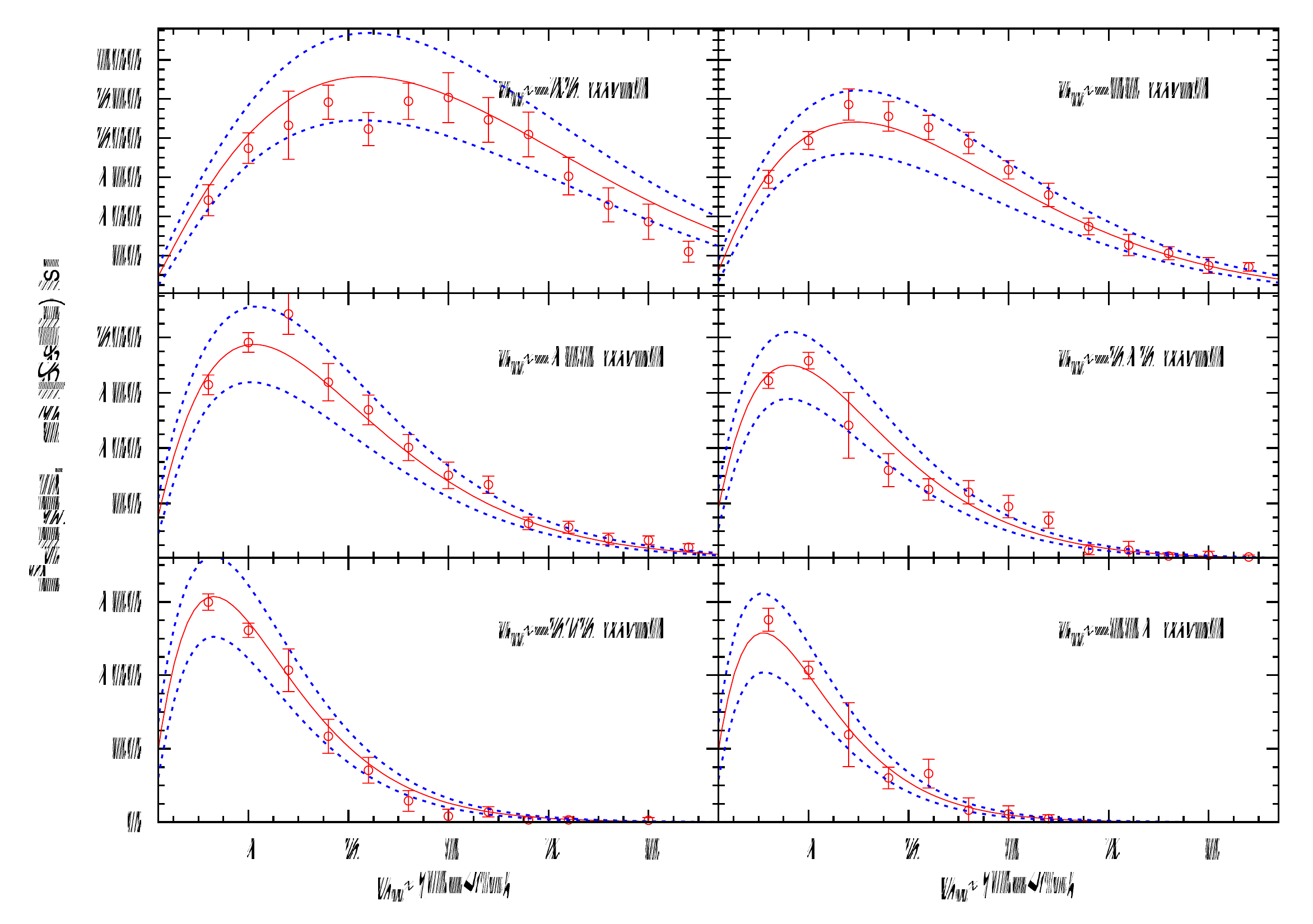}
 % e910_1norm_fracerr.eps: 0x0 pixel, 300dpi, 0.00x0.00 cm, bb=
     \caption{Comparison of E910 \pip production cross section data \cite{e910} (circles) 
		versus $p_\pi$ in bins of $\theta_\pi$ 
              from 12.3 \gevc $\mbox{p}$-Be interactions and best-fit SW model (solid lines). 
              The dashed lines represent the uncertainty band resulting from 
              varying the parameters within their correlated uncertainties, as described in the text.}
 \label{fig:e910fit12}
\end{center}
\end{figure}

\begin{figure}
\begin{center}
 \includegraphics[width=130mm]{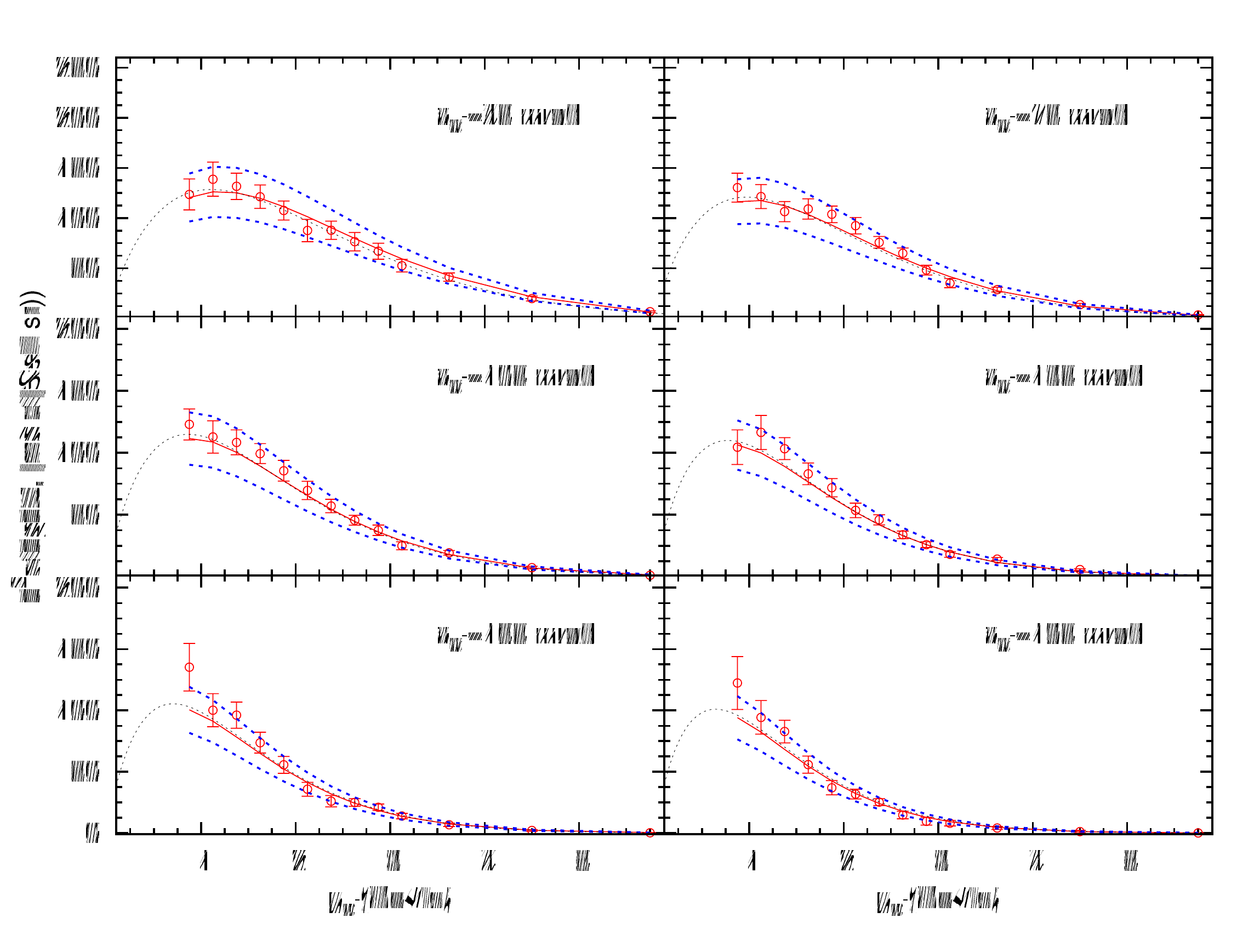}
 % e910_1norm_fracerr.eps: 0x0 pixel, 300dpi, 0.00x0.00 cm, bb=
     \caption{Comparison of HARP \pim production cross section data \cite{harp_pim}(circles) 
		versus $p_\pi$ in bins of $\theta_\pi$ 
              from 8.89 \gevc $\mbox{p}$-Be interactions and best-fit SW model (solid lines). 
              The dashed lines represent the uncertainty band resulting from 
              varying the parameters within their correlated uncertainties, as described in the text.}
 \label{fig:harpfitpm}
\end{center}
\end{figure}

\begin{figure}
\begin{center}
 \includegraphics[width=150mm, angle=0]{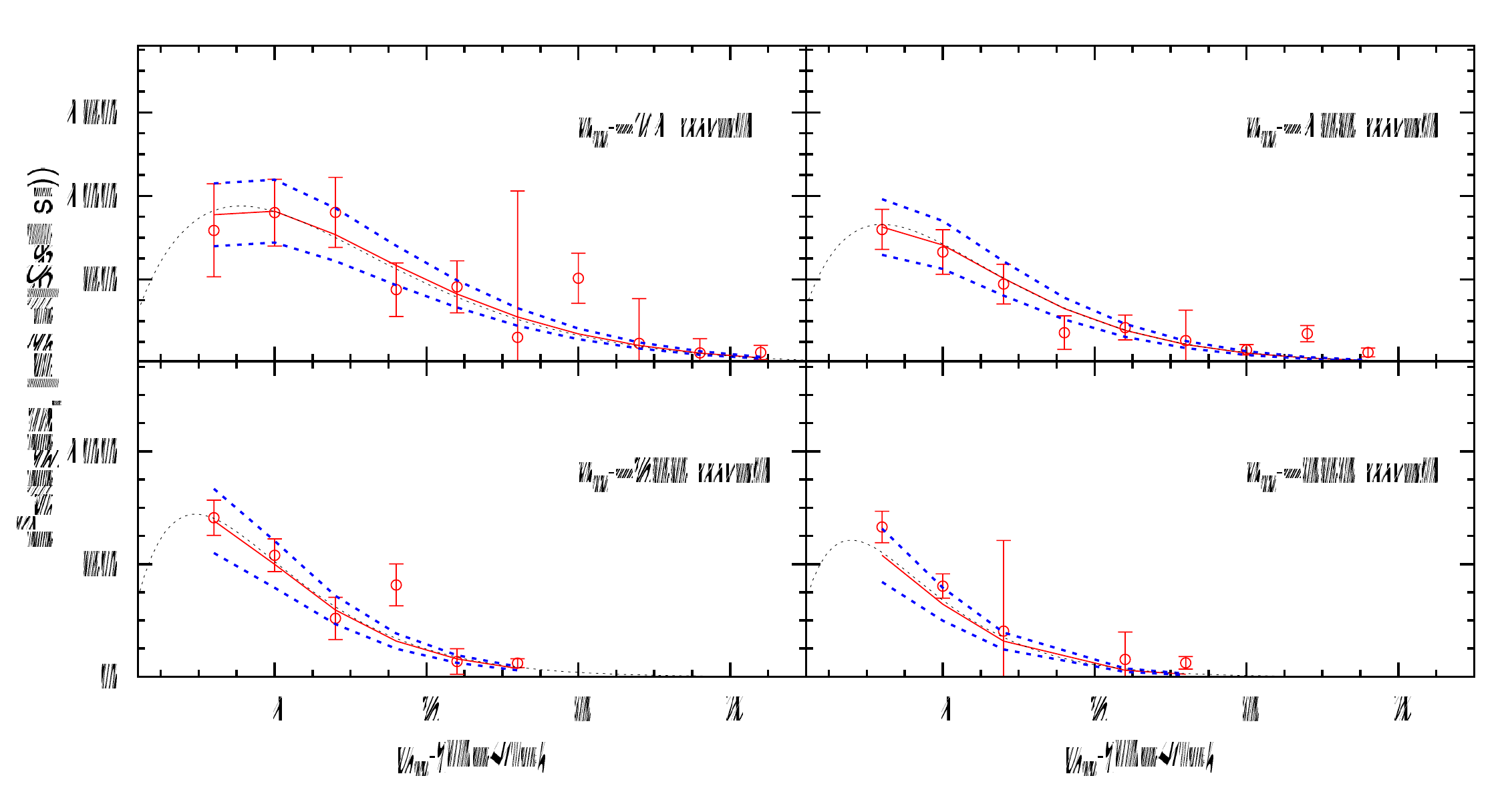}
 % e910_1norm_fracerr.eps: 0x0 pixel, 300dpi, 0.00x0.00 cm, bb=
     \caption{Comparison of E910 \pim production cross section data \cite{e910} (circles) 
		versus $p_\pi$ in bins of $\theta_\pi$ 
              from 6.4 \gevc $\mbox{p}$-Be interactions and best-fit SW model (solid lines). 
              The dashed lines represent the uncertainty band resulting from 
              varying the parameters within their correlated uncertainties, as described in the text.}
 \label{fig:e910fit6pm}
\end{center}
\end{figure}

\begin{figure}
\begin{center}
 \includegraphics[width=150mm, angle=0]{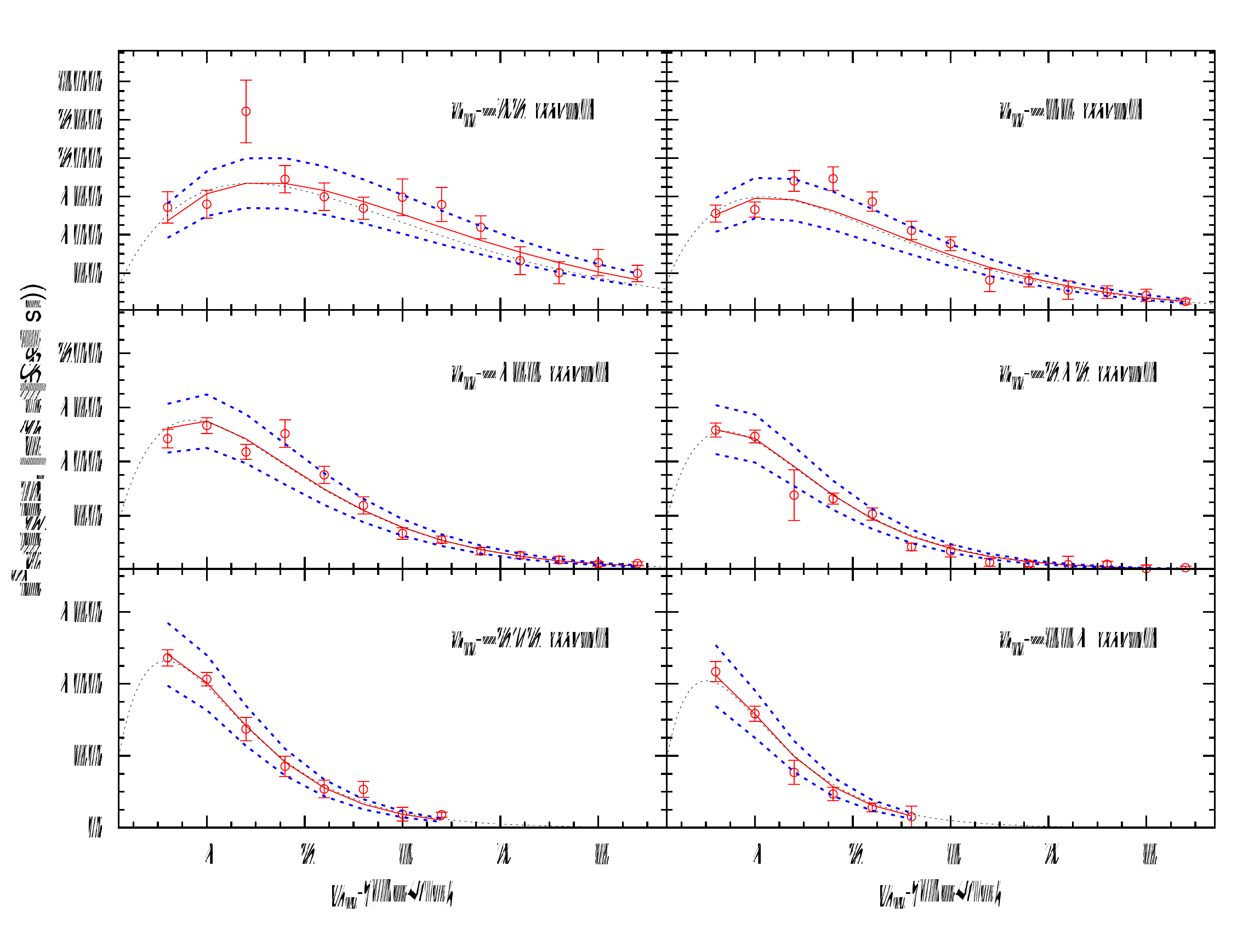}
 % e910_1norm_fracerr.eps: 0x0 pixel, 300dpi, 0.00x0.00 cm, bb=
     \caption{Comparison of E910 \pim production cross section data \cite{e910} (circles) 
		versus $p_\pi$ in bins of $\theta_\pi$ 
              from 12.3 \gevc $\mbox{p}$-Be interactions and best-fit SW model (solid line). 
              The dashed lines represent the uncertainty band resulting from 
              varying the parameters within their correlated uncertainties.}
 \label{fig:e910fit12pm}
\end{center}
\end{figure}

Likewise, the fit to the HARP \cite{harp_pim} and E910 $\pim$ production data\cite{e910} with nominal errors resulted in a
best-fit $\chi^2$ of 1.16/DOF. The experimental uncertainties are scaled by this factor to achieve a $\chi^2$/DOF
of unity. The resulting parameters and covariance matrix are shown in Table \ref{tab:swpimpar}, where the
covariances are shown in the upper right triangle of the matrix (including the diagnonal terms) and the
correlation coefficients are shown in the lower left triangle of the matrix.
The parametrizations using these best-fit parameters, along with the expected variation
due to the parameter uncertainties, are shown along with the production data in
 Figures \ref{fig:harpfit}-\ref{fig:e910fit12pm}.

\begin{table}
\begin{tabular}{c|rrrrrrrr}\hline\hline
       & $c_1$& $c_2$   & $c_3$ & $c_4$   & $c_5$  & $c_6$  & $c_7$  & $c_8$  \\ \hline
Value  & 220.7 & 1.080  & 1.000 & 1.978   & 1.32   &  5.572 & 0.0868 & 9.686     \\ \hline
$c_1$  &1707.2 & 1.146  & --    & -17.646 & -15.968& -8.81  & -0.7347& -60.816  \\
$c_2$  &0.139  & 0.0396 & --    & -0.1072 & -0.0993& 0.0325 & 0.0007 & -0.0777  \\
$c_3$  & --    & --     & --    & --      & --     & --     & --     &   --   \\
$c_4$  &-0.554 & -0.699 & --    & 0.5945  & 0.5049 & 0.0655 & 0.0025 & 0.198  \\  
$c_5$  &-0.582 & -0.751 & --    & 0.986   & 0.4411 & 0.0568 & 0.0025 & 0.2271  \\  
$c_6$  &-0.469 & 0.359  & --    & 0.187   & 0.188  & 0.2066 & 0.0047 & 0.1031  \\	  
$c_7$  &-0.795 & 0.157  & --    & 0.145   & 0.168  & 0.462  & 0.0005 & 0.0641  \\	  
$c_8$  &-0.368 & -0.098 & --    & 0.064   & 0.085  & 0.057  & 0.716  & 16.0189  \\  \hline
\end{tabular}
\vskip 0.5 cm
\caption{\label{tab:swpippar}Extracted Sanford-Wang parameters $c_{1-8}$ (first row),
covariance matrix (upper right triangle including diagonals terms), and correlation
coefficients (lower left triangle), for $\pip$ secondary production in $\mbox{p}$-Be interactions. 
There are no entries in the covariance matrix for parameter $c_3$, which is fixed in the fit 
due to its large correlation with $c_5$.}

%\begin{table}
%\begin{tabular}{c|rrrrrrrr}\hline\hline
%     & $c_1$& $c_2$& $c_3$& $c_4$& $c_5$& $c_6$& $c_7$& $c_8$ \\ \hline
%Value& 220.7 & 1.080    & 1.000 & 1.978& 1.32     &  5.572      & 0.0868     & 9.686     \\ \hline
%$c_1$&  1707.2  &  1.1460  & -- & -17.646  & -15.968  &  -8.8100  &  -0.7347  &  -60.816 \\
%$c_2$&  1.1460  &  0.0396  & -- & -0.1072  & -0.0993  &   0.0325  &   0.0007  &  -0.0777  \\
%$c_3$&    --    &    --    & -- &  --      &    --    &  --       &    --     &    --         \\
%$c_4$&-17.646  & -0.1072  & -- & 0.5945  &  0.5049  &   0.0655  &   0.0025  &   0.1980  \\
%$c_5$&-15.968  & -0.0993  & -- & 0.5049  &  0.4411  &   0.0568  &   0.0025  &   0.2271  \\
%$c_6$&-8.8100  &  0.0325  & -- & 0.0655  &  0.0568  &   0.2066  &   0.0047  &   0.1031  \\
%$c_7$&-0.7347  &  0.0007  & -- & 0.0025  &  0.0025  &   0.0047  &   0.0005  &   0.0641  \\
%$c_8$&-60.816  & -0.0777  & --&  0.1980  &  0.2271  &   0.1031  &   0.0641  &   16.0189 \\ \hline
%\end{tabular}
%\vskip 0.5 cm
%\caption{\label{tab:swpippar}Extracted Sanford-Wang parameters $c_{1-8}$ (first row) and
%the covariance matrix for $\pip$ secondary production in $\mbox{p}$-Be interactions. 
%There are no entries in the covariance matrix for parameter $c_3$, which is fixed in the fit due 
%to its large correlation with $c_5$.}
\vskip 0.5 cm

\begin{tabular}{c|rrrrrrrr}\hline\hline
      & $c_1$ & $c_2$   & $c_3$  & $c_4$   & $c_5$   & $c_6$   & $c_7$   & $c_8$  \\ \hline
Value & 213.7  & 0.9379 &  5.454 & 1.210   & 1.284   & 4.781   & 0.07338 & 8.329  \\ \hline
$c_1$ &3688.9  & 7.61   &   --   & -15.666 & -17.48  & -11.329 & -0.9925 & -91.4  \\
$c_2$ &0.636   & 0.0388 &   --   & -0.0437 & -0.0509 & 0.0102  & -0.0009 & -0.1957\\
$c_3$ &  --    & --     &   --   & --      &   --    &  --     &  --     &   --   \\
$c_4$ & -0.889 & -0.765 &   --   & 0.0841  & 0.0895  & 0.0301  & 0.0029  & 0.2588 \\ 
$c_5$ & -0.917 & -0.823 &   --   & 0.983   & 0.0986  & 0.0375  & 0.0033  & 0.3141 \\ 
$c_6$ & -0.467 & 0.130  &   --   & 0.260   & 0.299   & 0.1595  & 0.0051  & 0.1933 \\	 
$c_7$ & -0.731 & -0.204 &   --   & 0.447   & 0.470   & 0.571   & 0.0005  & 0.064  \\	   
$c_8$ & -0.362 & -0.239 &   --   & 0.215   & 0.241   & 0.117   & 0.689   & 17.242 \\ \hline    
\end{tabular}
\vskip 0.5 cm
\caption{\label{tab:swpimpar}Extracted Sanford-Wang parameters $c_{1-8}$ (first row),
the covariance matrix (upper right triangle including diagonal terms), and correlation 
coefficients (lower left triangle) for $\pim$ secondary production in $\mbox{p}$-Be interactions. 
 There are no entries in the covariance matrix for parameter $c_3$, which is fixed in the fit due to its 
large correlation with $c_5$.}

%\begin{tabular}{c|rrrrrrrr}\hline\hline
% 	  & $c_1$ & $c_2$   & $c_3$ & $c_4$  & $c_5$      & $c_6$    & $c_7$    & $c_8$ \\ \hline
%Value            & 213.7  & 0.9379  &  5.454 & 1.210    & 1.284     & 4.781     & 0.07338 & 8.329 \\ \hline
%$c_1$& 3688.9&  7.6100 &   --       & -15.666&  -17.480 &  -11.329  &  -0.9925  &  -91.400\\ 
%$c_2$&7.6100 & 0.0388  &  --        & -0.0437& -0.0509  & 0.0102    &  -0.0009  &-0.1957 \\
%$c_3$& --    & --      &  --        & --     & --       & --        &    --     &   --        \\
%$c_4$&-15.666& -0.0437 &  --       &  0.0841 &  0.0895  &  0.0301   &   0.0029  &  0.2588 \\
%$c_5$& -17.480&-0.0509 &  --       & 0.0895  &  0.0986  &  0.0375  &  0.0033  &  0.3141 \\
%$c_6$& -11.329& 0.0102 &  --       & 0.0301  &   0.0375 &  0.1595  &  0.0051  & 0.1933 \\
%$c_7$& -0.9925& -0.0009& --        & 0.0029  &    0.0033&  0.0051 &  0.0005   &0.0640 \\
%$c_8$& -91.400 & -0.1957&  --      &  0.2588 &  0.3141   &   0.1933&  0.0640  &  17.242 \\ \hline
%\end{tabular}
%\vskip 0.5 cm
%\caption{\label{tab:swpimpar}Extracted Sanford-Wang parameters $c_{1-8}$ (first row) and
%the covariance matrix for $\pim$ secondary production in $\mbox{p}$-Be interactions. There are no 
%entries in the covariance matrix for parameter $c_3$, which is fixed in the fit due to its large 
%correlation with $c_5$.}
\end{table}

%\vskip 0.5 cm

\begin{table}
\begin{center}
\begin{tabular}{l|c|rccccr}\hline\hline
Dataset	 					& $P_{beam} $ & $P_{\Kp}$	& $\theta_{\Kp}$  & $x_{F}$      & $p_T$       & $\sigma_{N}$ \\
			 					& $(\gevc)$	& $(\gevc)$	& (degrees)         &                      & $(\gevc)$ &                        \\ \hline
Abbott	 \cite{abbott}	& 14.6		& 2--8 		& 20--30               & -0.12--0.07 & 0.2-0.7     & 10\% \\
Aleshin	 \cite{aleshin}&  9.5		& 3--6.5		& 3.5                    & 0.3--0.8      & 0.2-0.4      & 10\% \\
Allaby 	 \cite{allaby}	& 19.2		&3--16		& 0--7                   & 0.3--0.9        & 0.1--1.0    & 15\% \\
Dekkers \cite{dekkers}& 18.8, 23.1	& 4--12		& 0, 5                    & 0.1--0.5      & 0.0--1.2     & 20\% \\
Eichten \cite{eichten}& 24.0		& 4--18		& 0--6               & 0.1--0.8        &0.1--1.2     & 20\% \\
Lundy	\cite{lundy}	& 13.4		& 3--6		& 2, 4, 8              & 0.1--0.6       &  0.1--1.2    & 20\% \\ 
Marmer\cite{marmer} & 12.3		& 0.5-1.0		& 0,  5, 10           & -0.3--1.0      & 0.15-0.5   & 20\% \\       
Vorontsov	 \cite{vorontsov}& 10.1		& 1--4.5		& 3.5                    & 0.03--0.5      & 0.1--0.25  & 25\% \\ \hline
\end{tabular}
\vskip 0.5 cm
\caption{\label{tab:kplusprod}Summary of $\Kp$ production measurements in $\mbox{p}$-Be interactions used
to characterize $\Kp$ production in the BNB. The table includes $P_{beam}$, the primary proton momenta
in the measurement, the momentum and angular ranges of the measurements, as well as the corresponding ranges of the Feynman scaling variable $x_F$ and transverse momentum $p_T$. Finally, the quoted overall normalization
uncertainty $\sigma_N$ is listed.}
\vskip 0.5 cm
\end{center}
\end{table}

\subsection{$\Kp$ Production Measurements}
For charged kaons, whose decays result in a significant contribution to the $\num$
flux at high energies as well as the $\nue$ flux through the $K_{e3}$ decay mode,
there are no measurements from the HARP or BNL E910. As a result, measurements
reported by other experiments measuring $\Kp$ production in $\mbox{p}$-Be interactions at primary beam
momenta close to $8.89\gevc$ are used \cite{abbott,aleshin,allaby,dekkers,eichten,lundy,marmer,vorontsov}.
The measurements are summarized in Table \ref{tab:kplusprod}.

\begin{figure}
\begin{center}
 \includegraphics[width=150mm]{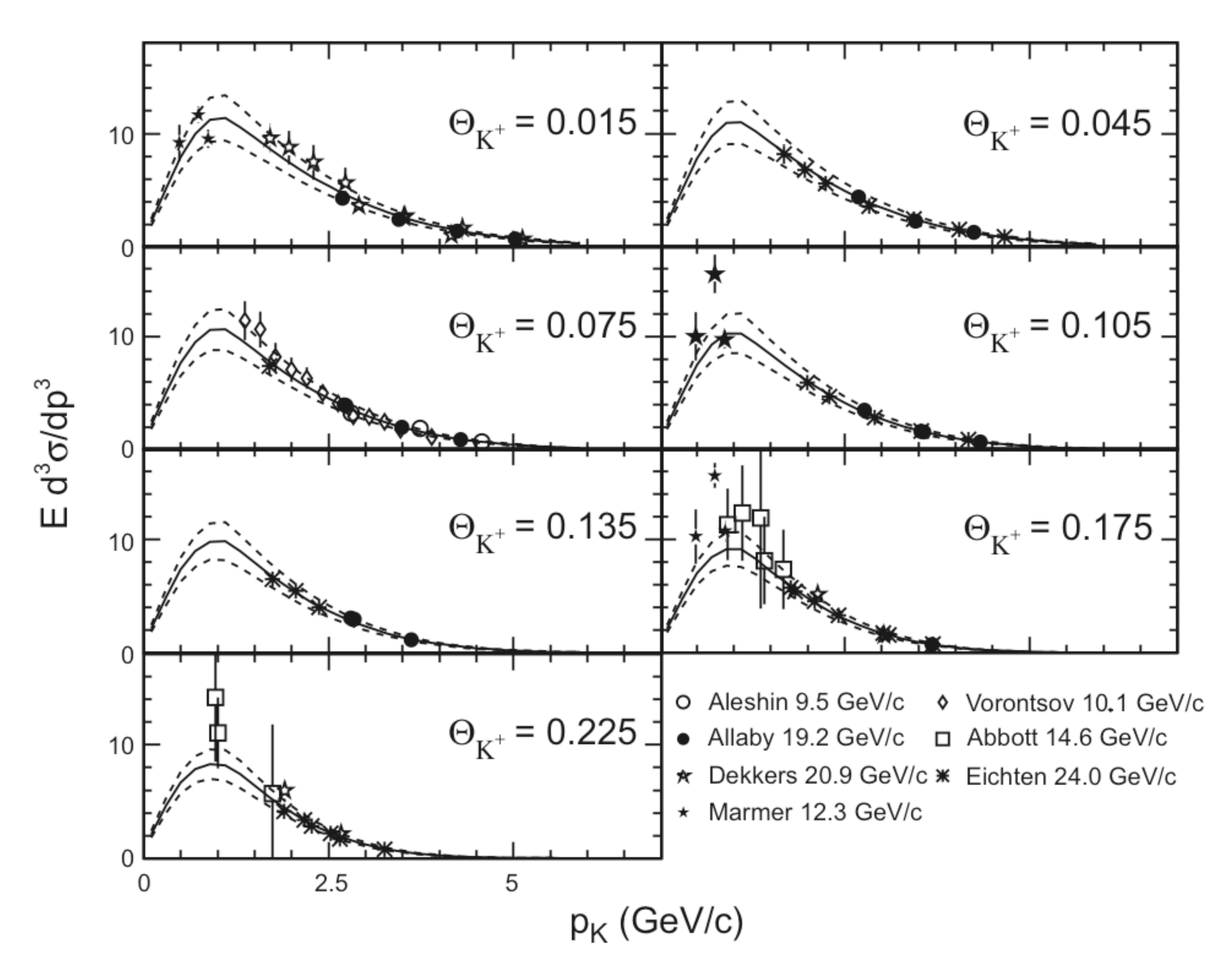}
 % e910_1norm_fracerr.eps: 0x0 pixel, 300dpi, 0.00x0.00 cm, bb=
     \caption{Comparison of $\Kp$ invariant cross section data (points) as a function
     	of $p_{\Kp}$, the $\Kp$ momentum, in bins of $\Theta_{\Kp}$, the $\Kp$
	production angle (in radians), with 
              the Feynman-scaling based parametrization with best-fit parameters shown as a solid line.
              The scaling has been used to relate the measurements
              at different primary beam momenta to the $8.89\gevc$ primary momentum in the BNB.
		Normalization factors from the best-fit are  applied to the data.
              The dashed lines represent the uncertainty band resulting from 
              varying the parameters within their correlated uncertainties. The uncertainty
bands include the factor 1.5 error inflation to set $\chi^2/\mbox{DOF}$ = 1.}
 \label{fig:kpfit}
\end{center}
\end{figure}

Since no measurements of $\Kp$ exist at the $8.89\gevc$ BNB primary momentum, we employ the Feynman scaling hypothesis
to relate $\Kp$ production measurements at different primary energies to the expected production at $8.89\gevc$. Theoretically, Feynman scaling should be a better model for comparing data from
different primary beam momenta. This is born out by comparisons of data scaled to the BNB momentum of $8.89\gevc$ as shown in Figure \ref{fig:kpfit}.
The hypothesis states that the invariant cross section is
a function of only two variables, namely $x_F$ and $p_T$, where
\begin{equation}
x_F = \frac{p^{cm}_{\parallel}}{p^{\mbox{\small{max}},cm}_{\parallel}}
\end{equation}
is the Feynman scaling variable, defined as the ratio of the parallel component of the momentum of the
produced particle in the center-of-mass frame and the maximum possible value of this quantity for
the given reaction, and $p_T$ is the transverse component of the momentum of the produced particle.
In calculating $x_F$,  the $p^{\mbox{\small{max}},cm}_{\parallel}$ value is taken from the exclusive channel $\mbox{p}+(\mbox{p}/\mbox{n})\to \mathrm{\Lambda}^0 + (\mbox{p}/\mbox{n}) + \Kp$. A more
complete description of the Feynman scaling fit procedure and results can be found in Reference 
\cite{fs}.

The Feynman scaling is used as a basis for parametrizing the production data using the variables $x_F$ and $p_T$.  
This motivates a six-parameter model given by:
\begin{equation}
\label{eq:fs}
\begin{array}{ll}
\frac{d^2\sigma}{dp d\mathrm{\Omega} } = \frac{p^2_{\Kp}}{E_{\Kp}} \left( E_{\Kp}\frac{d^{3}\sigma}{dp^3_{\Kp}}\right) = & 
 \frac{p^2_{\Kp}}{E_{\Kp}} \times c_1(1-|x_F|) \times \\
  &  \exp{\left[ -c_2 p_T  -c_3 |x_F|^{c_4}  - c_5 p_T^2 -c_7 |p_T \times x_F|^{c_6} \right]}
\end{array} 
 \end{equation}
The model is basically a translation of the Feynman scaling hypothesis where the invariant
cross section is only a function of $x_F$ and $p_T$.
It incorporates an exponentially falling $p_T$ distribution, correlations between $p_T$ and $x_F$,
a flat rapidity plateau at $x_F=0$ and zero cross section  as $x_F \to 1$. The kinematic threshold constraint
is imposed by setting the function equal to zero for $|x_F|>1$. Figure \ref{fig:kpfit}  shows the
momentum distribution of the data scaled to $8.89\gevc$ primary momentum in bins of scaled $\Kp$ production  angle. The data include normalizations factors obtained from the fit procedure described below. The right plot of Figure \ref{fig:prodcoverage} shows as boxes the $x_F$ versus $p_T$ distribution
of $\Kp$ produced in the target that produce neutrinos at the MiniBooNE detector. The colored points indicate the kinematic coverage of the various measurements 
in these two variables.

The $c_i$ parameters are determined in a $\chi^2$ fit to the production data for $1.2 <p^{BNB}_{\Kp} < 5.5$, where $p^{BNB}_{\Kp}$ is the kaon momentum translated to the BNB primary energy using Feynman scaling. The
$p^{BNB}_{\Kp}$ requirement  eliminates most of the data at negative $x_F$, where nuclear effects are expected to be dominant. 
The $\chi^2$ takes the same form as in Equation \ref{eq:swchi2}, where the covariance matrix from the
the experiments is diagonal, and the quoted normalization uncertainties are used to constrain
the normalization factors. The Vorontsov data \cite{vorontsov} has indications of an error in the normalization outside of their quoted
uncertainties. As a result, a large normalization uncertainty (500\%) was assigned to these measurements with the effect that
the measurements from this experiment contributes only ``shape'' information without any normalization constraint. The discrepancy is not apparent in Figure \ref{fig:kpfit} since the fitted normalization factor is applied to the measured cross sections.
The Lundy  data \cite{lundy} were excluded from the fit due to inconsistencies with the other measurements,
while the Marmer measurements \cite{marmer} were excluded by the $p^{BNB}_{\Kp}$ requirement.

The $\chi^2/{\mbox{DOF}}$ for the fit is 2.28. The errors in the measurements are inflated by a factor of $\sqrt{2.28}=1.5$  to bring the
$\chi^2/{\mbox{DOF}}$ to unity. The covariance matrix of the parameters is extracted in the same way as employed
for the pion production fits. Table \ref{tab:kpfs} summarizes the results, with the first row listing the seven best-fit parameters,
and the $7\times7$ matrix below it listing the covariance. While the parametrization represents the measurements quite well (as shown in Figure \ref{fig:kpfit}),
the uncertainties are inflated by a further factor of four for the $\num\to\nue$ oscillation analysis to account for 
inconsistencies within the production data, the use of the Feynman-scaling hypothesis to relate the production
measurements from experiments with proton momenta different from the $8.89\gevc$ used at the BNB, and
a possible discrepancy in the rate of $\num$ events observed in the MiniBooNE detector compared with the predictions
based on the beam line simulation\cite{ccqe}. The covariance matrix
in Table \ref{tab:kpfs} and the uncertainty bands in Figure \ref{fig:kpfit} do not include this further  inflation of the uncertainties by a factor of four.

\begin{figure}[tb]
\begin{center}
\includegraphics[width=150mm]{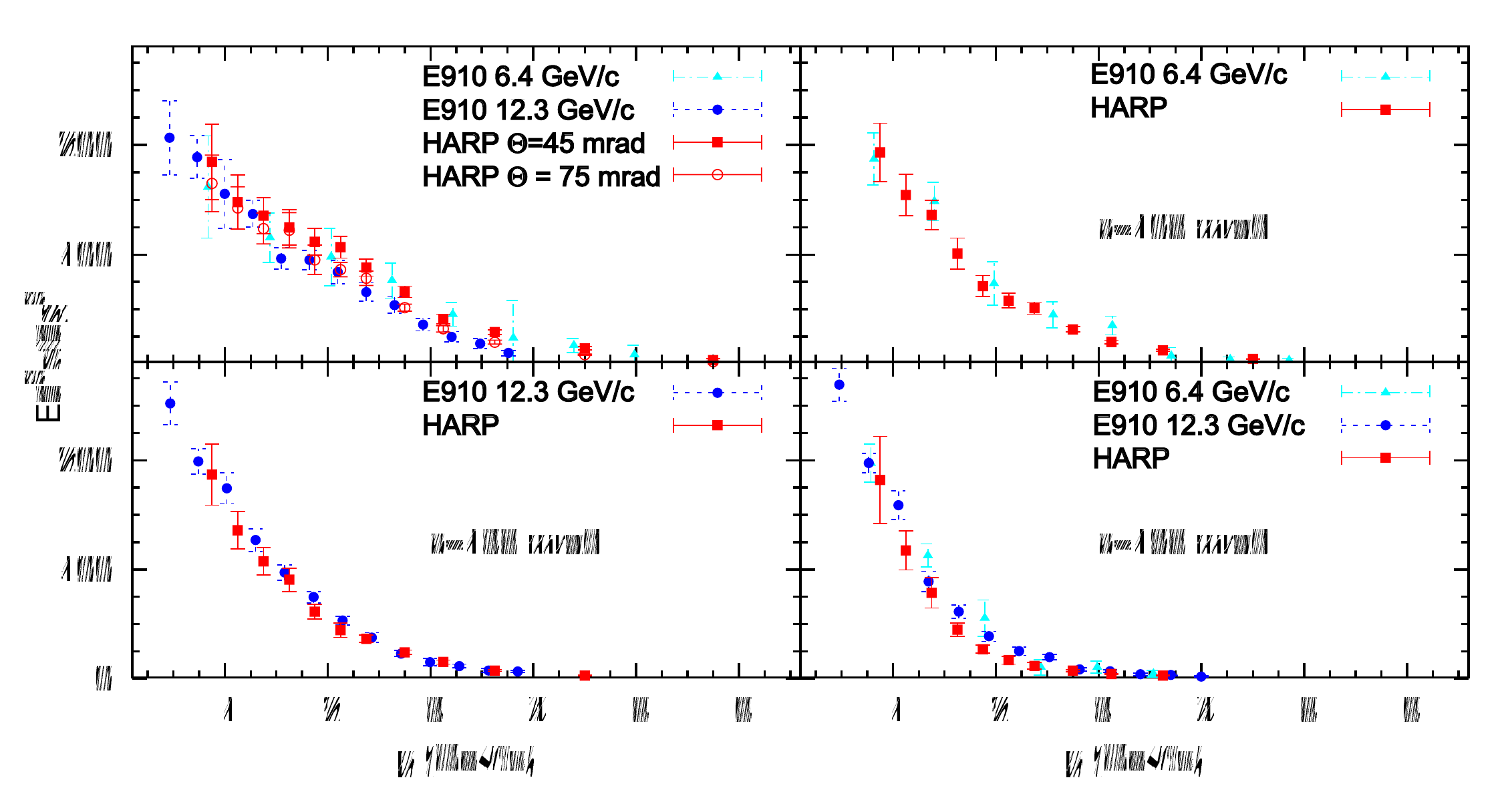}
\caption{Invariant pion production cross section from HARP and E910 versus $p_\pi$ in bins of $\theta_\pi$. 
The E910 measurements are rescaled to $p_B=8.89\gevc$.}
 \label{fig:pipfsdata}
\end{center}
\end{figure}

Figure \ref{fig:pipfsdata} shows a similar comparison of the E910 and HARP $\pip$
production data where the E910 data have been scaled to $8.89\gevc$ primary momentum.
The data indicates that the $\pip$ production is also consistent with the Feynman scaling
hypothesis. When fit to Equation \ref{eq:fs}, a slightly poorer $\chi^2$ results than in the SW fit. As
a result, the Feynman scaling model is not used in the $\pip$ production model.

\begin{table}
\begin{center}
%{\scriptsize
\begin{tabular}{c|rrrrrrr}\hline\hline
      & $c_1$  & $c_2$	 & $c_3$     & $c_4$	    & $c_5$      & $c_6$       & $c_7$       \\ \hline
Value & 11.70  &  0.88	 & 4.77	      &  1.51	    &  2.21      & 2.17        & 1.51        \\ \hline
$c_1$ & ~~~~1.094  &  ~~~0.0502 & $~~2.99\mil$ & -0.0332     & -0.0375    & 0.125       & 0.0743      \\
$c_2$ & 0.378  &  0.0161 & $1.39\mil$ & $-1.44\mil$ & -0.0126    & 0.0322      & 0.022       \\
$c_3$ & 0.033  &  0.127  & $7.47\mil$ & $~~2.06\mil$  & $~~1.93\mil$ & 0.0135      & $~~-3.34\mil$ \\
$c_4$ &-0.540  & -0.193  & 0.405      & $3.46\mil$  & $2.03\mil$ & $~~-4.11\mil$ & $~~-6.28\mil$ \\
$c_5$ &-0.297  & -0.822  & 0.185      &  0.286      &  0.0146    & -0.0154     & -0.0244     \\
$c_6$ & 0.280  &  0.595  & 0.366      & -0.164      & -0.299     & 0.182       &  0.126      \\
$c_7$ & 0.178  &  0.435  & -0.097     & -0.268      & -0.506     & 0.741       &  0.159      \\ \hline
\end{tabular}
%	  & $c_1$	  & $c_2$		&$c_3$	& $c_4$	& $c_5$	& $c_6$ & $c_7$ \\ \hline
%Value & 11.70	      & 0.88		& 4.77		& 1.51	& 2.21 & 2.17    & 1.51   \\ \hline
%$c_1$ &1.094	      & 0.0502  	& $2.99\mil$ 	& -0.0332 & -0.0375 & 0.125 & 0.0743 \\
%$c_2$ & 0.0502     &  0.01610	& $1.39\mil$ 	&$-1.44\mil$ & -0.0126 & 0.0322 & 0.0220 \\
%$c_3$ & $2.99\mil$& $1.39\mil$ & $7.47\mil$ 	& $2.06\mil$ & $1.93\mil$ & 0.0135 & $-3.34\mil$ \\
%$c_4$ & -0.0332    & $-1.44\mil$& $2.06\mil$ 	& $3.46\mil$ & $2.03\mil$ & $-4.11\mil$ & $-6.28\mil$ \\
%$c_5$ & -0.0375     & -0.0126    & $1.93\mil$ 	& $2.03\mil$ & $0.0146$ & -0.0154 & -0.0244 \\
%$c_6$ & 0.0125      & 0.0322	& 0.0135		& $-4.11\mil$ & -0.0154 & 0.0182 & 0.0126 \\
%$c_7$ & 0.0743      & 0.0220	& $-3.34\mil$ & $-6.28\mil$ & -0.0244 & 0.126 & 0.159 \\ \hline
%\end{tabular}
%}
\vskip 0.5 cm
\caption{\label{tab:kpfs}Best-fit Feynman scaling model parameters $c_i$ from a fit
to $\Kp$ production data (first row). The covariance matrix for the parameters with uncertainties inflated by a factor of 1.5 to set $\chi^2/\mbox{DOF} =1$ is in the upper right triangle (including diagonal terms) 
of the table below the parameters, but does not include the
additional inflation by a factor of four described in the text. The lower left triangle shows the correlation
coefficients.}
\end{center}
\vskip 0.5 cm
\end{table}

\begin{table}
\begin{center}
%{\scriptsize
\begin{tabular}{c|ccccccccc}\hline\hline
       & $c_1$ & $c_2$   & $c_3$    & $c_4$    &  $c_5$   & $c_6$   &  $c_7$   & $c_8$    & $c_9$   \\\hline
Value  &15.130 & 1.975   &  4.084   &  0.928   &  0.731   & 4.362   & 0.048    & 13.300   & 1.278   \\\hline
$c_1$  &32.3   & -0.09687 & 0.8215  & -0.1018  & -0.2124  & -0.8902 & -0.1333  & 16.55    & -1.789  \\
$c_2$  &-0.055 &  0.09574 & 0.03248 &  0.00131 & -0.01303 & 0.08836 & -0.00031 & -1.536   & -0.2156 \\
$c_3$  & 0.199 &  0.144   & 0.5283  & -0.01922 &  0.02267 & -0.0033 & -0.00236 & 0.0391   & -0.08017\\
$c_4$  &-0.062 &  0.015   & -0.091  &  0.08442 &  0.00405 & 0.00071 & -0.00037 & -0.01443 & -0.07301\\
$c_5$  &-0.377 & -0.425   & 0.315   &  0.141   &  0.00982 & 0.00287 & 0.00028  & -0.05777 & 0.02966 \\
$c_6$  &-0.261 &  0.476   & -0.008  &  0.004   &  0.048   & 0.3599  & 0.00385  & -4.751   & -0.1577 \\
$c_7$  &-0.724 & -0.031   & -0.100  & -0.039   &  0.087   & 0.198   & 0.00105  & 0.05806  & 0.00686 \\
$c_8$  & 0.255 & -0.435   & 0.005   & -0.004   & -0.051   & -0.694  & 0.157    & 130.2    & 1.222   \\
$c_9$  &-0.183 & -0.406   & -0.064  & -0.146   &  0.174   & -0.153  & 0.123    & 0.062    & 2.948   \\\hline
\end{tabular}
%\begin{tabular}{c|ccccccccc}\hline\hline
%      & $c_1$	 & $c_2$       &$c_3$	& $c_4$	& $c_5$ & $c_6$ &  $c_7$ & $c_8$  & $c_9$\\ \hline
%Value & 15.130   & 1.975       & 4.084  & 0.928    &  0.731    & 4.362  & 0.048  & 13.300 & 1.278 \\ \hline
%$c_1$ & 32.30    & -0.09687  & 0.8215   & -0.1018 & -0.2124 &-0.8902& -0.1333 & 16.55 & -1.789 \\
%$c_2$ & -0.09687 & 0.09574  & 0.03248 & 0.00131 & -0.01303 & 0.08836 & -0.00031 & -1.536 & -0.2156 \\
%$c_3$ & 0.8215   & 0.03248   & 0.5283  & -0.01922 & 0.02267 &  -0.00330 & -0.00236 & 0.03910 & -0.08017 \\
%$c_4$ & -0.1018  & 0.00131   & -0.01922& 0.08442 & 0.00405 & 0.00071 & -0.00037 & -0.01443 & -0.07301 \\
%$c_5$ &-0.2124   & -0.01303  & 0.02267 & 0.00405 & 0.00982 & 0.00287 & 0.00028  & -0.05777 & 0.02966\\
%$c_6$ &-0.8902   &0.08836   & -0.00330& 0.00071 & 0.00287 &  0.3599  & 0.00385 & -4.751 & -0.1577 \\
%$c_7$ &-0.1333   & -0.00031  & -0.00236&-0.00037 &0.00028  & 0.00385 & 0.00105 & 0.05806 & 0.00686 \\
%$c_8$ &16.55     &-1.536       &0.03910  &-0.01443 & -0.05777 &-4.751     & 0.05806 & 130.2 & 1.222 \\
%$c_9$ & -1.789   &-0.2156    &-0.08017 & -0.07301& 0.02966  & -0.1577 &0.00686  & 1.222 & 2.948 \\ \hline
%\end{tabular}
%}
\vskip 0.5 cm
\caption{\label{tab:kssw}Best-fit Sanford-Wang model parameters $c_i$ from a fit to $\KS$ production 
data (first row).
The covariance matrix for the parameters  is in the upper right triangle (including diagonal terms) of the 
table  below the parameters, while the bottom left corner of the table reports the corresponding correlation 
coefficients.}
\end{center}
\vskip 0.5 cm
\end{table}

\subsection{Production of $\Kz$, $\Km$ and other particles}
A scheme similar to that used to parametrize the $\pip$ and $\Kp$ production data is used for neutral kaon production, for which  the $K_{e3}$ decay mode of the $\KL$ is a source of background $\nue$. Since the 
kaons are produced in strong interactions as $\Kz$ and $\Kzb$ (primarily the former), the 
kaons have equal content as $\KS$ and $\KL$.  As a result, the production properties
 of neutral kaons decaying as $\KL$ can be obtained by measuring the $\KS$ production properties. While the $\KS$ can contribute to the neutrino flux via the
decay of the charged pions produced in the $\KS\to\pip+\pim$ decay, the most important consideration
is the production of $\nue$ from the decay of the $\KL$. The long life time of the $\KL$, together
with the fact that they are not focussed, lead to the expectation that
the contribution of neutrinos for this source will be small relative to the $\Kp$. 

The primary source of data for the parametrization comes from two measurements of $\KS$ production in $\mbox{p}$-Be
interactions in the BNL E910 experiment ($p_{beam}=12.3$ and $17.5\gevc$) and the measurements of Abe {\em et al.} \cite{abe} 
($12.3\gevc$) at KEK.  
Since the neutral kaons are not focused by the magnetic field of the horn, the forward production $
(<5^{\circ})$ is particularly relevant for predicting the BNB neutrino flux.  While the production data from the BNL E910 and KEK measurements do not cover this region, the combination of the two data sets are sufficient
to constrain the production cross section in this forward region via the Sanford-Wang parametrization.
The extracted parameter values and covariance matrix are summarized in Table \ref{tab:kssw}.

For $\Km$ production, the scarcity of production measurements in the relevant kinematic regions motivated
the use of the MARS hadronic interaction package \cite{mars} to determine
the absolute double differential cross sections. The cross sections are obtained by simulating
$8.89\gevc$ $\mbox{p}$-Be interactions on a thin beryllium target and recording the rate and spectrum
of outgoing $\Km$. The expected relative contribution of neutrinos of all species from $\Km$ decays
is expected to be small. Neutrino flux contributions from semileptonic hyperon decays ({\em e.g.} 
$\mathrm{\Lambda}, \mathrm{\Sigma}$, etc.), estimated using a FLUKA\cite{fluka} simulation, are also negligible.

Secondary protons and neutrons emerging from the $\mbox{p}$-Be inelastic interactions are simulated based
on the predictions of the MARS model, with the exception of quasi-elastic scattering, in which case
the final state proton kinematics are handled by a custom model. The production of all other particle
species is handled by the default Geant4 hadronic model.

\begin{table}[tb]
\begin{center}
\begin{tabular}{c|ccc} \hline\hline
Particle & Multiplicity   & $\langle p\rangle$  & $\langle\theta\rangle$ \\ 
         & per reaction&       $(\gevc)$     & (mrad)                    \\ \hline
$\mbox{p}$ & 1.5462 & 2.64 & 441 \\
$\mbox{n}$ & 1.3434 & 1.59 & 586  \\
$\pi^-$ & 0.9004* & 0.82 & 556 \\
$\pi^+$ & 0.8825* & 1.11 & 412 \\
$K^+$ & 0.0689 & 1.69 & 332 \\
$\Kz$ & 0.0241 & 1.34 & 414 \\
$K^-$ & 0.0024 & 1.26 & 259 \\ \hline
Total & 4.7679 & 1.69 & 496 \\ \hline

\end{tabular}
\end{center}
\caption{\label{tab:partprod}{
Average multiplicity per particle-producing reaction
for secondary particles produced in the inelastic collisions
of $8.89\gevc$ primary protons on beryllium, as well as average
momentum $\langle p\rangle$ and angle $\langle\theta\rangle$ with respect to the primary
proton direction. Multiplicities and average kinematics refer to particles 
produced in the forward hemisphere in the laboratory frame and with transverse
momentum less than $1\gevc$. *see comment in text.}}
\end{table}

The properties of the particle production model are summarized in Table \ref{tab:partprod}.
The table shows the average multiplicity per $\mbox{p}-\mbox{Be}$ reaction (defined
as inelastic interactions excluding quasi-elastic scattering), along with
the mean momentum and production angle. The $\pip$ and $\pim$ production occur
with similar multiplicities, though the former tends to be harder and more forward
directed. The larger overall multiplicity for the $\pim$ is due to the extrapolation of the
cross sections to large angles that are not covered in the HARP and E910 measurements. Since the
contribution to the neutrino flux from such pions is small, the impact of uncertainty in this
extrapolation is suppressed. 
The kaon production is an order of magnitude smaller than the pion
production, with $\Km$ production particularly suppressed relative to $\Kp$ and
$\Kz$ production.

\section{Prediction of Neutrino Flux at MiniBooNE}
\label{sec:neutrinoflux}
\begin{figure}[thp]
\begin{center}
\includegraphics[width=13 cm]{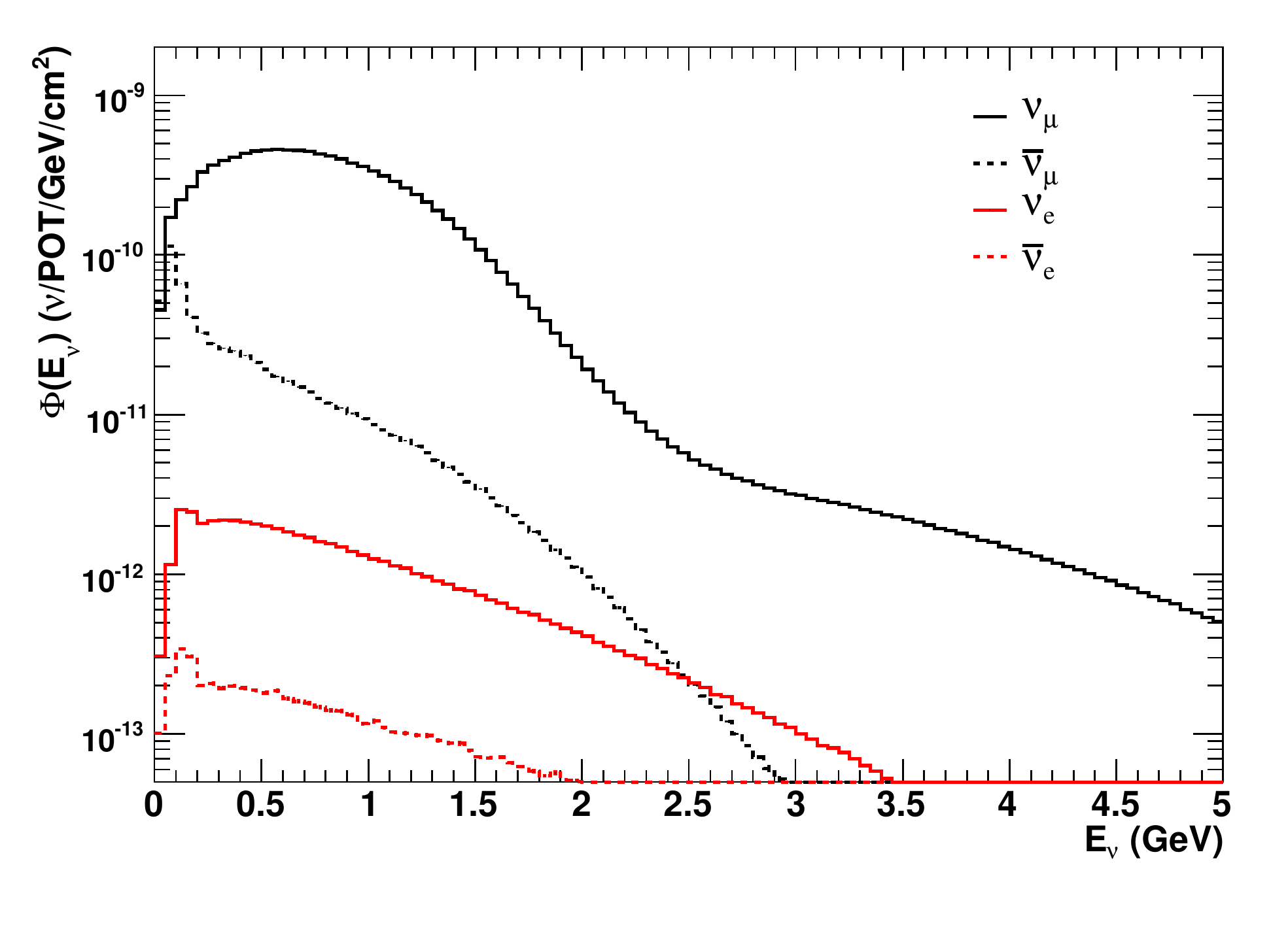}
\caption{ \label{fig:flux_nu} Total predicted flux at the MiniBooNE detector by neutrino species with horn
in neutrino mode.}
\vskip 0.5 cm
\includegraphics[width=13 cm]{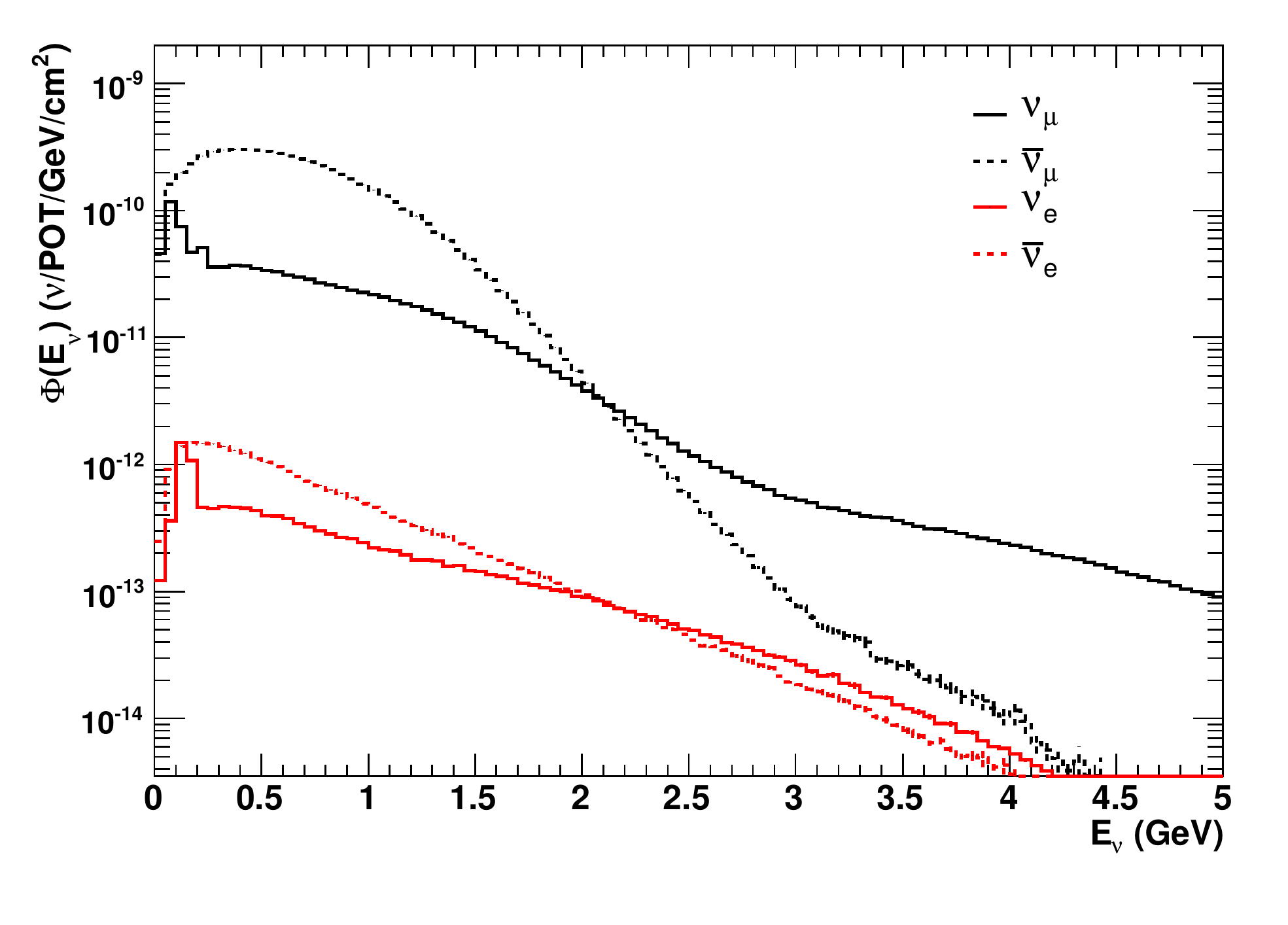}
\caption{ \label{fig:flux_nub} Total predicted flux at the MiniBooNE detector by neutrino species with
horn in anti-neutrino mode.}
\end{center}
\end{figure}

The results of the simulation are summarized in Figures \ref{fig:flux_nu} and \ref{fig:flux_nub},
which show the total predicted flux of each neutrino species at the 
MiniBooNE detector in neutrino mode and anti-neutrino modes, respectively.
In each case, the $\nue/\nueb$ contribution is less than $1\%$ at the peak of the $\num/\numb$ flux, though it rises at higher energies. 
As shown, the predicted fluxes exhibit many features that are better understood by analyzing the 
sources of each  component of the flux.

The integrated contribution of each (anti-)neutrino species, along
with their dominant decay chains, are shown in Table \ref{tab:numodeflux} 
for the neutrino-mode horn configuration, and Table \ref{tab:nubarmodeflux} 
for the anti-neutrino-mode horn configuration. The dominant contribution from decay chains in which
the parent meson is produced by a nucleon is separated from those in which it is produced by a meson interaction.
This is due to the qualitatively different level of systematic understanding for the two processes. For the former,
the production cross sections are based on the particle production experiments described in 
Section \ref{sec:secondary}, with systematic uncertainties propagated from the uncertainties reported by
these experiments. For the latter, the simulation relies on the default Geant4 hadronic interaction model
to provide the production cross sections. Fortunately, the latter is a small contribution to the flux
in all cases.

%Num mode 
%Num  flux 5.19343e-10
%Numb flux 3.26e-11
%Nue  flux 2.86911e-12
%Nueb flux 2.99862e-13
\begin{table}[tbhp]
\begin{center}
\begin{tabular}{l|rr|rr}  \hline\hline
		& \multicolumn{2}{c}{$\num$}   & \multicolumn{2}{c}{$\numb$} \\ \hline  
Flux ($\nu/\cm^2/\mbox{POT}$) 
		&  		& $5.19\times 10^{-10}$ &   	& $3.26\times 10^{-11}$ \\
Frac. of Total  &             	&  $93.6\%$        &              &  5.86\% \\ \hline  
Composition 	& $\pip$:& 96.72\%            &$\pim$:          & 89.74\% \\	       
		& $\Kp$: & 2.65\%             &$\pip\to\mup$:   &  4.54\% \\	       
		& $\Kp\to\pip$: & 0.26\%      & $\Km$:          &  0.51\%   \\ 	       
		& $\Kz\to\pip$: & 0.04\%      & $\Kz$:          &  0.44\%    \\	       
		& $\Kz$:        & 0.03\%      & $\Kz\to\pim$:   &  0.24\%   \\	       
		& $\pim\to\mun$:& 0.01\%      & $\Kp\to\mup$:   &  0.06\%   \\	       
		& Other:        & 0.30\%      & $\Km\to\pim$:  &  0.03\%   \\ 	       
		&               &             & Other:          &  4.43\%   \\ \hline      
\end{tabular}
%\caption{\label{tab:numodenumuflux}Predicted $\num$ and $\numb$ fluxes at the MiniBooNE detector with horn in neutrino mode.
%The contribution of flux from meson decays where the parent particle in the decay chain is
%produced by proton or neutron interaction. The ``other'' category includes channels with
%contributions less than those shown, along with cases where the parent particle in the
%decay chain is produced by a meson interaction.}
\vskip 0.2 cm
\begin{tabular}{l|rr|rr}  \hline\hline
 		& \multicolumn{2}{c}{$\nue$} & \multicolumn{2}{c}{$\nueb$}  \\ \hline   
Flux ($\nu/\cm^2/\mbox{POT}$)	&    & $2.87\times 10^{-12}$ & & $3.00\times 10^{-13}$ \\          
Frac. of Total	&               & 0.52\%    &               &      0.05\%  \\ \hline  
Composition 	& $\pip\to\mup$:& 51.64\%   & $\KL$:        &     70.65\%  \\	       
		& $\Kp$:        & 37.28\%   & $\pim\to\mun$ &     19.33\%  \\	       
		& $\KL$:        & 7.39\%    & $\Km$:        &      4.07\%  \\ 	       
		& $\pip$:       & 2.16\%    & $\pim$:       &      1.26\%  \\	       
		& $\Kp\to\mup$: & 0.69\%    & $\Km\to\mun$: &      0.07\%  \\	       
		& Other:        & 0.84\%    & Other:        &      4.62\%  \\ \hline	       
\end{tabular}
\end{center}
\vskip 0.2 cm
\caption{\label{tab:numodeflux}Predicted $\num/\numb$ (top) and $\nue/\nueb$ (bottom) fluxes at the MiniBooNE detector with horn in neutrino mode.
The contribution of flux from meson decays where the parent particle in the decay chain is
produced by proton or neutron interaction. The ``other'' category includes channels with
contributions less than those shown, along with cases where the parent particle in the
decay chain is produced by a meson interaction.}
\end{table}
% Numb mode
%Num  flux 5.42275e-11
%Numb flux 2.92817e-10
%Nue  flux 6.70729e-13
%Nueb flux 1.26625e-12
\begin{table}[tbhp]
\begin{center}
\begin{tabular}{l|rr|rr}  \hline\hline
		& \multicolumn{2}{c}{$\num$}   & \multicolumn{2}{c}{$\numb$} \\ \hline  
Flux ($\nu/\cm^2/\mbox{POT}$) 
		&  		& $5.42\times 10^{-11}$ &   	& $2.93\times 10^{-10}$ \\
Frac. of Total  &             	&  15.71\%        &              & 83.73\% \\ \hline  
Composition 	& $\pip$:       & 88.79\%	  & $\pim$:		& 98.4\% \\	       
		& $\Kp$:        &  7.53\%	  & $\Km$:		&  0.18\%\\	       
		& $\pim\to\mun$:&  1.77\%	  & $\Kz\to\pim$:   	&  0.05\%\\ 	       
		& $\Kz$:        &  0.26\%    	  & $\Kz$:         	&  0.05\%\\	       
		& Other:        &  2.00\%         & $\pip\to\mup$:  	&  0.03\%\\	       
		&  	        &                 & $\Km\to\pim$:   	&  0.02\%\\	
		&		&		  & Other:		&  1.30\% \\ \hline
\end{tabular}
%\caption{\label{tab:nubarmodenumuflux}Predicted $\num$ and $\numb$ flux at the MiniBooNE 
%detector with horn in anti-neutrino mode.
%The contribution of flux from meson decays where the parent particle in the decay chain is
%produced by proton or neutron interaction. The ``other'' category includes channels with
%contributions less than those shown, along with cases where the parent particle in the
%decay chain is produced by a meson interaction.}
\vskip 0.2 cm
\begin{tabular}{l|rr|rr}  \hline\hline
 		& \multicolumn{2}{c}{$\nue$} & \multicolumn{2}{c}{$\nueb$}  \\ \hline   
Flux ($\nu/\cm^2/\mbox{POT}$)	&    & $6.71\times 10^{-13}$ & & $1.27\times 10^{-12}$ \\          
Frac. of Total	&                & 0.2\%     &               &      0.4\%  \\ \hline  
Composition 	& $\Kp$  :       &   51.72\% & $\pim\to\mun$: &  75.67\% \\	       
		& $\Kz$  :       &   31.56\% & $\Kz$:         &  16.51\% \\	       
		& $\pip\to\mup$  &   13.30\% & $\Km$:         &   3.08\% \\ 	       
		& $\pip$  :      &    0.83\% & $\pim$:        &   2.58\% \\	       
		& $\Kp\to\mup$:  &    0.41\% & $\Km\to\mun$:  &   0.06\% \\	       
		& Other:         &    2.17\% & Other          &   2.10\% \\ \hline	       
\end{tabular}
\end{center}
\vskip 0.2 cm
\caption{\label{tab:nubarmodeflux}Predicted $\num/\numb$ (top) and $\nue/\nueb$ 
(bottom) fluxes at the MiniBooNE 
detector with horn in anti-neutrino mode.
The contribution of flux from meson decays where the parent particle in the decay chain is
produced by proton or neutron interaction. The ``other'' category includes channels with
contributions less than those shown, along with cases where the parent particle in the
decay chain is produced by a meson interaction.}

\end{table}

\begin{figure}[pth]
\begin{center}
\includegraphics[width=14 cm]{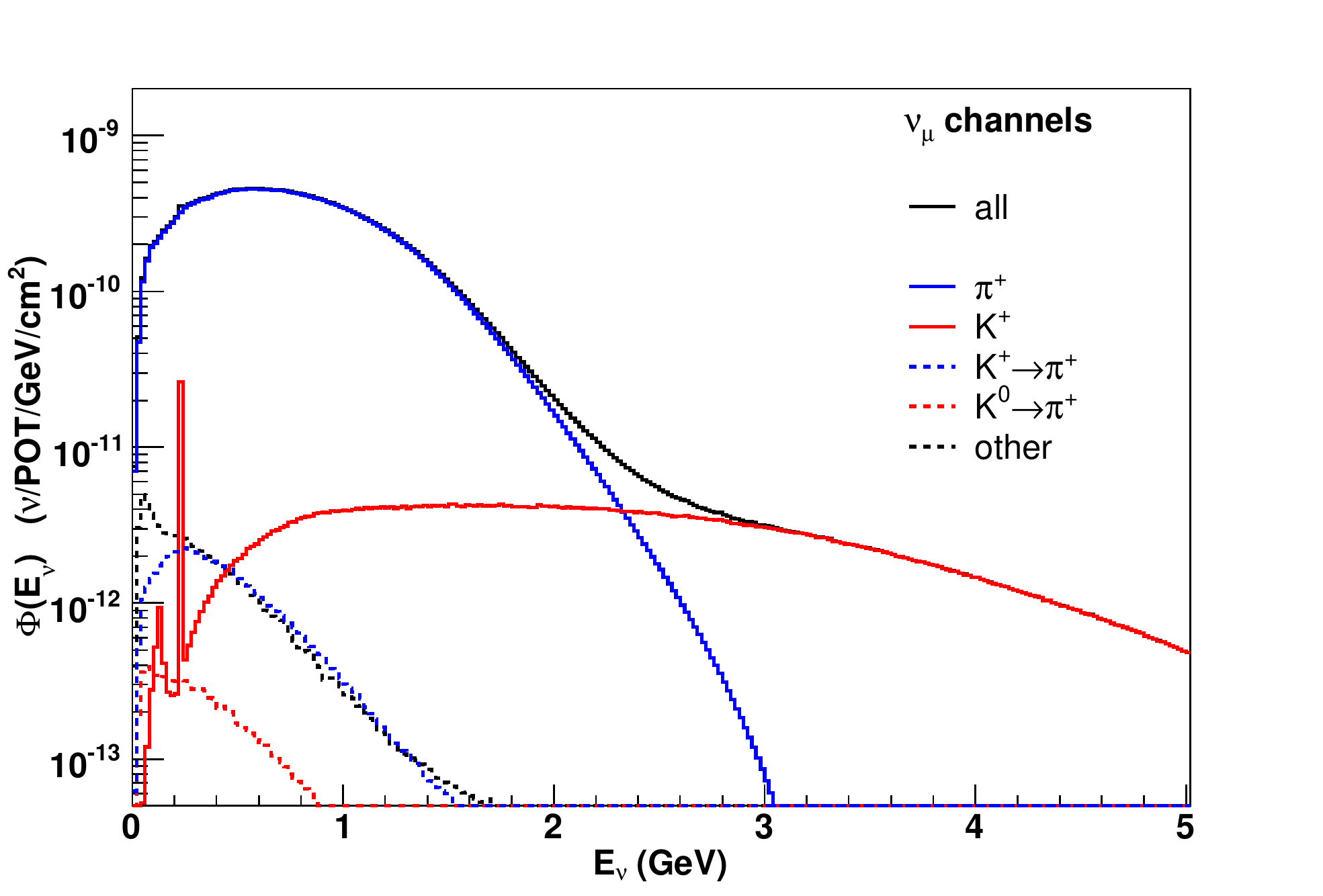}
\includegraphics[width=14 cm]{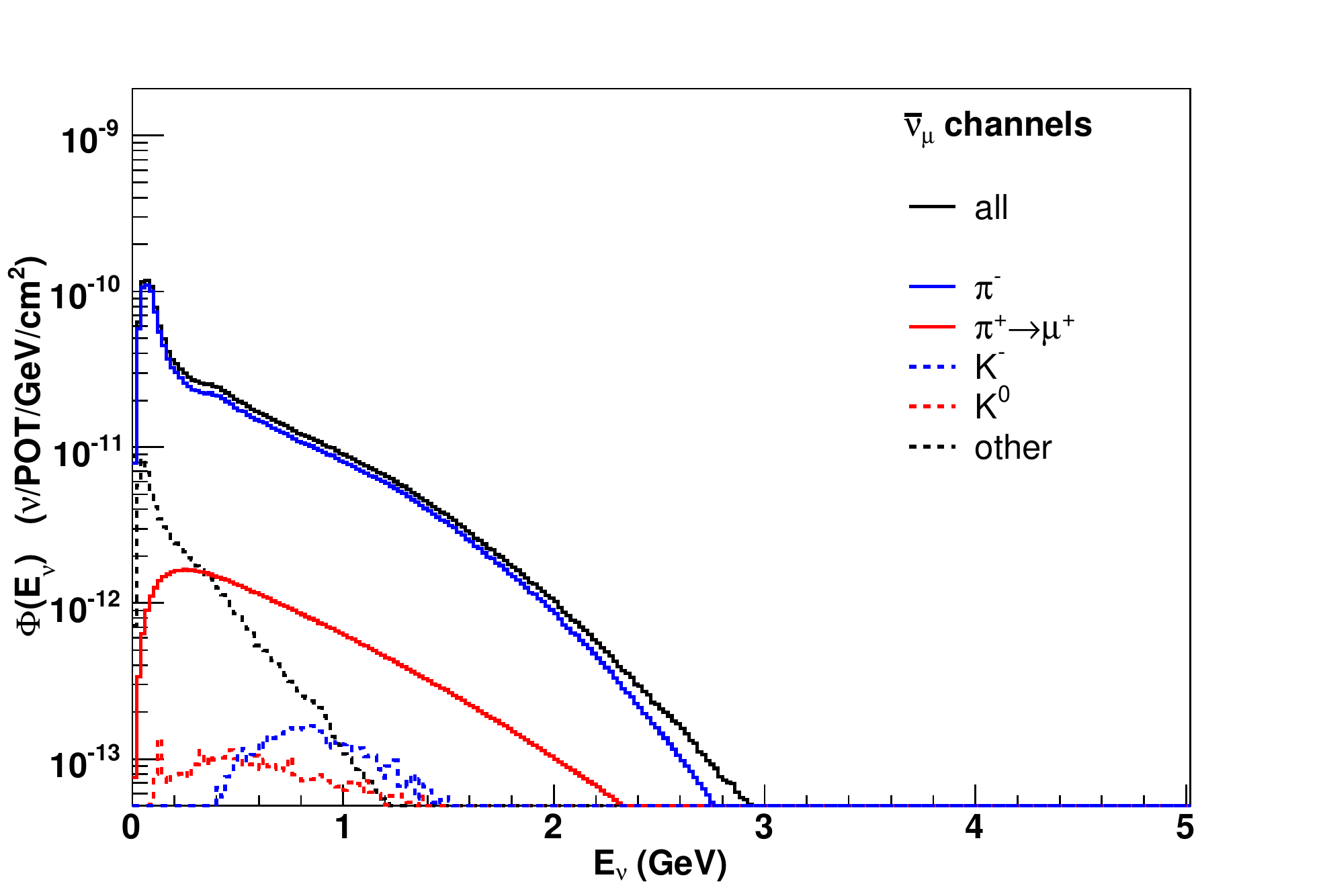}
\caption{ \label{fig:flux_nu_numu} Predicted $\num$ (top) and $\numb$ (bottom) fluxes at the MiniBooNE detector  by parent meson species with horn in neutrino mode. The black line is the total predicted flux, while  all the subcomponents apart from the dashed black are  from nucleon-induced meson production of the indicated decay chains. The dashed black histogram includes all other contributions, primarily  from  meson decay chains initiated by meson-nucleus interactions.}
\end{center}
\end{figure}

\begin{figure}[pth]
\begin{center}
\includegraphics[width=14 cm]{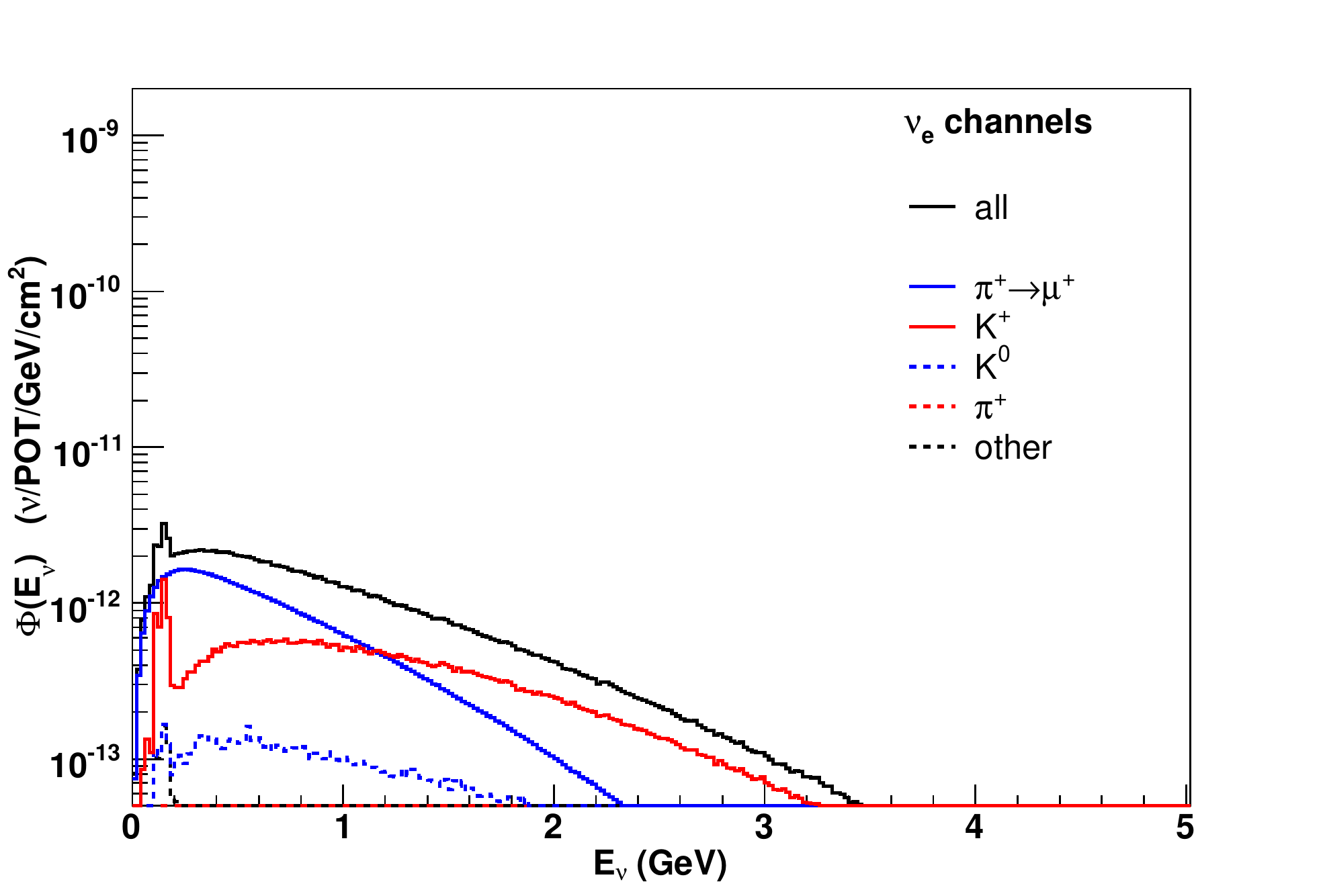}
\includegraphics[width=14 cm]{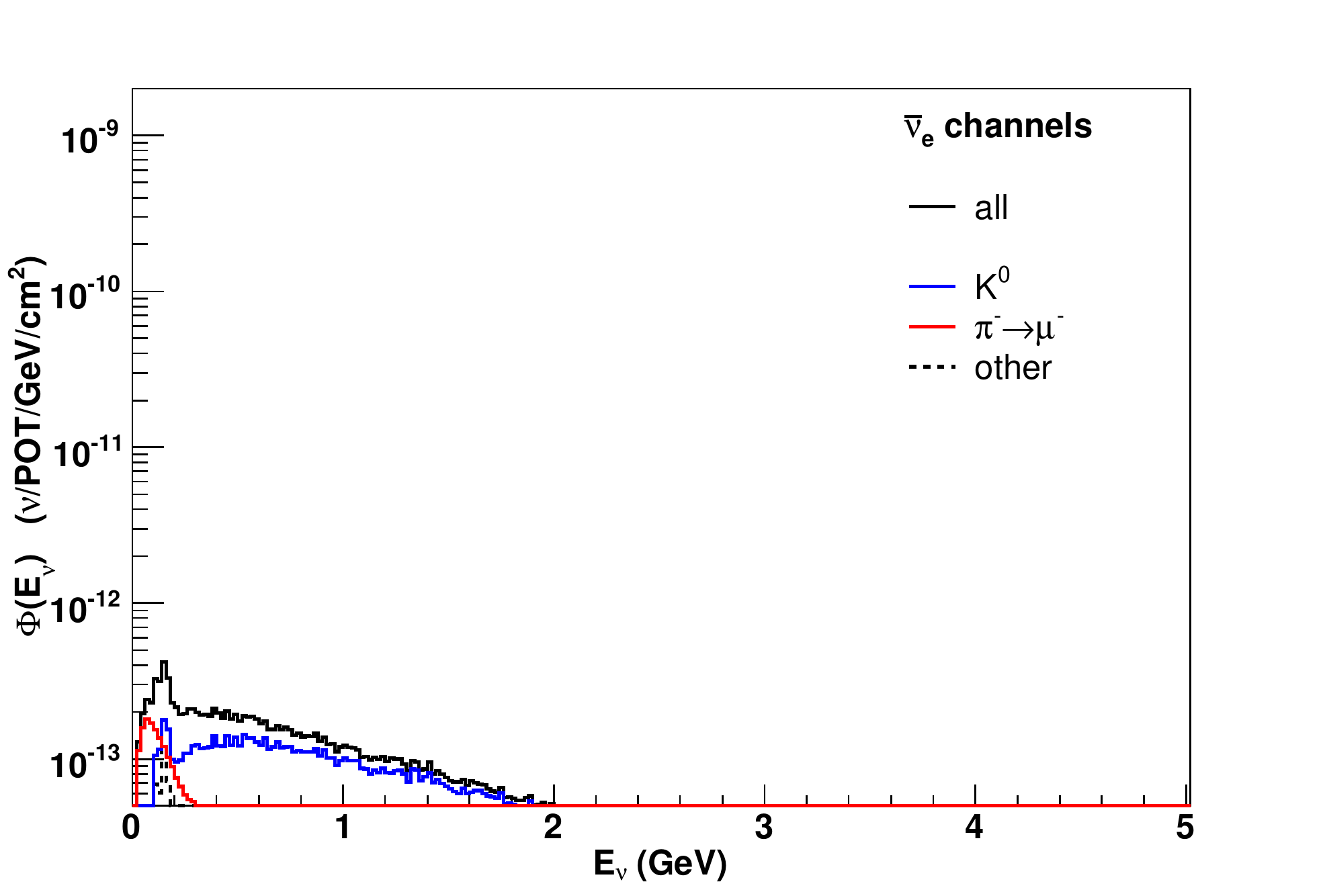}
\caption{ \label{fig:flux_nu_nue} Predicted $\nue$ (top) and $\nueb$ (bottom)
flux at the MiniBooNE detector by parent meson species with horn in neutrino mode 
The black line is the total predicted flux, while all the subcomponents apart from
the dashed black are  from nucleon-induced meson production
of the indicated decay chains. The dashed black histogram includes all other contributions, primarily 
from  meson decay chains initiated by meson-nucleus interactions.}
\end{center}
\end{figure}

Figure \ref{fig:flux_nu_numu} shows the channels through which the $\num$ and $\numb$
are produced in neutrino mode. For the $\num$ flux, the $\pip\to\num$ contribution is 
dominant for energies less than $2\gev$, while the $\Kp\to\num$ flux become dominant at higher energies. 
The two peaks in the $\Kp$ flux at low energies are from two- and three-body $\Kp$ decays at rest. Due to 
the
relative size of the $\pip$ flux, however, they are not visible in the total $\num$ flux.
There is a small contribution to the flux from pions produced in the decay of kaons, and a similar
contribution from tertiary meson-induced production of other mesons that decay to produce $\num$. 

For the $\numb$ in neutrino mode, $\pim\to\numb$ flux is dominant at all energies. The 
next largest contribution comes from the $\pip\to\mup\to\numb$ decay chain.  For the $\num$ flux, 
the analogous contribution from the $\pim\to\mu^-\to\num$ decay chain is suppressed by the defocusing 
of the $\pim$. The kaon contribution is suppressed by the lower rate of $\Km$ production relative 
to $\Kp$ production. Apart from low energies ($<200\mev$) the predicted  $\numb$ flux is typically
$\sim 6\%$ of the $\num$ flux.

The channels through which $\nue$ and $\nueb$ are produced in neutrino mode are shown in 
Figure \ref{fig:flux_nu_nue}. For the $\nue$ flux, the two dominant components are the
$\pip\to\mup\to\nue$ decay and three-body $\Kp\to\nue$ decay, where the former is dominant at 
low energies ($<1\gev$) and the latter is dominant at higher energies. The peak in the $\Kp\to\nue$
spectrum at low energies is from the decay of $\Kp$ at rest (the peak from two-body decay
is much smaller due to helicity suppression). For $\nueb$, the $\pim\to\mu^-\to\nueb$ flux contributes
only at lower energies due to the defocusing of the $\pim$, and the $\Km\to\nueb$ contribution is suppressed
both by the lower production rates and the defocusing. The rest of the spectrum is dominated
by $\KL$ decay. As in the $\num/\numb$ case, the predicted $\nueb$ flux is $\sim 10\%$ of the $\nue$
flux.

\begin{figure}[pth]
\begin{center}
\includegraphics[width=14 cm]{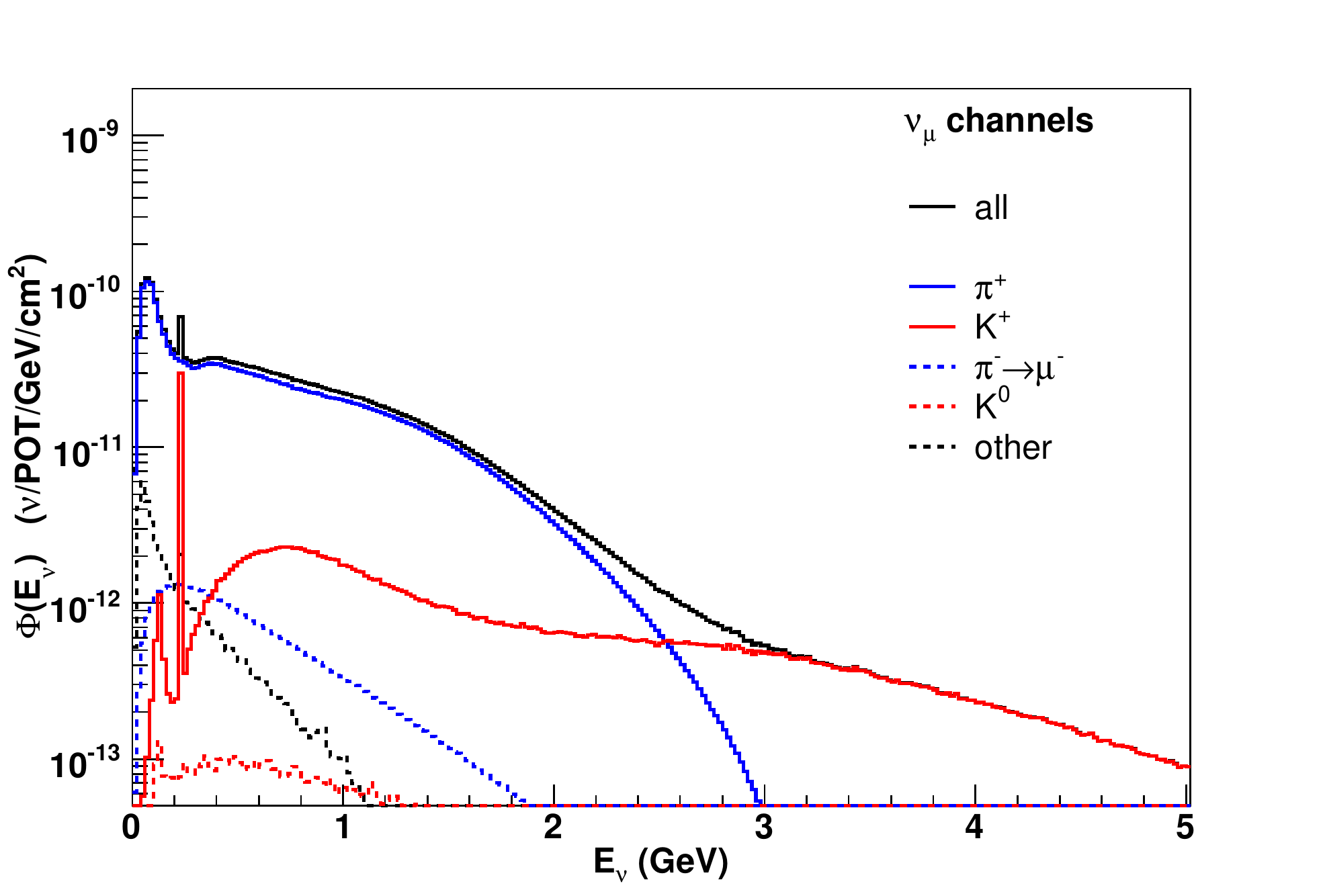}
\includegraphics[width=14 cm]{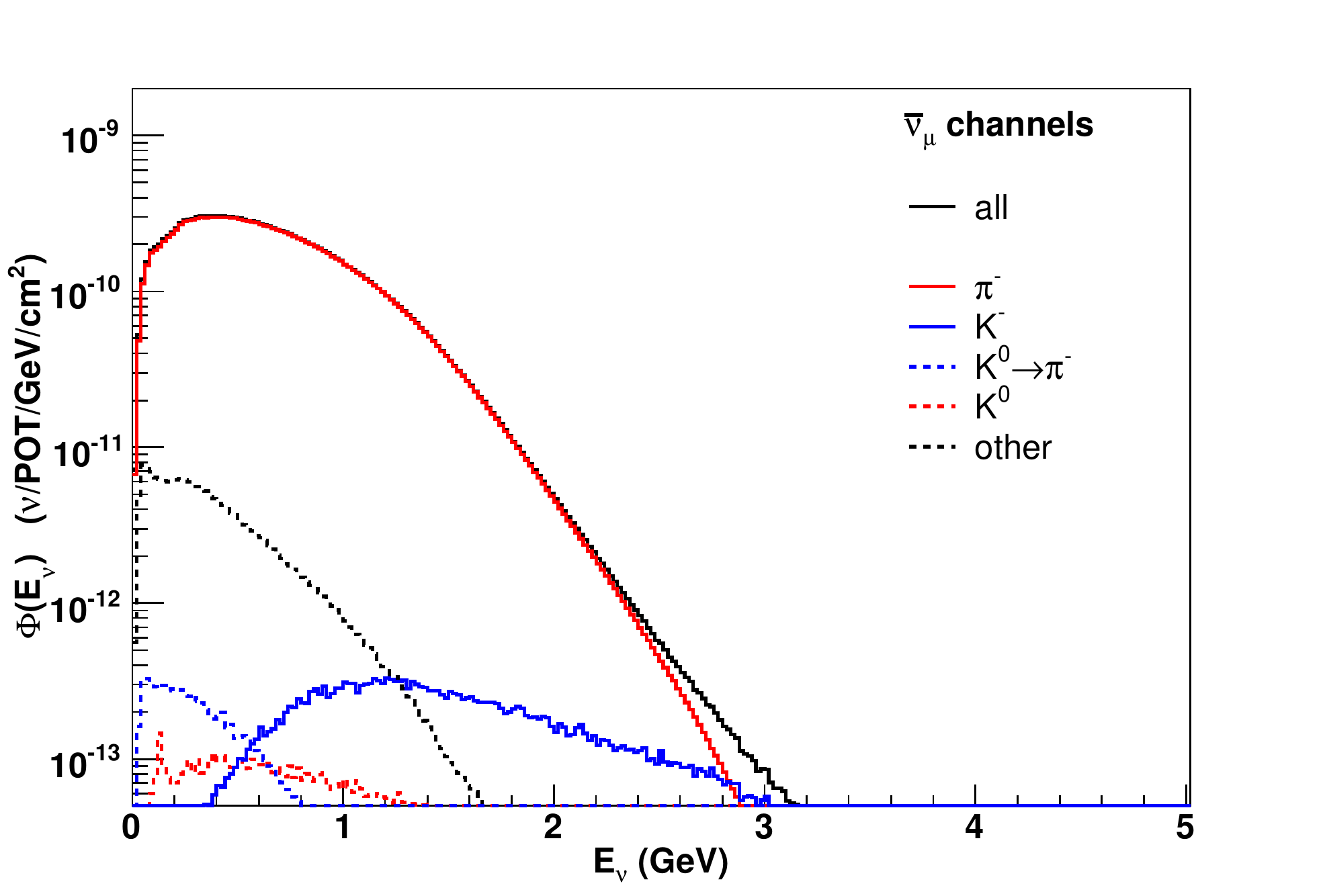}
\caption{ \label{fig:flux_nub_numu} Predicted $\num$ (top) and $\numb$ (bottom) fluxes at the MiniBooNE detector 
by parent meson species with horn in anti-neutrino mode. The black line is 
the total predicted flux, while all the subcomponents apart from
the dashed black are  from nucleon-induced meson production.
 The dashed black histogram includes all other contributions, primarily 
from  meson decay chains initiated by meson-nucleus interactions.}
\end{center}
\end{figure}

\begin{figure}[pth]
\begin{center}
\includegraphics[width=14 cm]{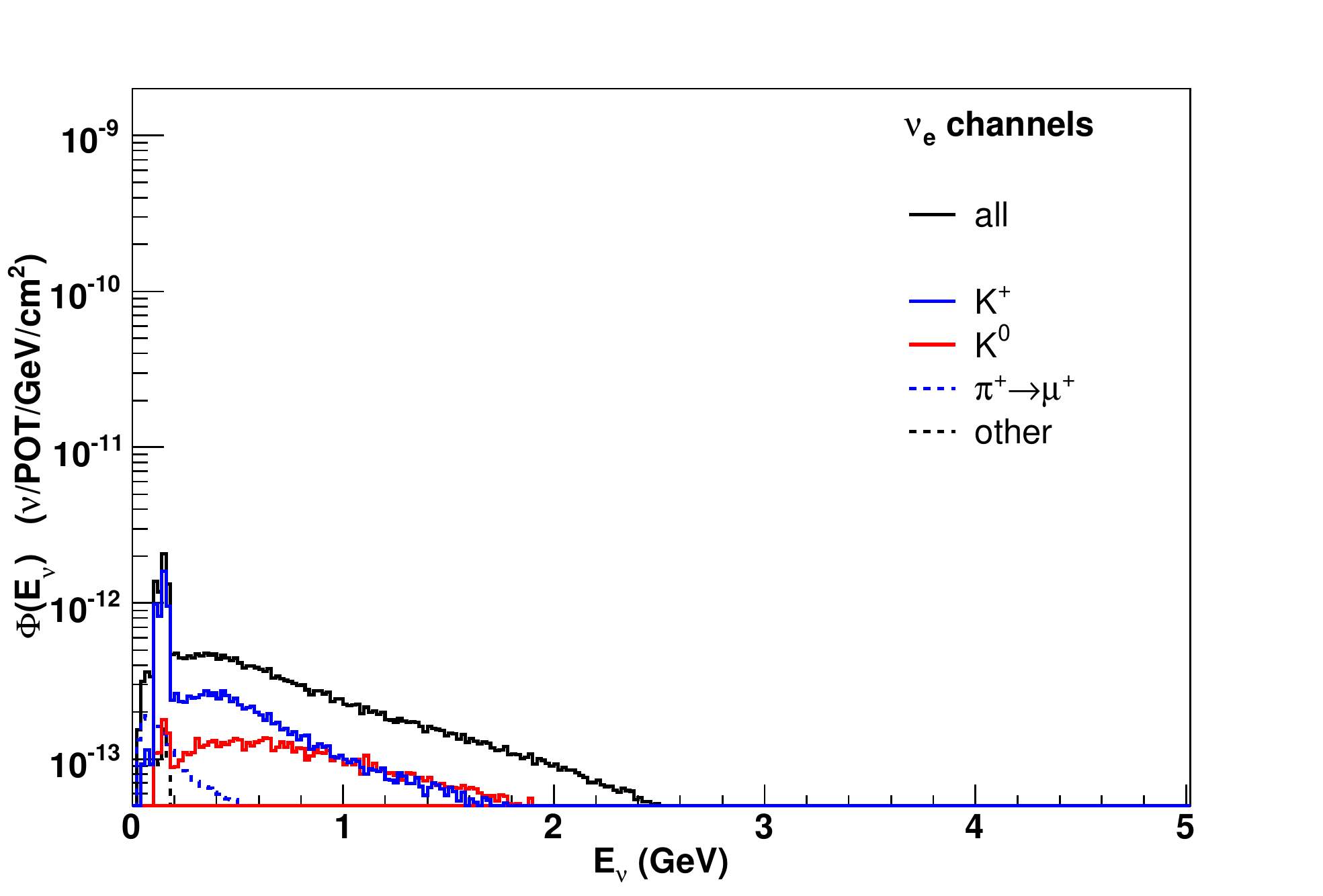}
\includegraphics[width=14 cm]{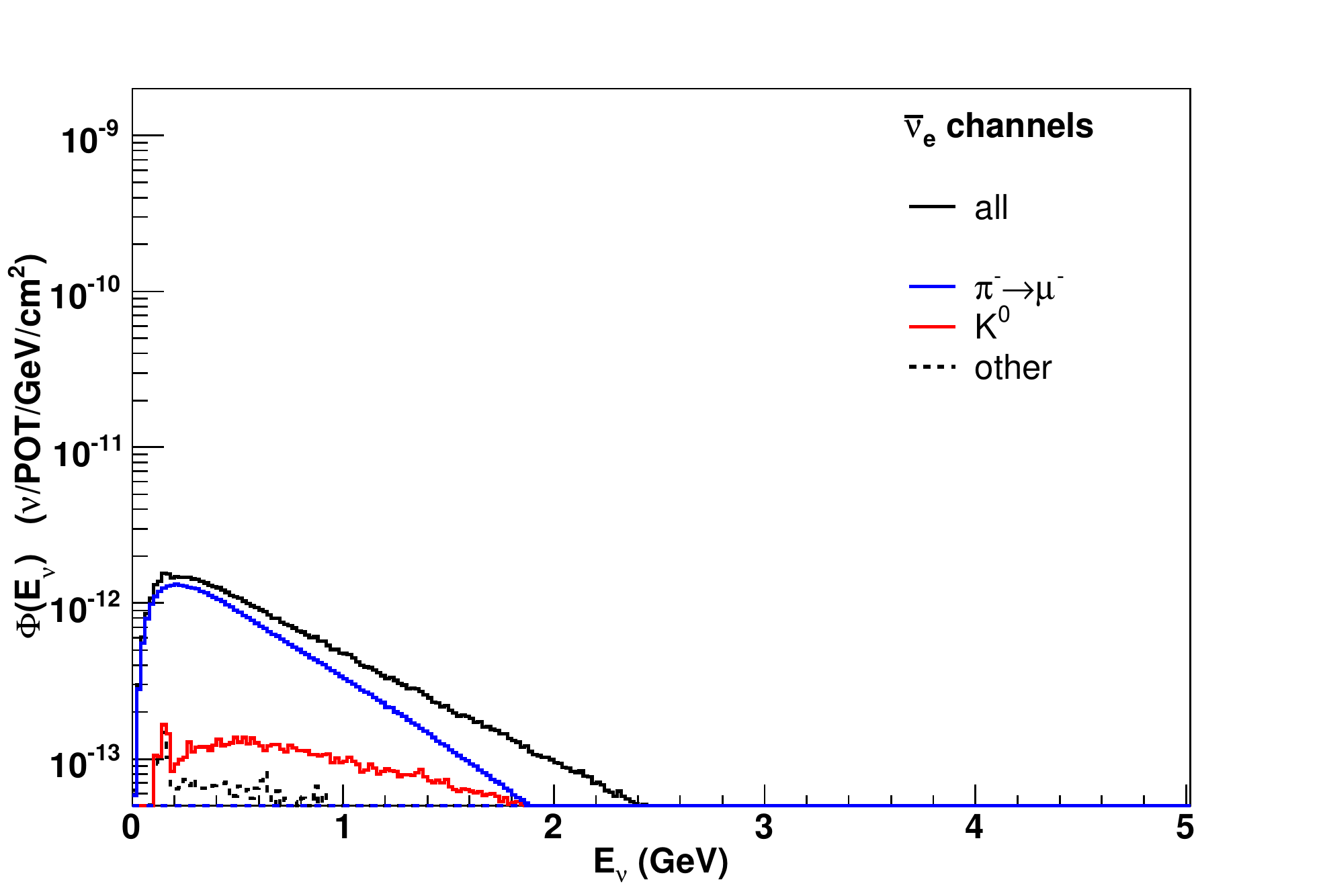}
\caption{ \label{fig:flux_nub_nue} Predicted $\nue$ (top) and $\nueb$ (bottom) fluxes at the 
MiniBooNE detector by parent meson species with horn in anti-neutrino mode. The 
black line is the total predicted flux, while all the subcomponents apart from
the dashed black are  from nucleon-induced meson production.
 The dashed black histogram includes all other contributions, primarily 
from  meson decay chains initiated by meson-nucleus interactions.}
\end{center}
\end{figure}

Figures \ref{fig:flux_nub_numu} and \ref{fig:flux_nub_nue} show a similar composition
for the predicted anti-neutrino mode flux.  The $\numb$ flux is dominated at all energies
by $\pim\to\numb$ decays; the suppressed production of $\Km$ results in the contribution
of $\Km\to\numb$ being much smaller than the corresponding $\Kp\to\num$ contribution in
neutrino mode. Furthermore, since the $\Km$
that come to rest are captured, the $\numb$ flux does not show the peaks from
two-body and three-body decay-at-rest at low energies that are found in the $\num$ from
$\Kp$ decay in both neutrino and anti-neutrino mode. 
It can also be seen that the high energy flux of $\num$ is not substantially suppressed relative to the $\numb$. In fact,
despite the  defocusing of $\Kp$, the $\Kp\to\num$ flux is larger than 
that of the $\Km\to\numb$ decay.  This is due to the relative production rates and, at high energies, 
the leading particle effect
where $\pip$ and $\Kp$ have a harder momentum spectrum relative to their negatively-charged
counterparts. The high momentum of the particles that produce these neutrinos, along
with their forward angular distribution, result in less defocusing from the horn for the wrong-sign component (positive (negative) particles for (anti-)neutrino mode). A similar effect is seen for
the $\nue/\nueb$ components in anti-neutrino mode: while the $\nueb$  are dominated by
$\mu^-$ decays at energies below 2 GeV, the $\Kp\to\nue$ flux is larger
than the $\Km\to\nueb$ flux. A related observation is the fact that
while the absolute rate of $\nue/\nueb$ from $\KL$ is unchanged from neutrino mode, the relative contribution is
 much stronger in anti-neutrino mode.

\section{Systematic Uncertainties}
\label{sec:systematic}
The systematic uncertainties in the neutrino flux prediction come from several sources: 
\begin{itemize}
\item Proton delivery: The simulation determines the rate and spectrum of
neutrinos per proton-on-target. This information is combined with the
number of protons delivered to the target to determine the number of neutrinos
passing through the MiniBooNE detector over the data collection period. As a result,
the predicted number of neutrino interactions in the detector varies directly
with the uncertainty in the number of protons-on-target. A related
uncertainty arises from the optics of the proton beam which can change the expected
number of protons interacting in the target (or elsewhere), changing the neutrino flux.
\item Particle Production: The uncertainties in the rate and spectrum of secondary
particles produced in the p-Be interactions likewise affect the rate and spectrum
of the neutrinos they produce. This is the dominant uncertainty.
\item Hadronic Interactions: The rate of hadronic interactions affect many aspects
of the neutrino production, including the rate of p-Be interactions as well as the 
probability for mesons to survive possible hardonic interactions in the target or horn 
and decay to produce neutrinos. Uncertainties
in the rate of these interactions affect both the rate and shape of the flux.
\item Horn magnetic field: The focusing properties of the horn change with the current
as well as the distribution of the magnetic field within the conducting elements. Uncertainties
in these properties result in spectral distortions of the neutrino flux.
\item Beamline geometry: Misalignments or displacements of the beamline components from
their expected orientation and locations can affect the neutrino flux in many ways. For example, a misalignment
can result in the detector being exposed to a different part of the neutrino flux than expected. A 
displacement of the target with respect to the horn can result in a variation in the focusing properties.
\end{itemize}

\subsection{Proton delivery}
\label{sec:primarysys}
The systematic uncertainties associated with the delivery of the primary proton beam
to the beryllium target can be divided into two parts: the uncertainty in the
number of protons delivered to the beamline and the uncertainty in the
number which actually strike the target. Having entered the target, there
are further uncertainties associated with how often the protons will interact
to produce secondary particles based on the assumed hadronic cross sections; 
we consider these uncertainties in Section \ref{sec:hadint}.

As mentioned in Section \ref{sec:bnb}, the protons delivered to the BNB are measured
by two toroids upstream of the target. The systematic uncertainty in the resulting
spill-by-spill measurements has been estimated to be $2\%$ based on uncertainties
in the toroid circuit elements and uncertainties in the calibration procedure. Since the
overall neutrino flux scales with the delivered protons, this source or error can be treated
as an overall normalization uncertainty. The toroid measurements have been cross
checked by measuring the activation on a gold foil inserted into the beam. The number of
protons striking the foil inferred from this measurement agree with the toroid measurements 
within the $\sim 10\%$ uncertainty of the measurement.

The effect of uncertainties in the primary beam optics, most notably the transverse profile
and focusing and divergence properties, have been estimated by simulating the effects introduced
by perturbing the default beam parameters. A number of different configurations, including
varying the focal point across the length of the target, a ``pin'' configuration with no transverse
spread or angular divergence, and a ``pencil'' configuration with transverse spread but no
angular divergence, have been considered.
 The resulting changes to the number of protons expected to interact in the target is less than $1\%$, which is taken as a
systematic uncertainty in the overall normalization of the neutrino flux.

\begin{figure}[p]
\begin{center}
\includegraphics[width=15 cm]{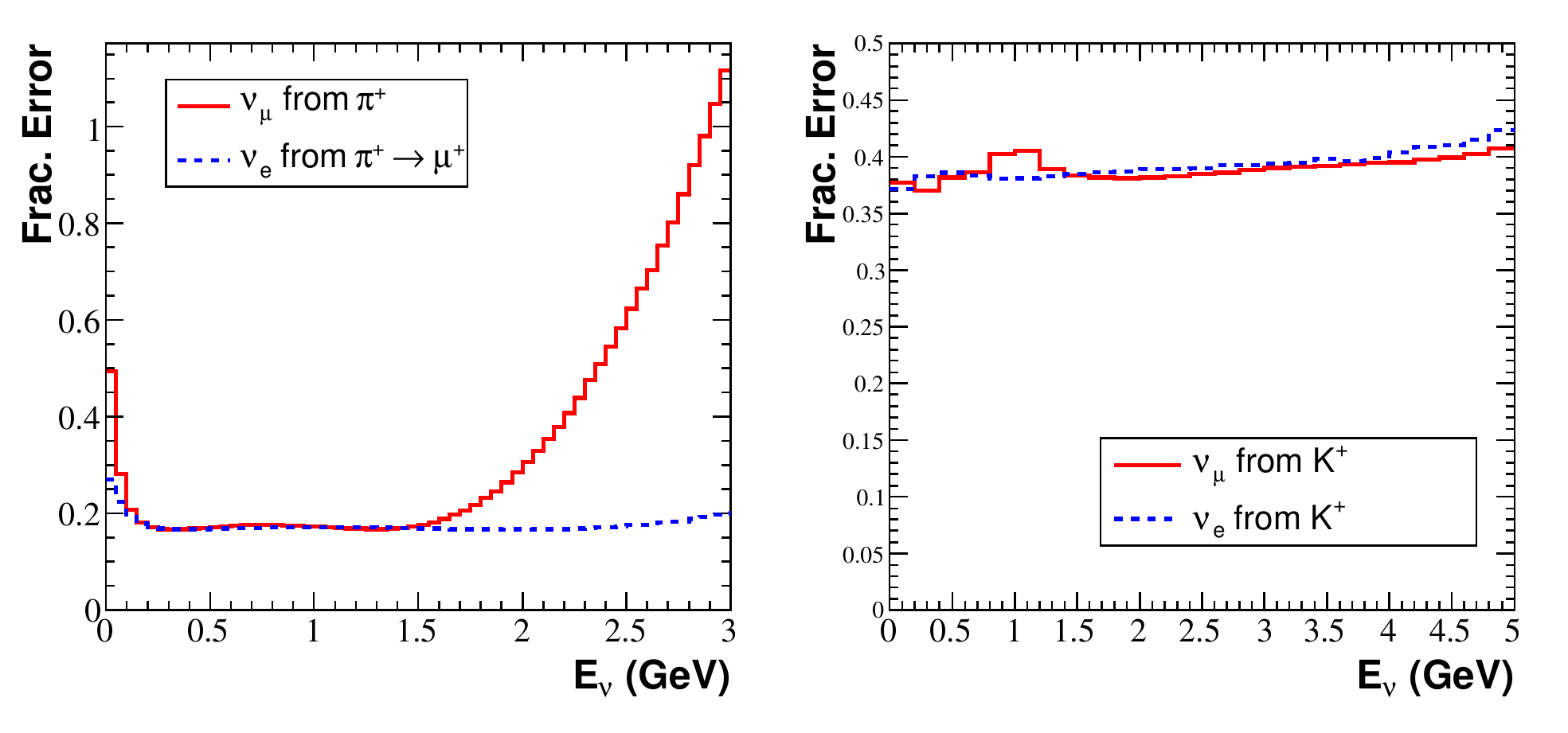}
\caption{\label{fig:proderr}Left: The fractional uncertainties in the $\pip\to\num$  and $\pip\to\mup\to\nue$
flux  with the horn in neutrino mode due to uncertainties in the $\pip$ production in $\mbox{p}$-Be interactions.
Right: Same for the $\Kp\to\num$ and $\Kp\to\nue$ flux from uncertainties in the $\Kp$ production
in $\mbox{p}$-Be interactions.}
\vskip 0.5 cm
\includegraphics[width=15 cm]{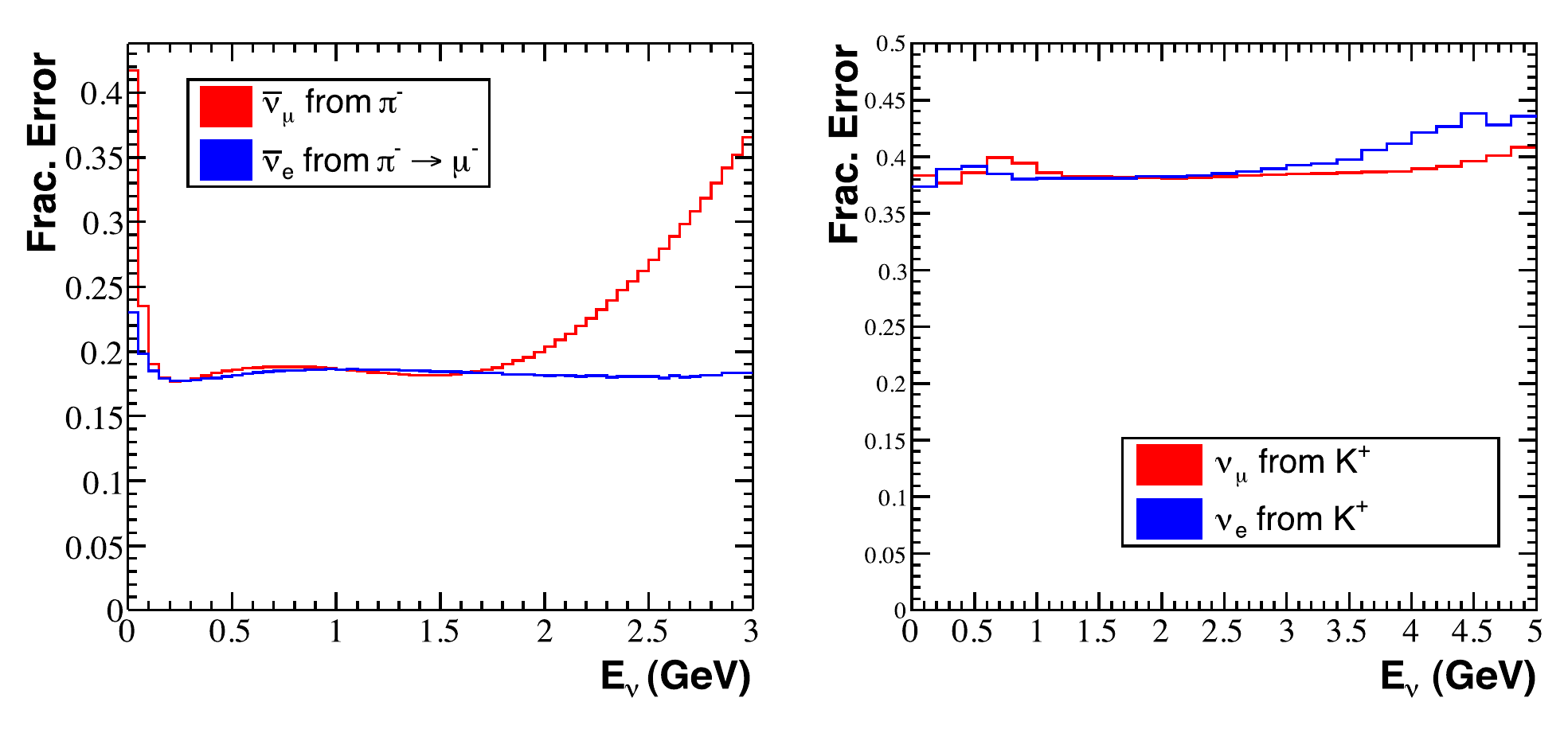}
\caption{\label{fig:proderr_nubar}Left: The fractional uncertainties in the $\pim\to\numb$  and $\pim\to\mun\to\nueb$ flux  with the horn in anti-neutrino mode due to uncertainties in the $\pim$ production in $\mbox{p}$-Be interactions.
Right: Same for the $\Kp\to\num$ and $\Kp\to\nue$ flux from uncertainties in the $\Kp$ production
in $\mbox{p}$-Be interactions.}
\end{center}
\end{figure}

\begin{figure}[t]
\begin{center}
\includegraphics[width=15 cm]{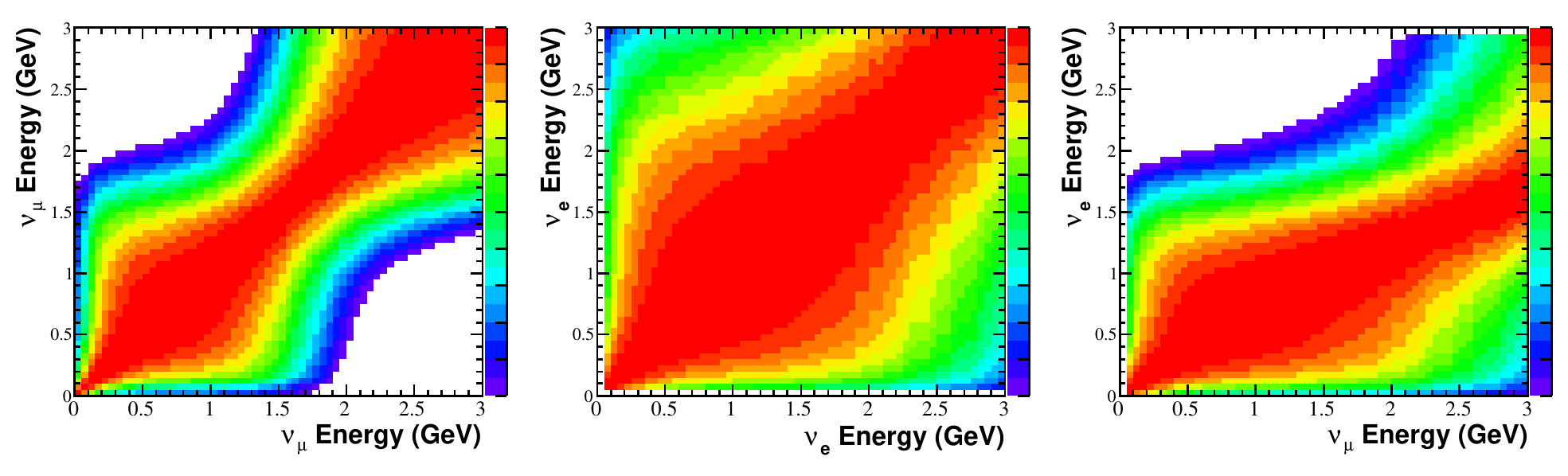}
\vskip 0.2 cm
\caption{\label{fig:pipcorr}Left: Correlation matrix for variations in the $\pip\to\num$ flux
due to uncertainties in $\pip$ production in $\mbox{p}$-Be interactions with the horn in neutrino mode. Center/Right: Same for variations
in the $\pip\to\mup\to\nue$ flux (center) and correlations between the $\pip\to\num$ flux
and the $\pip\to\mup\to\nue$ flux (right). The color scale on each plot ranges from 0 to 1.}
\vskip 0.25 cm
\includegraphics[width=15 cm]{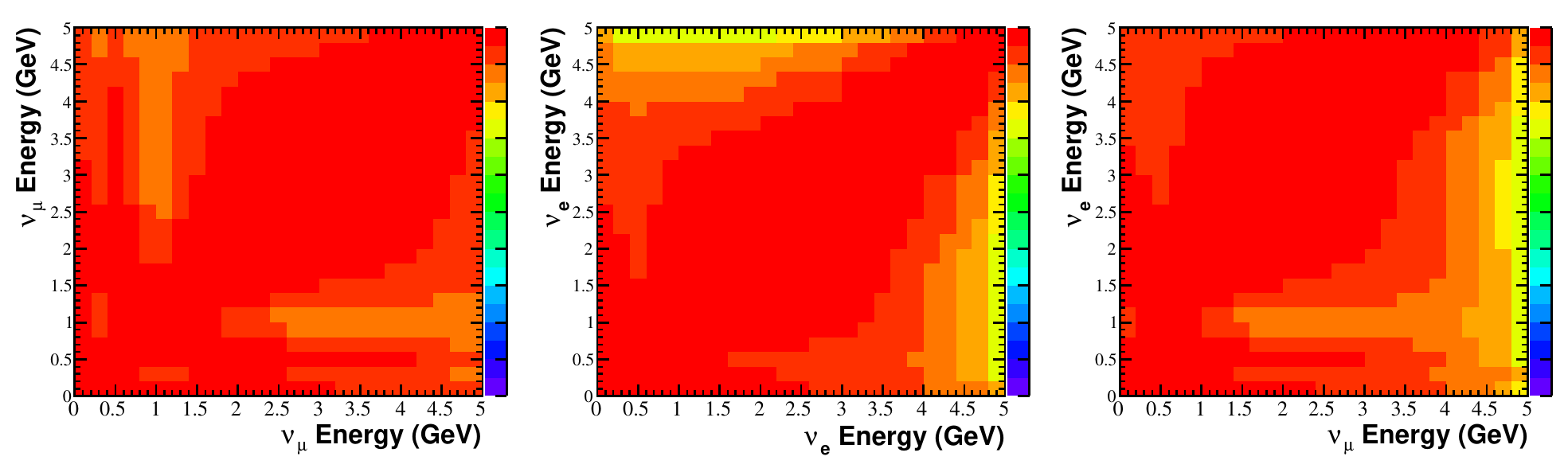}
\vskip 0.2 cm
\caption{\label{fig:kpcorr} Left: Correlation matrix for variations in the $\Kp\to\num$ flux
due to uncertainties in $\Kp$ production in $\mbox{p}$-Be interactions with the horn in neutrino mode. Center/Right: Same for variations
in the $\Kp\to\nue$ flux (center) and correlations between the $\Kp\to\num$ flux
and the $\Kp\to\nue$ flux (right). The color scale on each plot ranges from 0 to 1.}
\end{center}
\end{figure}

\begin{figure}[t]
\begin{center}
\includegraphics[width=15 cm]{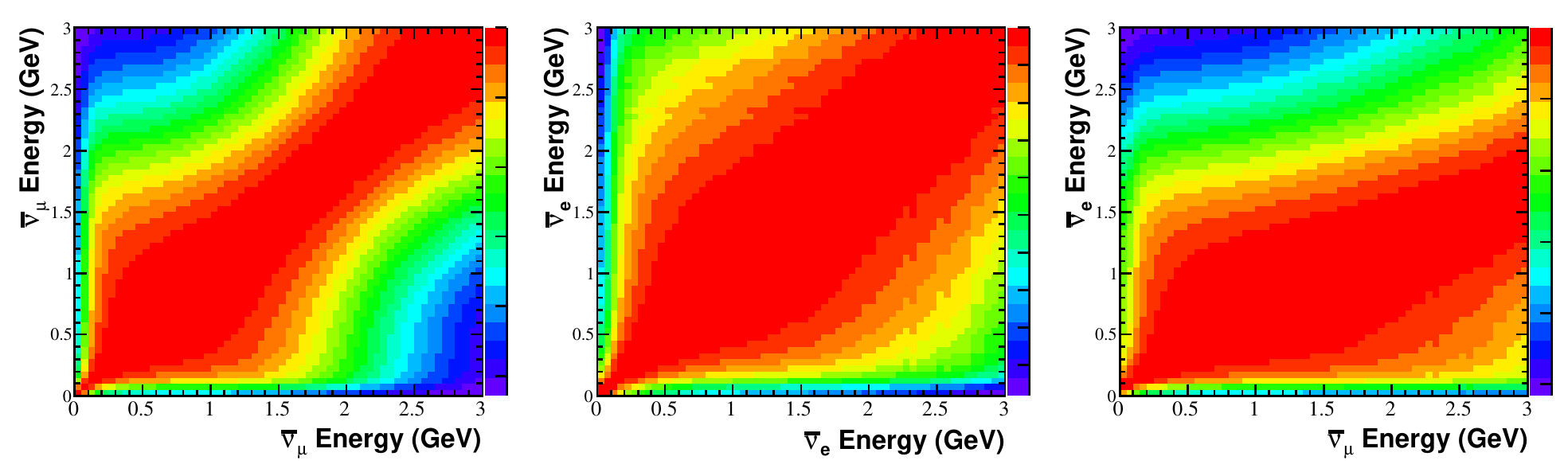}
\vskip 0.2 cm
\caption{\label{fig:pimcorr_nubar}Left: Correlation matrix for variations in the $\pim\to\numb$ flux
due to uncertainties in $\pim$ production in $\mbox{p}$-Be interactions with the horn in
anti-neutrino mode. Center/Right: Same for variations in the $\pim\to\mun\to\nueb$ flux (center) and correlations between the $\pim\to\numb$ flux
and the $\pim\to\mun\to\nueb$ flux (right). The color scale on each plot ranges from 0 to 1.}
\vskip 0.25 cm
\includegraphics[width=15 cm]{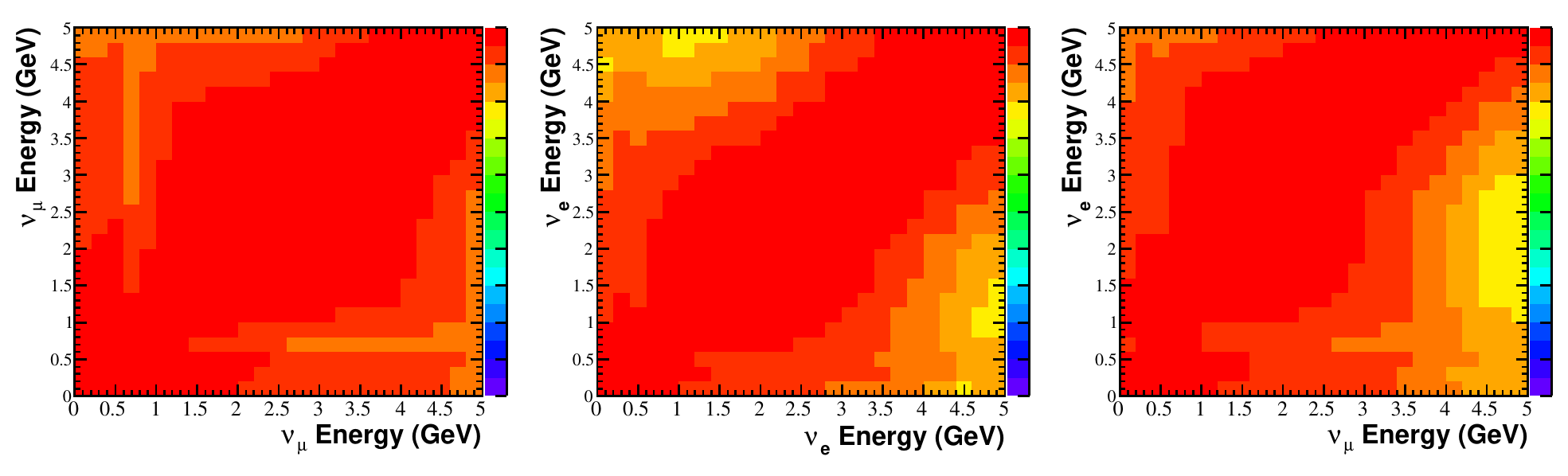}
\vskip 0.2 cm
\caption{\label{fig:kpcorr_nubar} Left: Correlation matrix for variations in the $\Kp\to\num$ flux
due to uncertainties in $\Kp$ production in $\mbox{p}$-Be interactions with the horn in anti-neutrino mode. Center/Right: Same for variations
in the $\Kp\to\nue$ flux (center) and correlations between the $\Kp\to\num$ flux
and the $\Kp\to\nue$ flux (right). The color scale on each plot ranges from 0 to 1.}
\end{center}
\end{figure}

\subsection{Particle production}
\label{sec:prodsys}
The uncertainties in the particle production are summarized as a covariance matrix
in the fitted parameters of the functions parametrizing the double differential cross section
as described in Section \ref{sec:secondary}. The effect of these uncertainties
is propagated to the neutrino flux by drawing random parameter vectors according to the
covariance matrix via the Cholesky decomposition \cite{pdg}. The resulting variation
in the double differential meson production cross section at any point in $(p,\theta)$ can be evaluated
with respect to the default value. The change in the neutrino flux can then
be recalculated by assigning a weight corresponding to the ratio of the double differential
cross section of the secondary particle producing the neutrino with the varied
and default parameters. 

In this way, the flux resulting from different production distributions
summarized by alternate parameters can be calculated without re-running the flux
simulation. By accumulating the covariance of the flux distribution as the parameters are
varied according to their covariance matrix, the uncertainties are propagated into the
neutrino flux. This procedure is repeated for each parent particle species ($\pip$, $\pim$, 
$\Kp$, $\KL$), and for each neutrino species ($\num$, $\nue$, $\numb$, $\nueb$)
to obtain the total flux uncertainty, accounting for the correlated variations in the 
different neutrino species. This results in a covariance matrix for the predicted flux 
of each neutrino species from each of the meson species.

Figure \ref{fig:proderr} shows the fractional uncertainty in the neutrino flux from $\pip$
and $\Kp$ production uncertainties, corresponding to the square-root of the diagonal entries 
of the covariance matrix resulting from the procedure described above divided by the predicted
flux.  In the left plot, the solid histogram shows the fractional uncertainty in the flux of
neutrinos at the MiniBooNE detector from $\pip\to\num$ produced in  $\mbox{p}$-Be interactions due to the uncertainties
in the $\pip$ production. The strong correlation between the energy of the $\num$ and the 
energy of the $\pip$ which decayed to produce it results in a large rise  in the fractional 
uncertainty at neutrino energies greater than $2\gev$ reflecting the large uncertainties in  
high-momentum pion production. Likewise, the uncertainty rises at low neutrino  energies 
($<200\mev$) due to the
rise in the uncertainties for low momentum $\pip$ production. Fortunately, relatively few neutrinos 
are produced in this region by the $\pip$ decays; in the region below $1\gev$ where the $\pip\to\num$ 
contribution is dominant, the uncertainty is approximately $17\%$.

The dashed histogram shows the fractional uncertainty in the $\nue$ flux from the $\pip\to\mup\to\nu$
decay chain resulting from the uncertainties in the $\pip$ production, the primary channel
for low energy $(<1\gev)$ $\nue$ flux. Since the correlation
between the energy of the $\nue$  and the $\pip$ which produces it is weak due to the three-body
decay of the muon, the uncertainties are more uniform as a function of energy. 

The right plot in Figure \ref{fig:proderr} likewise shows the fractional uncertainty for the flux of $\num$
and $\nue$ resulting from the decay of $\Kp$ produced in $\mbox{p}$-Be interactions due
to the uncertainties in the $\Kp$ production. These channels are the primary contribution
for $\num$ with energy greater than $2.3\gev$ and $\nue$ with energy greater than $1.2\gev$. 
Due to the larger $\Kp$ mass, the correlation between the momentum of the  $\Kp$ and the 
neutrinos from its decay is also weak. 

Figure \ref{fig:proderr_nubar} shows the corresponding plots for the horn in anti-neutrino mode,
where the $\pim\to\numb$ flux is dominant. While the corresponding charged kaon channel
would be $\Km\to(\numb/\nueb)$, the $\Kp\to(\num/\nue)$ uncertainties are shown instead, since the contribution of this channel is larger. 

Figure \ref{fig:pipcorr} shows the bin-to-bin correlations in the uncertainties related
to the pion production. The left and center plots show the correlation matrix associated
with the fractional uncertainties in the $\pip\to\num$ and $\pip\to\mup\to\nue$ flux
in  $\mbox{p}$-Be interactions, respectively. The $\pip\to\num$ flux exhibit correlations that are strongest between nearby bins, with the correlations steadily weakening for bins separated by
more than several hundred $\mev$. The $\pip\to\mup\to\nue$ flux, however, shows correlation
between energies which are more widely separated, as would be expected from the three-body
decay of the $\mup$ that produces this flux. The right plot in Figure \ref{fig:pipcorr}
shows the correlations between the uncertainties in the two components of the $\pip$ flux.
As expected, the $\pip\to\num$ flux at a given energy is most strongly correlated with 
$\pip\to\mup\to\nue$ flux at lower energies.

Figure \ref{fig:kpcorr} shows similar correlations for the $\Kp\to\num$ (left) and $\Kp\to\nue$
(center) fluxes. The situation is quite different from the $\pip$ flux; the uncertainties
are correlated across energies for each flux. Likewise, the right plot, which shows the 
correlations between the uncertainties in the two sources of neutrinos, is also strongly
correlated. This reflects the large normalization uncertainty assigned to the $\Kp$ production;
the variations in the $\Kp$ production correspond mainly to shifts in the overall rate.

Figures \ref{fig:pimcorr_nubar} and \ref{fig:kpcorr_nubar} show the corresponding plots for the horn in anti-neutrino mode. Once again, the correlations for the $\Kp\to(\num/\nue)$ uncertainties are shown instead of the corresponding $\Km\to(\numb/\nueb)$. The results show  a similar pattern of correlations to that observed in the neutrino mode.

Due to the large uncertainties in the flux prediction from particle production, which result not from
uncertainties in the measured particle production but from the parameterizations used to describe
the measurements, an alternative description of the measurements was investigated. In this method, the particle
production data are interpolated as splines using the DCSPLN routine from CERNLIB \cite{cernlib}. The 
two-dimensional measurements of the double differential cross section in $p$ and $\theta$ are interpolated
as splines first in $\theta$ at fixed values of $p$. The resulting splines in $\theta$ are then interpolated
to produce values as a function of $p$. The splines also extrapolate the double differential
cross section into region where measurements do not exist, most notably at low pion momentum ($<700\mevc$).

Uncertainties in the particle production are derived by varying the measured double differential cross
sections according to the $78\times 78$ covariance matrix describing the uncertainties in the measurements. 
Each variation results in an alternate set of double differential cross sections that is interpolated. 
By repeating this process (in practice one thousand times) a covariance matrix describing the pion production 
variations can be derived that can be used to propagate the uncertainties into the predicted neutrino flux. As a 
result of this procedure, the pion
production uncertainties can be derived in a manner that is directly connected with the experimental 
uncertainties, circumventing the difficulties associated with parameterizing the double differential
cross section and reducing the uncertainties in the neutrino flux significantly.
In the core of the $\num$ energy distribution ($0.5-1.0 \gev$) the $\num$ flux uncertainty resulting from
this method is $5-7\%$. At low energies ($<500\mev$), where measurements do not exist, the uncertainties 
rise to $\sim 20\%$. At higher energies $>1 \gev$, where the measurements have larger uncertainties, the 
$\num$ flux uncertainty rises to $\sim 10\%$. This method was not used for the neutrino oscillation results 
reported in Reference \cite{oscillation} but illustrates how the additional uncertainties associated 
with the parameterization of particle production measurements can be avoided in future analyses.

\begin{table}[t]
\begin{center}
\begin{tabular}{l|cccccc} \hline\hline
		 &  \multicolumn{2}{c}{ $\Delta\stot$ (mb)}	&  \multicolumn{2}{c}{ $\Delta\sine$ (mb)}	&  \multicolumn{2}{c}{ $\Delta\sqel$ (mb)}	\\
			& Be & Al & Be & Al & Be & Al \\ \hline
$(p/n)$-(Be/Al)	 & $~~\pm 15.0$ & $\pm 25.0~~$ & $~~\pm 5$    & $\pm 10~~$ &  $~~\pm 20$  & $\pm 45~~$   \\
$\pipm$-(Be/Al)&$~~\pm 11.9$ & $\pm 28.7~~$ 	& $~~\pm 10$ & $\pm 20~~$ &  $~~\pm 11.2$ & $\pm 25.9~~$ \\ \hline
\end{tabular}
\caption{\label{tab:beamuni} Cross section variations for systematic studies. For each hadron-nucleus
cross section type, the momentum-dependent cross section is offset by the amount shown.}

\end{center}

\end{table}

%\begin{figure}[t]
%\begin{center}
%\includegraphics[width=14 cm]{pipInelastic.pdf}
%\includegraphics[width=14 cm]{pimInelastic.pdf}
%\caption{ \label{fig:pipminesys}  $\pip$-nucleus (top) and $\pim$-nucleus (bottom)
% (Be on left, Al on right) inelastic  cross sections used in the flux simulation. Measured cross 
% sections from various experiments are shown as points. The default parametrization  is in solid black line, while the  systematic variations are shown as dotted red lines.}
%\end{center}
%\end{figure}

\begin{figure}[t]
\begin{center}
\includegraphics[width=15 cm]{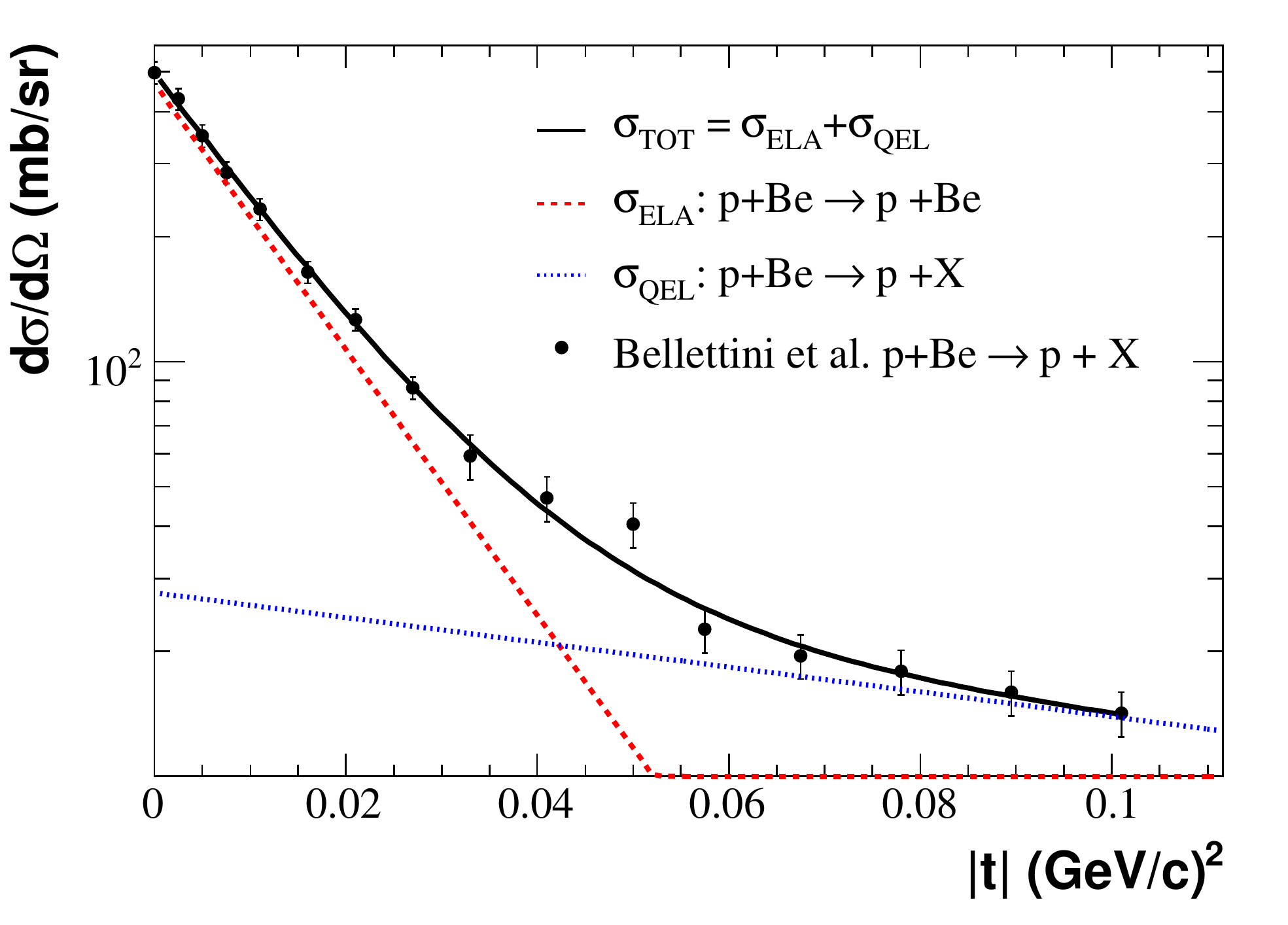}
\caption{ \label{fig:beq2} Observed $q^2$ distribution from Bellettini {\em et al.} \cite{bellettini}
for $20\gevc$ $\mbox{p}$-Be scattering. The solid line represents the fit to two exponential
distributions with the dashed and dotted lines showing the elastic and quasi-elastic
contributions, respectively.}
\end{center}
\end{figure}

\subsection{Hadronic Interactions}
\label{sec:hadint}
Uncertainties due to hadronic interactions are considered by varying the
 components of the hadronic cross sections. First, the total hadronic 
cross section $\stot$, the total inelastic $\sine$ and the quasi-elastic $\sqel$
cross sections are separately varied for nucleons on beryllium and aluminum. 
Second, the same is done for the pion cross sections. In each case, the variations 
are a flat, momentum-independent offset. Due to the various relations between the cross sections, the variation of $\stot$ results
in a variation of $\sela$ ($\sine$ is fixed). When $\sine$ is varied,
the balance between $\sela$ and $\sine$ is changed, while keeping their total at 
$\stot$. Finally, when $\sqel$ is varied, the relative proportion of $\sqel$ to the cross section
for all other inelastic processes is changed, while keeping their sum ($\sine$) fixed.

The variations for $\stot$ are based on the agreement of the Glauber model calculations
with the available $n$-nucleus measurements, shown in Figure \ref{fig:nuctotxsec}.
The deviations of the measurements from the model are used to set the magnitude of the
variation. These variations are also applied to $\stot$ for $\pipm$-nucleus measurements
by scaling the variations from the nuclear case by the ratio of $\stot$ at high momentum in 
the two cases. Since there are no $\pipm$-nucleus  $\stot$ data above the $\Delta(1232)$
resonance to verify the model, this assumes that the model  works as well for $\pipm$-nucleus
interactions as it does for nucleon-nucleus interactions. 

The variations for $\sine$ are similarly set by the deviations of the measurements from
the parametrization. Here the deviations are smaller since the parametrizations are
derived directly from the data; the variations are intended to incorporate the uncertainties
in the measurements.
								  
The uncertainties for $\sqel$ are set by comparing the calculated cross section to 
the inferred $\sqel$ from the measured  $q^2$ distribution in $\mbox{p}$-Be scattering from
Reference \cite{bellettini}
shown in Figure \ref{fig:beq2}. This distribution is fit to the sum of two exponentials corresponding
to the elastic and quasi-elastic scattering components. The fitted slope of the quasi-elastic
component is consistent with free nucleon-nucleon elastic scattering and leads a ratio $\sqel/\sela$
of 0.6. Since the free nucleon cross sections and $\stot$ do not change appreciably from
$8.89\gevc$ to $20\gevc$, the value and uncertainty of $\sela$ at $8.89\gevc$ can be used
to obtain $\sqel=44\pm 9$ mb at the same beam momentum. This can be compared to the 34.9 mb 
obtained from the shadowed scattering model. An uncertainty of $20$ mb is assigned to $\sqel$ 
to account for both the difference and the uncertainty from the Bellettini measurement. The 
uncertainty for nucleon-aluminum scattering is obtained
by scaling this uncertainty by the ratio of the predicted $\sqel$ in aluminum and beryllium.
The variation for $\sqel$ in  $\pipm$-nucleus scattering is obtained by scaling the
variation in nucleon-nucleus scattering by the ratio of $\sqel$ for $\pipm$ and nucleons.

Table \ref{tab:beamuni} summarizes the variations in all six hadron-nucleus combinations.
Of these variations, the variations in $\sqel$ have the largest effect.

\subsection{Horn Magnetic Field Modeling}
\label{sec:hornsys}
Uncertainties on two properties of the horn magnetic field result in systematic uncertainties
in the neutrino flux. The first is the horn current. The commercial current transformers
(Stangenes Industries 3-0.002) have a rated accuracy of $0.5\%$. The effect of a 1 kA variation
in the nominal horn current ($174\pm1$ kA) is simulated to set the systematic uncertainty. 

A second source of uncertainty arises from the modeling of the current within the inner cylinder due to the so-called ``skin effect''. The skin effect allows for the magnetic field to penetrate into the conductor,
increasing the effective magnetic field experienced by particles traversing through the inner conductor
into the horn cavity. This is  important for particles produced at small angles (particularly high momentum) that barely penetrate into the horn cavity before exiting the front of the
horn into the collimator region. For these particles, the bulk of the magnetic field seen by
the particle in this trajectory may come from the field within the inner conductor.

The expected current distribution in a cylindrically symmetric configuration was numerically evaluated 
and found to be well-approximated by a current density exponentially decreasing from the outer 
surface of the inner conductor with a decay length set by the skin depth (1.4 mm). The magnetic field
configuration corresponding to this current density is simulated and taken as the default configuration.
The simulation is also run without the skin effect, simulating the situation where
the current density lies entirely on the outer surface of the inner conductor of the horn, resulting
in no magnetic field penetration. The difference between these two configurations, a few percent in the predicted neutrino flux for $\num$ with energies $<1\gev$ and up to $18\%$ for $\num$ with energies  
$\sim 2\gev$, is considered a systematic uncertainty.

\subsection{Geometry Uncertainties}
Variations in the geometric configuration of the beamline are simulated to investigate their
effect on the neutrino flux. These variations include moving the target position relative to 
the rest of the beamline (in particular the horn), varying the radius of the decay pipe, and moving
the collimator along the beam axis and changing its aperture. The magnitudes of the geometric
perturbations which are required to effect a substantial $(>1\%)$ change in the flux are
well outside of what are considered the tolerances and precision of the constructed beamline.
As a result, no significant systematic uncertainty is assigned to the beamline geometry.

\begin{table}[t]
\begin{center}
\begin{tabular}{l|rrrr}\hline\hline
Source of Uncertainty 	& \num	& \numb	& \nue	& \nueb 	\\ \hline
Proton delivery		&  2.0\%& 2.0\%	&  2.0\%&  2.0\%\\ 
Proton optics		&  1.0\%& 1.0\%	&  1.0\%&  1.0\%\\
$\pip$ production	& 14.7\%& 1.0\%	&  9.3\%&  0.9\%\\
$\pim$ production	&  0.0\%& 16.5\%&  0.0\%&  3.5\%\\
$\Kp$ production	&  0.9\%& 0.2\%	& 11.5\%&  0.3\%\\
$\Kz$ production	&  0.0\%& 0.2\% & 2.1\%	&  17.6\%\\
Horn field		& 2.2\%	& 3.3\%	& 0.6\%	& 0.8\%	\\
Nucleon cross sections	& 2.8\%	& 5.7\%	& 3.3\%	& 5.6\%	\\
Pion cross sections	& 1.2\%	& 1.2\%	& 0.8\%	& 0.7\%	\\ \hline
\end{tabular}
%
%      	nue     		nueb    		numu    		numub
%Pi+     	9.3044  		0.91096 		14.7028 		0.989329
%Pi-     	0.0174007       	3.53862 		0.031324        	16.4595
%Kp      	11.5378 		0.327014        	0.858606        	0.245736
%K0      	2.06939 		17.5877 		0.0195409       	0.163181
\caption{\label{tab:nusys}Variations in the total flux of
each neutrino species in neutrino mode due to the systematic uncertainties.}
\vskip 0.25 cm
\begin{tabular}{l|rrrr}\hline\hline
Source of Uncertainty 	& \num	& \numb	& \nue	& \nueb \\ \hline
Proton delivery		&  2.0\%& 2.0\%	&  2.0\%&  2.0\%\\ 
Proton optics		&  1.0\%& 1.0\%	&  1.0\%&  1.0\%\\
$\pip$ production	& 13.8\%& 0.1\%	& 2.1\%	& 0.1\%	\\
$\pim$ production	& 0.5\%	&17.5\%	& 0.0\%	& 13.6\%\\
$\Kp$ production	& 3.1\%& 0.0\%	& 22.3\%& 0.4\%	\\
$\Kz$ production	& 0.1\%	& 0.0\%	& 6.1\%	& 3.9\%	\\
Horn field		& 1.5\%	& 1.0\%	& 3.2\%	& 1.5\%	\\
Nucleon cross sections	& 6.2\%	& 2.1\%	& 6.2\%	& 2.5\%	\\
Pion cross sections	& 1.5\%	& 1.2\%	& 1.6\%	& 1.5\%	\\ \hline
\end{tabular}
\caption{\label{tab:nubarsys}Variations in the total flux of
each neutrino species in anti-neutrino mode due to the systematic uncertainties.}
\end{center}
%       	 nue     	nueb    	numu    		numub
%Pi+     	2.06202 	0.145477	13.836  		0.146064
%Pi-     	0.027409 13.6028 	0.451784        	17.5474
%Kp      	22.2959 	0.43417 	3.09555 		0.033463
%K0      	6.12381 	3.88413 	0.110485        	0.0357433
\end{table}

\begin{figure}[t]
\begin{center}
\includegraphics[width=14 cm]{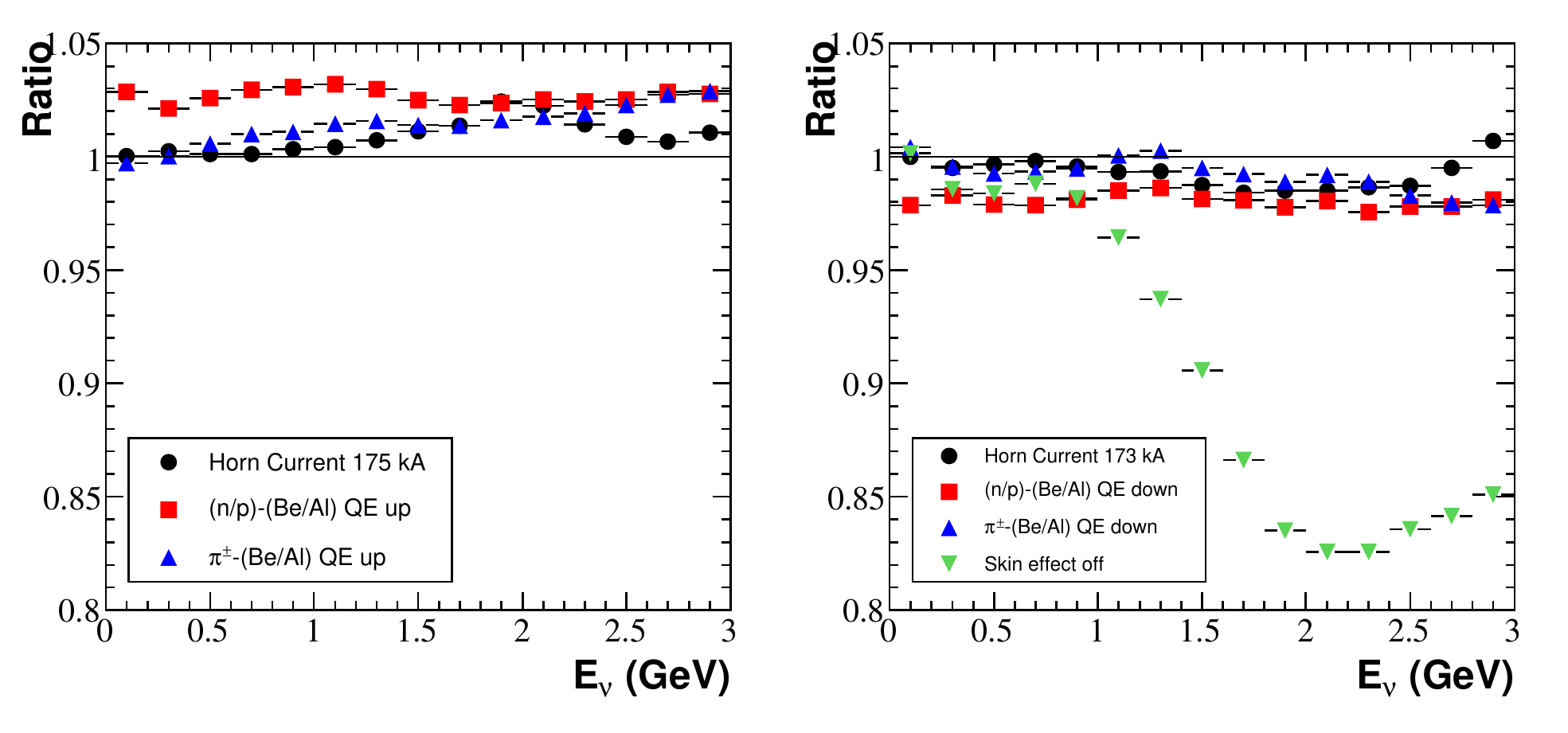}
\vskip 0.25 cm
\includegraphics[width=15 cm]{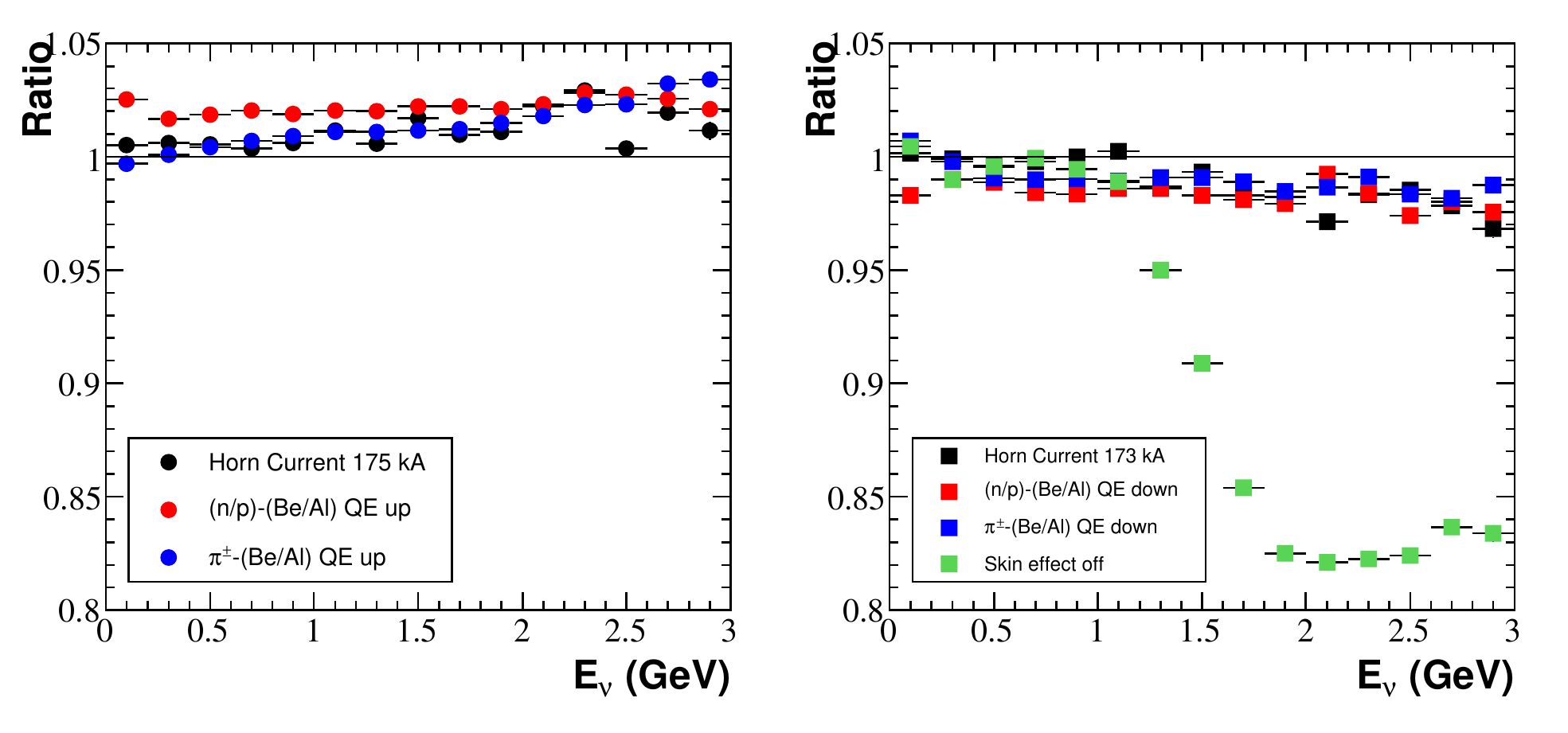}
\caption{ \label{fig:domsys} Top: Change in the $\num$ flux from $\pip$ due to the 
dominant sources of systematic uncertainty apart from particle production. Left: Ratio
of flux from increased horn current (circles), increased nucleon-nucleus quasi-elastic
cross section (squares) and increased pion-nucleus quasi-elastic cross section (upward triangles) to the
default flux. Right: Ratio of flux from decreased horn current (circles), decreased
nucleon-nucleus quasi-elastic cross section (squares), decreased pion-nucleus quasi-elastic
cross section (upward triangles), and turning off the skin effect (downward trianges), to the default flux.
Bottom: Same for predicted $\numb$ flux from $\pim$ in antineutrino mode.}
\end{center}
\end{figure}

\subsection{Summary of Systematic Uncertainties}
Tables \ref{tab:nusys} and \ref{tab:nubarsys} summarize the variations in the total flux 
for each neutrino species resulting from the systematic uncertainties discussed in this Section.
By far the largest uncertainty arises from the particle production uncertainties. Much of
this uncertainty arises not from the accuracy of the measurements, but from the parametrizations
used to summarize the double differential cross sections. These latter contributions manifest as
inflated uncertainties on the parameters resulting from the $\chi^2/\mbox{DOF}$ at the best-fit values
and other considerations such as the dependence on the choice of parameterization.

The flux-averaged uncertainties provide a rough gauge to the relative size of the various uncertainties; they 
are not used in the $\num\to\nue$ oscillation analysis. 
As seen in  Section \ref{sec:prodsys}, the uncertainties can vary significantly with energy and
exhibit correlations across energies and neutrino species. In the $\num\to\nue$ oscillation
analysis, uncertainties are propagated with covariance matrices where
the energy-dependent variations in uncertainties and  correlations are taken into account.

The top two plots of Figure \ref{fig:domsys} illustrate the effects from the largest sources of 
systematic uncertainty, apart from the particle production uncertainties which have already been discussed,
on the predicted $\pip\to\num$ flux at the MiniBooNE 
detector with the horn in neutrino mode.  The largest effect comes from the presence or absence of the 
skin effect in the conduction of the horn current
along the inner conductor of the horn.  The effect is particularly large for high energy neutrinos
($>1\gev$) due to the correlation of the pion momentum with angle (higher momentum $\pip$
tend to be produced in the forward direction). As mentioned in Section \ref{sec:hornsys},
these particles will usually have the largest change in the amount of magnetic field experienced
in traversing from the target, through the horn, and into the decay region. However, for very
high momentum $\pip$ ($>4\gevc$), the production is collimated to such an extent that 
an increasing part of the production never enters the horn, and instead travels alongside
the target into the decay region without traversing the inner conductor. For these $\pip$ the
skin effect is irrelevant; as a result, the effect diminishes for the high energy neutrinos associated
with the decay of these pions.

The next largest source of systematic uncertainty  is from the magnitude of the hadron-nucleus
quasi-elastic cross section. This effect is investigated for nucleon-nucleus and pion-nucleus
cross sections separately. As discussed in Section \ref{sec:hadint}, in the nucleon-nucleus case, larger  quasi-elastic cross section results in more protons emerging from the inelastic interactions with energies
close to the primary energies. The interactions of these secondary protons is much like that of the primary protons. As a result, there is an overall increase in the particle production and the neutrino flux.
For pions, an increase in the quasi-elastic cross section increases the effective hadronic transparency
of the material which intervenes between the production of the pion and its decay (primarily the
target and the horn). The effect is largest for forward particles (which tend to be at higher momentum) which traverse more material. As a result, an increase in the pion quasi-elastic cross section increases the neutrino flux with an energy dependence that favors high energy neutrinos. These trends for
nucleon and pion quasi-elastic cross section variations are evident in Figure  \ref{fig:domsys}.
The bottom plot of Figure \ref{fig:domsys} shows the same summary for the predicted
$\pim\to\numb$ flux at MiniBooNE with the horn in anti-neutrino 
mode. The pattern of systematic uncertainties is similar to that observed for the
$\pip\to\num$ flux in neutrino mode.

\begin{figure}[t]
\begin{center}
 \includegraphics[width=140mm]{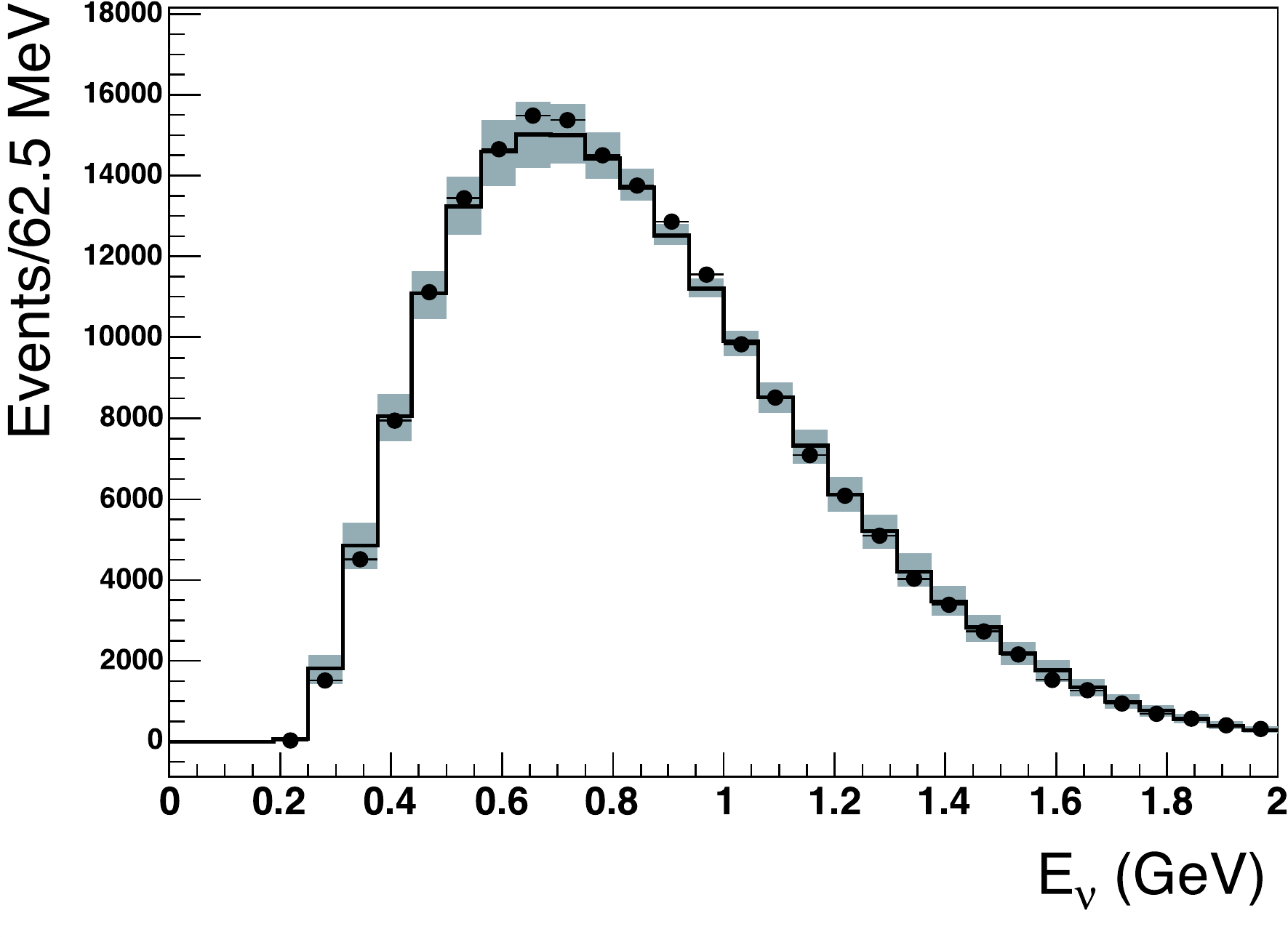}
     \caption{
 \label{fig:comparison} Comparison of the observed (points) and predicted (histogram) energy distribution
in $\num$ CCQE events selected in the MiniBooNE data. A normalization factor of 1.21
has been applied to the predicted distribution as described in the text. The error bars on the predicted
distribution are the estimated uncertainties in the shape of the spectrum once the normalization has been fixed to match
the data.}
\end{center}
\end{figure}

\section{Comparison to Observed Neutrino Events}
The observed energy spectrum of $\num$ charged-current quasi-elastic (CCQE) events at MiniBooNE
in $5.6\times 10^{20}$ protons-on-target taken in neutrino-enhanced mode is compared to the 
predicted spectrum in Figure \ref{fig:comparison}.
The sample of $\num$ CCQE interactions is obtained by selecting events consistent with a single
muon-like Cherenkov ring in the detector with activity consistent with muon decay-at-rest following
the primary neutrino interaction\cite{boonenim}. The predicted neutrino energy spectrum results from generating
simulated neutrino interactions with the NUANCE neutrino event generator\cite{nuance} according
to the predicted neutrino fluxes and propagating the predicted final state particles through a
detailed Monte Carlo simulation of the MiniBooNE detector. The neutrino energy is calculated 
using the reconstructed muon energy and angle relative to the beam axis, assuming that the event
is CCQE  ($\num + n \to \mu^- + p$). Identical selection criteria are applied to the data. Details
of the event selection and tuning of the NUANCE event generator can be found in Reference \cite{ccqe}.
Systematic uncertainties in the shape of the spectrum from the neutrino flux prediction, cross section model, and detector model
are represented by the grey squares. The predicted purity of the selected events in $\num$ CCQE interactions is
$74\%$ with charged-current single pion production channels as the dominant source of background.

The observed rate of interactions is factor of $1.21\pm 0.24$ higher than the rate predicted by the nominal
simulation; the predicted spectrum in Figure \ref{fig:comparison} has been scaled by this factor
to facilitate the comparison of the spectrum shapes. While this is a sizable discrepancy, the 
uncertainties resulting from the predicted flux, the neutrino cross section model and the detector simulation
are such that the scale factor is compatible with unity as indicated by its uncertainty. We note that
the neutrino cross section parameters measured in Reference \cite{ccqe} are derived from the $Q^2$
distribution and not the rate or energy spectrum of the observed events and that the neutrino flux prediction
has not be adjusted in any way in response to the observed neutrino data apart from this normalization factor.

\section{Conclusions}
The neutrino flux at MiniBooNE is predicted by a detailed Geant4-based neutrino
flux simulation. The Geant4 framework allows for a realistic representation of the
beamline geometry and accounting of the  electromagnetic and hadronic effects
experienced by particles as they traverse the beamline.
The software framework incorporates a number of custom features that have
been tailored to the needs of the analyses at MiniBooNE. In particular, the properties
of key hadronic processes, most notably the cross sections of nucleons and pions on
beryllium and aluminum, and the particle production properties of p-Be interactions,
have been tuned based on external measurements wherever possible. These have been
summarized in a number of parametrizations that describe the momentum dependence
of the overall cross sections, as well as the multiplicity and kinematic properties
 of the relevant secondary particle species in the primary p-Be interactions.
The simulation also accounts for the measured properties of the primary proton beam.

A separate model controls the kinematics of neutrinos resulting from the decays of
mesons propagated to their point of decay in the Geant4 simulation. This model
accounts for polarization effects as well as non-trivial decay form factors, and
reflects the latest knowledge of key kaon branching fractions. The geometric acceptance
of the neutrinos at the MiniBooNE detector is also handled. Both software
frameworks employ a number of statistical enhancement techniques that reduce the uncertainties
overall and enhance the statistical precision in kinematic regions and channels where 
important contributions to the flux may come from processes that are small in relation to the dominant channels.

The flexibility of the framework allows the determination of a number of
systematic uncertainties by varying parameters within the simulation. In this way,
the effect of varying hadronic cross sections and different horn currents can
be estimated. By recording the kinematics of the secondary mesons at production,
the uncertainties in the production differential cross sections can be propagated
through reweighting without rerunning the simulation. The study of systematic uncertainties
indicate that the dominant uncertainty arises from the particle production. These uncertainties
arise not only from the intrinsic uncertainties in the particle production measurements,
but also from the parametrizations used to model the differential cross sections.
The resulting neutrino flux predictions and uncertainties are a critical element of the Monte Carlo simulation
chain at MiniBooNE, where they are combined with the NUANCE neutrino event generator\cite{nuance}
and a detector Monte Carlo simulation \cite{boonenim} to determine the rate and properties of neutrino interactions
in the detector. Comparison of the predicted event rate and spectrum with a sample 
of $\num$ charged-current quasi-elastic events observed at MiniBooNE indicate that the
spectrum is reproduced well by the simulation while there is a sizable discrepancy in the overall 
rate, with the observed data rate a factor $1.21\pm 0.24$ higher than predicted. Due to the large
uncertainty in the predicted rates, the observed and predicted
rates are compatible.

\section{Acknowledgements}
The MiniBooNE collaboration acknowledges support from the Department of
Energy and the National Science Foundation of the United States. We are
grateful to Fermilab for hosting the experiment and thank the Accelerator
Division for the excellent  accelerator performance. We thank  Los Alamos National Laboratory for LDRD funding. We acknowledge Bartoszek Engineering for the design of 
the focusing horn. This research was done using resources provided by the Open Science Grid, which is supported by the  NSF and DOE-SC. We also acknowledge the use of the 
LANL PINK cluster and CONDOR software in the analysis of  the data. 

\bibliography{prdflux}

\begin{thebibliography}{66}
\expandafter\ifx\csname natexlab\endcsname\relax\def\natexlab#1{#1}\fi
\expandafter\ifx\csname bibnamefont\endcsname\relax
  \def\bibnamefont#1{#1}\fi
\expandafter\ifx\csname bibfnamefont\endcsname\relax
  \def\bibfnamefont#1{#1}\fi
\expandafter\ifx\csname citenamefont\endcsname\relax
  \def\citenamefont#1{#1}\fi
\expandafter\ifx\csname url\endcsname\relax
  \def\url#1{\texttt{#1}}\fi
\expandafter\ifx\csname urlprefix\endcsname\relax\def\urlprefix{URL }\fi
\providecommand{\bibinfo}[2]{#2}
\providecommand{\eprint}[2][]{\url{#2}}

\bibitem[{\citenamefont{Aguilar et~al.}(2001)}]{lsnd}
\bibinfo{author}{\bibfnamefont{A.}~\bibnamefont{Aguilar}} \bibnamefont{et~al.}
  (\bibinfo{collaboration}{LSND}), \bibinfo{journal}{Phys. Rev.}
  \textbf{\bibinfo{volume}{D64}}, \bibinfo{pages}{112007}
  (\bibinfo{year}{2001}), \eprint{hep-ex/0104049}.

\bibitem[{\citenamefont{Aguilar-Arevalo et~al.}(2007)}]{oscillation}
\bibinfo{author}{\bibfnamefont{A.~A.} \bibnamefont{Aguilar-Arevalo}}
  \bibnamefont{et~al.} (\bibinfo{collaboration}{MiniBooNE}),
  \bibinfo{journal}{Phys. Rev. Lett.} \textbf{\bibinfo{volume}{98}},
  \bibinfo{pages}{231801} (\bibinfo{year}{2007}), \eprint{0704.1500}.

\bibitem[{\citenamefont{Aguilar-Arevalo et~al.}(2008)}]{ccqe}
\bibinfo{author}{\bibfnamefont{A.~A.} \bibnamefont{Aguilar-Arevalo}}
  \bibnamefont{et~al.} (\bibinfo{collaboration}{MiniBooNE}),
  \bibinfo{journal}{Phys. Rev. Lett.} \textbf{\bibinfo{volume}{100}},
  \bibinfo{pages}{032301} (\bibinfo{year}{2008}), \eprint{0706.0926}.

\bibitem[{\citenamefont{Monroe}(2006)}]{jocelyn}
\bibinfo{author}{\bibfnamefont{J.~R.} \bibnamefont{Monroe}}
  (\bibinfo{year}{2006}), \bibinfo{note}{{FERMILAB-THESIS-2006-44}}.

\bibitem[{\citenamefont{Patterson}(2007)}]{ryan}
\bibinfo{author}{\bibfnamefont{R.~B.} \bibnamefont{Patterson}}
  (\bibinfo{year}{2007}), \bibinfo{note}{{FERMILAB-THESIS-2007-19}}.

\bibitem[{bnb()}]{bnbtdr}
\emph{\bibinfo{title}{{8 GeV Beam Technical Design Report}}},
  \eprint{http://www-boone.fnal.gov/publicpages/8gevtdr\_2.0.ps.gz}.

\bibitem[{\citenamefont{Kobilarcik}(2005)}]{tom}
\bibinfo{author}{\bibfnamefont{T.~R.} \bibnamefont{Kobilarcik}}
  (\bibinfo{collaboration}{MiniBooNE}) (\bibinfo{year}{2005}),
  \bibinfo{note}{presented at 5th International Workshop on Neutrino Beams and
  Instrumentation (NBI 2005), Batavia, Illinois, 7-11 Jul 2005}.

\bibitem[{\citenamefont{Moore et~al.}(2003)}]{moorepac2003}
\bibinfo{author}{\bibfnamefont{C.}~\bibnamefont{Moore}} \bibnamefont{et~al.}
  (\bibinfo{year}{2003}), \bibinfo{note}{{Particle Accelerator Conference (PAC
  03) 12-16 May 2003, Portland, Oregon}}.

\bibitem[{nub(2006)}]{nubarloi}
\emph{\bibinfo{title}{{Addendum to the MiniBooNE Run Plan: MiniBooNE Physics in
  2006}}} (\bibinfo{year}{2006}),
  \bibinfo{note}{http://www-boone.fnal.gov/publicpages/loi.ps.gz}.

\bibitem[{\citenamefont{Sorel}(2005)}]{michel}
\bibinfo{author}{\bibfnamefont{M.}~\bibnamefont{Sorel}} (\bibinfo{year}{2005}),
  \bibinfo{note}{{FERMILAB-THESIS-2005-07}}.

\bibitem[{\citenamefont{Nelson}(2006)}]{boblmc}
\bibinfo{author}{\bibfnamefont{R.~H.} \bibnamefont{Nelson}},
  \bibinfo{journal}{AIP Conf. Proc.} \textbf{\bibinfo{volume}{842}},
  \bibinfo{pages}{831} (\bibinfo{year}{2006}).

\bibitem[{\citenamefont{Agostinelli et~al.}(2003)}]{geant4}
\bibinfo{author}{\bibfnamefont{S.}~\bibnamefont{Agostinelli}}
  \bibnamefont{et~al.} (\bibinfo{collaboration}{GEANT4}),
  \bibinfo{journal}{Nucl. Instrum. Meth.} \textbf{\bibinfo{volume}{A506}},
  \bibinfo{pages}{250} (\bibinfo{year}{2003}).

\bibitem[{cer()}]{cernlib}
\emph{\bibinfo{title}{{The CERN Program Library (CERNLIB)}}},
  \bibinfo{note}{see cernlib.web.cern.ch/cernlib}.

\bibitem[{\citenamefont{Carey et~al.}(1998)\citenamefont{Carey, Brown, and
  Rothacker}}]{transport}
\bibinfo{author}{\bibfnamefont{D.~C.} \bibnamefont{Carey}},
  \bibinfo{author}{\bibfnamefont{K.~L.} \bibnamefont{Brown}}, \bibnamefont{and}
  \bibinfo{author}{\bibfnamefont{F.}~\bibnamefont{Rothacker}}
  (\bibinfo{year}{1998}), \bibinfo{note}{{SLAC-R-0530}}.

\bibitem[{\citenamefont{Chounet et~al.}(1972)\citenamefont{Chounet, Gaillard,
  and Gaillard}}]{Chounet:1971yy}
\bibinfo{author}{\bibfnamefont{L.~M.} \bibnamefont{Chounet}},
  \bibinfo{author}{\bibfnamefont{J.~M.} \bibnamefont{Gaillard}},
  \bibnamefont{and} \bibinfo{author}{\bibfnamefont{M.~K.}
  \bibnamefont{Gaillard}}, \bibinfo{journal}{Phys. Rept.}
  \textbf{\bibinfo{volume}{4}} (\bibinfo{year}{1972}).

\bibitem[{\citenamefont{Hagiwara et~al.}(2002)}]{Hagiwara:2002fs}
\bibinfo{author}{\bibfnamefont{K.}~\bibnamefont{Hagiwara}} \bibnamefont{et~al.}
  (\bibinfo{collaboration}{Particle Data Group}), \bibinfo{journal}{Phys. Rev.}
  \textbf{\bibinfo{volume}{D66}}, \bibinfo{pages}{010001}
  (\bibinfo{year}{2002}).

\bibitem[{\citenamefont{Gaisser}()}]{gaisser}
\bibinfo{author}{\bibfnamefont{T.~K.} \bibnamefont{Gaisser}},
  \emph{\bibinfo{title}{{Cosmic rays and particle physics}}},
  \bibinfo{note}{{Cambridge, UK: Univ. Pr. (1990) 279 p.}}

\bibitem[{\citenamefont{Eidelman et~al.}(2004)}]{pdg}
\bibinfo{author}{\bibfnamefont{S.}~\bibnamefont{Eidelman}} \bibnamefont{et~al.}
  (\bibinfo{collaboration}{Particle Data Group}), \bibinfo{journal}{Phys.
  Lett.} \textbf{\bibinfo{volume}{B592}}, \bibinfo{pages}{1}
  (\bibinfo{year}{2004}).

\bibitem[{\citenamefont{T.~Lasinski and Ukleja}(1972)}]{lasinski}
\bibinfo{author}{\bibfnamefont{B.~S.} \bibnamefont{T.~Lasinski},
  \bibfnamefont{R.~Levi~Setti}} \bibnamefont{and}
  \bibinfo{author}{\bibfnamefont{P.}~\bibnamefont{Ukleja}},
  \bibinfo{journal}{Nucl. Phys. B} \textbf{\bibinfo{volume}{37}},
  \bibinfo{pages}{1} (\bibinfo{year}{1972}).

\bibitem[{\citenamefont{Coffin et~al.}(1967)\citenamefont{Coffin, Dikmen,
  Ettlinger, Meyer, Saulys, Terwilliger, and Williams}}]{coffin}
\bibinfo{author}{\bibfnamefont{C.~T.} \bibnamefont{Coffin}},
  \bibinfo{author}{\bibfnamefont{N.}~\bibnamefont{Dikmen}},
  \bibinfo{author}{\bibfnamefont{L.}~\bibnamefont{Ettlinger}},
  \bibinfo{author}{\bibfnamefont{D.}~\bibnamefont{Meyer}},
  \bibinfo{author}{\bibfnamefont{A.}~\bibnamefont{Saulys}},
  \bibinfo{author}{\bibfnamefont{K.}~\bibnamefont{Terwilliger}},
  \bibnamefont{and} \bibinfo{author}{\bibfnamefont{D.}~\bibnamefont{Williams}},
  \bibinfo{journal}{Phys. Rev.} \textbf{\bibinfo{volume}{159}},
  \bibinfo{pages}{1169} (\bibinfo{year}{1967}).

\bibitem[{\citenamefont{Foley et~al.}(1963)\citenamefont{Foley, Lindenbaum,
  Love, Ozaki, Russell, and Yuan}}]{lindenbaum}
\bibinfo{author}{\bibfnamefont{K.~J.} \bibnamefont{Foley}},
  \bibinfo{author}{\bibfnamefont{S.~J.} \bibnamefont{Lindenbaum}},
  \bibinfo{author}{\bibfnamefont{W.~A.} \bibnamefont{Love}},
  \bibinfo{author}{\bibfnamefont{S.}~\bibnamefont{Ozaki}},
  \bibinfo{author}{\bibfnamefont{J.~J.} \bibnamefont{Russell}},
  \bibnamefont{and} \bibinfo{author}{\bibfnamefont{L.~C.~L.}
  \bibnamefont{Yuan}}, \bibinfo{journal}{Phys. Rev. Lett.}
  \textbf{\bibinfo{volume}{11}}, \bibinfo{pages}{425} (\bibinfo{year}{1963}).

\bibitem[{\citenamefont{{D. Harting {\em et al.}}}(1965)}]{cern}
\bibinfo{author}{\bibnamefont{{D. Harting {\em et al.}}}},
  \bibinfo{journal}{Nuovo Cimento} \textbf{\bibinfo{volume}{38}}
  (\bibinfo{year}{1965}).

\bibitem[{\citenamefont{Glauber}(1959)}]{glauber}
\bibinfo{author}{\bibfnamefont{R.~J.} \bibnamefont{Glauber}}
  (\bibinfo{year}{1959}), \bibinfo{note}{{in {\em Lectures in Theoretical
  Physics}, Volume 1, edited by W. E. Britten {\em et al.}}}

\bibitem[{\citenamefont{Franco}(1972)}]{franco}
\bibinfo{author}{\bibfnamefont{V.}~\bibnamefont{Franco}},
  \bibinfo{journal}{Phys. Rev. C} \textbf{\bibinfo{volume}{6}},
  \bibinfo{pages}{748} (\bibinfo{year}{1972}).

\bibitem[{\citenamefont{Woods and Saxon}(1954)}]{woodssaxon}
\bibinfo{author}{\bibfnamefont{R.~D.} \bibnamefont{Woods}} \bibnamefont{and}
  \bibinfo{author}{\bibfnamefont{D.~S.} \bibnamefont{Saxon}},
  \bibinfo{journal}{Phys. Rev.} \textbf{\bibinfo{volume}{95}},
  \bibinfo{pages}{577} (\bibinfo{year}{1954}).

\bibitem[{\citenamefont{Foley et~al.}(1965)\citenamefont{Foley, Gilmore,
  Lindenbaum, Love, Ozaki, Willen, Yamada, and Yuan}}]{foley}
\bibinfo{author}{\bibfnamefont{K.~J.} \bibnamefont{Foley}},
  \bibinfo{author}{\bibfnamefont{R.~S.} \bibnamefont{Gilmore}},
  \bibinfo{author}{\bibfnamefont{S.~J.} \bibnamefont{Lindenbaum}},
  \bibinfo{author}{\bibfnamefont{W.~A.} \bibnamefont{Love}},
  \bibinfo{author}{\bibfnamefont{S.}~\bibnamefont{Ozaki}},
  \bibinfo{author}{\bibfnamefont{E.~H.} \bibnamefont{Willen}},
  \bibinfo{author}{\bibfnamefont{R.}~\bibnamefont{Yamada}}, \bibnamefont{and}
  \bibinfo{author}{\bibfnamefont{L.~C.~L.} \bibnamefont{Yuan}},
  \bibinfo{journal}{Phys. Rev. Lett.} \textbf{\bibinfo{volume}{15}},
  \bibinfo{pages}{45} (\bibinfo{year}{1965}).

\bibitem[{\citenamefont{Clyde}()}]{clyde}
\bibinfo{author}{\bibfnamefont{R.}~\bibnamefont{Clyde}},
  \bibinfo{note}{{Thesis, UCRL 16275}}.

\bibitem[{\citenamefont{{S. Colleti {\em et al.}}}(1967)}]{colleti}
\bibinfo{author}{\bibnamefont{{S. Colleti {\em et al.}}}},
  \bibinfo{journal}{Nuovo Cimento} \textbf{\bibinfo{volume}{49A}},
  \bibinfo{pages}{479} (\bibinfo{year}{1967}).

\bibitem[{\citenamefont{Alexander et~al.}(1967)\citenamefont{Alexander, Benary,
  Czapek, Haber, Kidron, Reuter, Shapira, Simopoulou, and
  Yekutieli}}]{alexander}
\bibinfo{author}{\bibfnamefont{G.}~\bibnamefont{Alexander}},
  \bibinfo{author}{\bibfnamefont{O.}~\bibnamefont{Benary}},
  \bibinfo{author}{\bibfnamefont{G.}~\bibnamefont{Czapek}},
  \bibinfo{author}{\bibfnamefont{B.}~\bibnamefont{Haber}},
  \bibinfo{author}{\bibfnamefont{N.}~\bibnamefont{Kidron}},
  \bibinfo{author}{\bibfnamefont{B.}~\bibnamefont{Reuter}},
  \bibinfo{author}{\bibfnamefont{A.}~\bibnamefont{Shapira}},
  \bibinfo{author}{\bibfnamefont{E.}~\bibnamefont{Simopoulou}},
  \bibnamefont{and}
  \bibinfo{author}{\bibfnamefont{G.}~\bibnamefont{Yekutieli}},
  \bibinfo{journal}{Phys. Rev.} \textbf{\bibinfo{volume}{154}},
  \bibinfo{pages}{1284} (\bibinfo{year}{1967}).

\bibitem[{\citenamefont{Perl et~al.}(1970)\citenamefont{Perl, Cox, Longo, and
  Kreisler}}]{perl}
\bibinfo{author}{\bibfnamefont{M.~L.} \bibnamefont{Perl}},
  \bibinfo{author}{\bibfnamefont{J.}~\bibnamefont{Cox}},
  \bibinfo{author}{\bibfnamefont{M.~J.} \bibnamefont{Longo}}, \bibnamefont{and}
  \bibinfo{author}{\bibfnamefont{M.~N.} \bibnamefont{Kreisler}},
  \bibinfo{journal}{Phys. Rev. D} \textbf{\bibinfo{volume}{1}},
  \bibinfo{pages}{1857} (\bibinfo{year}{1970}).

\bibitem[{\citenamefont{Gibbard et~al.}(1970)\citenamefont{Gibbard, Jones,
  Longo, O'Fallon, Cox, Perl, Toner, and Kreisler}}]{gibbard}
\bibinfo{author}{\bibfnamefont{B.~G.} \bibnamefont{Gibbard}},
  \bibinfo{author}{\bibfnamefont{L.~W.} \bibnamefont{Jones}},
  \bibinfo{author}{\bibfnamefont{M.~J.} \bibnamefont{Longo}},
  \bibinfo{author}{\bibfnamefont{J.~R.} \bibnamefont{O'Fallon}},
  \bibinfo{author}{\bibfnamefont{J.}~\bibnamefont{Cox}},
  \bibinfo{author}{\bibfnamefont{M.~L.} \bibnamefont{Perl}},
  \bibinfo{author}{\bibfnamefont{W.~T.} \bibnamefont{Toner}}, \bibnamefont{and}
  \bibinfo{author}{\bibfnamefont{M.~N.} \bibnamefont{Kreisler}},
  \bibinfo{journal}{Phys. Rev. Lett.} \textbf{\bibinfo{volume}{24}},
  \bibinfo{pages}{22} (\bibinfo{year}{1970}).

\bibitem[{\citenamefont{{W. F. Schimmerling {\em et al.}}}(1971)}]{neutnuc1}
\bibinfo{author}{\bibnamefont{{W. F. Schimmerling {\em et al.}}}},
  \bibinfo{journal}{Phys. Letters} \textbf{\bibinfo{volume}{37B}},
  \bibinfo{pages}{177} (\bibinfo{year}{1971}).

\bibitem[{\citenamefont{Coor et~al.}(1955)\citenamefont{Coor, Hill, Hornyak,
  Smith, and Snow}}]{neutnuc2}
\bibinfo{author}{\bibfnamefont{T.}~\bibnamefont{Coor}},
  \bibinfo{author}{\bibfnamefont{D.~A.} \bibnamefont{Hill}},
  \bibinfo{author}{\bibfnamefont{W.~F.} \bibnamefont{Hornyak}},
  \bibinfo{author}{\bibfnamefont{L.~W.} \bibnamefont{Smith}}, \bibnamefont{and}
  \bibinfo{author}{\bibfnamefont{G.}~\bibnamefont{Snow}},
  \bibinfo{journal}{Phys. Rev.} \textbf{\bibinfo{volume}{98}},
  \bibinfo{pages}{1369} (\bibinfo{year}{1955}).

\bibitem[{\citenamefont{{W. L. Lakin {\em et al.}}}(1970)}]{neutnuc3}
\bibinfo{author}{\bibnamefont{{W. L. Lakin {\em et al.}}}},
  \bibinfo{journal}{Phys. Letters} \textbf{\bibinfo{volume}{31B}},
  \bibinfo{pages}{677} (\bibinfo{year}{1970}).

\bibitem[{\citenamefont{{E. F. Parker {\em et al.}}}(1970)}]{neutnuc4}
\bibinfo{author}{\bibnamefont{{E. F. Parker {\em et al.}}}},
  \bibinfo{journal}{Phys. Letters} \textbf{\bibinfo{volume}{31B}},
  \bibinfo{pages}{246} (\bibinfo{year}{1970}).

\bibitem[{\citenamefont{{J. Engler {\em et al.}}}(1968)}]{neutnuc5}
\bibinfo{author}{\bibnamefont{{J. Engler {\em et al.}}}},
  \bibinfo{journal}{Phys. Letters} \textbf{\bibinfo{volume}{27B}},
  \bibinfo{pages}{599} (\bibinfo{year}{1968}).

\bibitem[{\citenamefont{Carroll et~al.}(1976)}]{carroll}
\bibinfo{author}{\bibfnamefont{A.~S.} \bibnamefont{Carroll}}
  \bibnamefont{et~al.}, \bibinfo{journal}{Phys. Rev.}
  \textbf{\bibinfo{volume}{C14}}, \bibinfo{pages}{635} (\bibinfo{year}{1976}).

\bibitem[{\citenamefont{Gachurin et~al.}(1985)}]{gachurin}
\bibinfo{author}{\bibfnamefont{V.~V.} \bibnamefont{Gachurin}}
  \bibnamefont{et~al.} (\bibinfo{year}{1985}), \bibinfo{note}{{ITEP-59-1985}}.

\bibitem[{\citenamefont{Bobchenko et~al.}(1979)}]{bobchenko}
\bibinfo{author}{\bibfnamefont{B.~M.} \bibnamefont{Bobchenko}}
  \bibnamefont{et~al.}, \bibinfo{journal}{Sov. J. Nucl. Phys.}
  \textbf{\bibinfo{volume}{30}}, \bibinfo{pages}{805} (\bibinfo{year}{1979}).

\bibitem[{\citenamefont{Ashery et~al.}(1981)}]{ashery}
\bibinfo{author}{\bibfnamefont{D.}~\bibnamefont{Ashery}} \bibnamefont{et~al.},
  \bibinfo{journal}{Phys. Rev.} \textbf{\bibinfo{volume}{C23}},
  \bibinfo{pages}{2173} (\bibinfo{year}{1981}).

\bibitem[{\citenamefont{Allardyce et~al.}(1973)}]{allardyce}
\bibinfo{author}{\bibfnamefont{B.~W.} \bibnamefont{Allardyce}}
  \bibnamefont{et~al.}, \bibinfo{journal}{Nucl. Phys.}
  \textbf{\bibinfo{volume}{A209}}, \bibinfo{pages}{1} (\bibinfo{year}{1973}).

\bibitem[{\citenamefont{Barnett et~al.}(1996)}]{pdg1996}
\bibinfo{author}{\bibfnamefont{R.~M.} \bibnamefont{Barnett}}
  \bibnamefont{et~al.} (\bibinfo{collaboration}{Particle Data Group}),
  \bibinfo{journal}{Phys. Rev.} \textbf{\bibinfo{volume}{D54}},
  \bibinfo{pages}{1} (\bibinfo{year}{1996}).

\bibitem[{\citenamefont{Catanesi et~al.}(2007)}]{harp}
\bibinfo{author}{\bibfnamefont{M.~G.} \bibnamefont{Catanesi}}
  \bibnamefont{et~al.}, \bibinfo{journal}{Eur. Phys. J.}
  \textbf{\bibinfo{volume}{C52}}, \bibinfo{pages}{29} (\bibinfo{year}{2007}),
  \bibinfo{note}{{\\(Note: The HARP data used in the MiniBooNE analysis result
  from a preliminary analysis of the data prior to publication. The cross
  sections therefore differ slightly from those reported in this
  publication.)}}, \eprint{hep-ex/0702024}.

\bibitem[{\citenamefont{Chemakin et~al.}(2008)}]{e910}
\bibinfo{author}{\bibfnamefont{I.}~\bibnamefont{Chemakin}} \bibnamefont{et~al.}
  (\bibinfo{collaboration}{E910}), \bibinfo{journal}{Phys. Rev.}
  \textbf{\bibinfo{volume}{C77}}, \bibinfo{pages}{015209}
  (\bibinfo{year}{2008}), \eprint{0707.2375}.

\bibitem[{\citenamefont{Cho et~al.}(1971)\citenamefont{Cho, Derrick, Marmer,
  Wangler, Day, Kalbaci, Marshak, Randolph, and Key}}]{chopi}
\bibinfo{author}{\bibfnamefont{Y.}~\bibnamefont{Cho}},
  \bibinfo{author}{\bibfnamefont{M.}~\bibnamefont{Derrick}},
  \bibinfo{author}{\bibfnamefont{G.}~\bibnamefont{Marmer}},
  \bibinfo{author}{\bibfnamefont{T.~P.} \bibnamefont{Wangler}},
  \bibinfo{author}{\bibfnamefont{J.~L.} \bibnamefont{Day}},
  \bibinfo{author}{\bibfnamefont{P.}~\bibnamefont{Kalbaci}},
  \bibinfo{author}{\bibfnamefont{M.~L.} \bibnamefont{Marshak}},
  \bibinfo{author}{\bibfnamefont{J.~K.} \bibnamefont{Randolph}},
  \bibnamefont{and} \bibinfo{author}{\bibfnamefont{A.~W.} \bibnamefont{Key}},
  \bibinfo{journal}{Phys. Rev. D} \textbf{\bibinfo{volume}{4}},
  \bibinfo{pages}{1967} (\bibinfo{year}{1971}).

\bibitem[{\citenamefont{Vorontsov et~al.}(1983)}]{vorontsovpi}
\bibinfo{author}{\bibfnamefont{I.~A.} \bibnamefont{Vorontsov}}
  \bibnamefont{et~al.} (\bibinfo{year}{1983}), \bibinfo{note}{{ITEP-83-085}}.

\bibitem[{\citenamefont{Allaby et~al.}(1969)\citenamefont{Allaby, Diddens,
  Glauber, Klovning, Kofoed-Hansen, Sacharidis, SchlŸpmann, Thorndike, and
  Wetherell}}]{allabypi}
\bibinfo{author}{\bibfnamefont{J.}~\bibnamefont{Allaby}},
  \bibinfo{author}{\bibfnamefont{A.}~\bibnamefont{Diddens}},
  \bibinfo{author}{\bibfnamefont{R.}~\bibnamefont{Glauber}},
  \bibinfo{author}{\bibfnamefont{A.}~\bibnamefont{Klovning}},
  \bibinfo{author}{\bibfnamefont{O.}~\bibnamefont{Kofoed-Hansen}},
  \bibinfo{author}{\bibfnamefont{E.~J.} \bibnamefont{Sacharidis}},
  \bibinfo{author}{\bibfnamefont{K.}~\bibnamefont{SchlŸpmann}},
  \bibinfo{author}{\bibfnamefont{A.~M.} \bibnamefont{Thorndike}},
  \bibnamefont{and} \bibinfo{author}{\bibfnamefont{A.~M.}
  \bibnamefont{Wetherell}}, \bibinfo{journal}{Phys. Lett. B}
  \textbf{\bibinfo{volume}{30}}, \bibinfo{pages}{549} (\bibinfo{year}{1969}).

\bibitem[{\citenamefont{Marmer and Lundquist}(1971)}]{marmerpi}
\bibinfo{author}{\bibfnamefont{G.~J.} \bibnamefont{Marmer}} \bibnamefont{and}
  \bibinfo{author}{\bibfnamefont{D.~E.} \bibnamefont{Lundquist}},
  \bibinfo{journal}{Phys. Rev. D} \textbf{\bibinfo{volume}{3}},
  \bibinfo{pages}{1089} (\bibinfo{year}{1971}).

\bibitem[{\citenamefont{Aleshin et~al.}(1977)\citenamefont{Aleshin, Drabkin,
  and Kolesnikov}}]{aleshin}
\bibinfo{author}{\bibfnamefont{Y.~D.} \bibnamefont{Aleshin}},
  \bibinfo{author}{\bibfnamefont{I.~A.} \bibnamefont{Drabkin}},
  \bibnamefont{and} \bibinfo{author}{\bibfnamefont{V.~V.}
  \bibnamefont{Kolesnikov}} (\bibinfo{year}{1977}),
  \bibinfo{note}{{ITEP-80-1977}}.

\bibitem[{\citenamefont{Ahn et~al.}(2006)}]{k2kprd}
\bibinfo{author}{\bibfnamefont{M.~H.} \bibnamefont{Ahn}} \bibnamefont{et~al.}
  (\bibinfo{collaboration}{K2K}), \bibinfo{journal}{Phys. Rev.}
  \textbf{\bibinfo{volume}{D74}}, \bibinfo{pages}{072003}
  (\bibinfo{year}{2006}), \eprint{hep-ex/0606032}.

\bibitem[{\citenamefont{Sanford and Wang}(1967)}]{swpar}
\bibinfo{author}{\bibfnamefont{J.~R.} \bibnamefont{Sanford}} \bibnamefont{and}
  \bibinfo{author}{\bibfnamefont{C.~L.} \bibnamefont{Wang}}
  (\bibinfo{year}{1967}), \bibinfo{note}{{BNL Note 11299}}.

\bibitem[{\citenamefont{{HARP Collaboration}}()}]{harp_pim}
\bibinfo{author}{\bibnamefont{{HARP Collaboration}}}, \bibinfo{note}{{Paper in
  preparation}}.

\bibitem[{\citenamefont{Abbott et~al.}(1992)\citenamefont{Abbott, Akiba,
  Beavis, Bloomer, Bond, Chasman, Chen, Chu, Cole, Costales et~al.}}]{abbott}
\bibinfo{author}{\bibfnamefont{T.}~\bibnamefont{Abbott}},
  \bibinfo{author}{\bibfnamefont{Y.}~\bibnamefont{Akiba}},
  \bibinfo{author}{\bibfnamefont{D.}~\bibnamefont{Beavis}},
  \bibinfo{author}{\bibfnamefont{M.~A.} \bibnamefont{Bloomer}},
  \bibinfo{author}{\bibfnamefont{P.~D.} \bibnamefont{Bond}},
  \bibinfo{author}{\bibfnamefont{C.}~\bibnamefont{Chasman}},
  \bibinfo{author}{\bibfnamefont{Z.}~\bibnamefont{Chen}},
  \bibinfo{author}{\bibfnamefont{Y.~Y.} \bibnamefont{Chu}},
  \bibinfo{author}{\bibfnamefont{B.~A.} \bibnamefont{Cole}},
  \bibinfo{author}{\bibfnamefont{J.~B.} \bibnamefont{Costales}},
  \bibnamefont{et~al.}, \bibinfo{journal}{Phys. Rev. D}
  \textbf{\bibinfo{volume}{45}}, \bibinfo{pages}{3906} (\bibinfo{year}{1992}).

\bibitem[{\citenamefont{{J. V. Allaby, {\em et al.}}}()}]{allaby}
\bibinfo{author}{\bibnamefont{{J. V. Allaby, {\em et al.}}}},
  \bibinfo{note}{{CERN Report No. CERN 70-12 (unpublished)}}.

\bibitem[{\citenamefont{Dekkers et~al.}(1965)\citenamefont{Dekkers, Geibel,
  Mermod, Weber, Willitts, Winter, Jordan, Vivargent, King, and
  Wilson}}]{dekkers}
\bibinfo{author}{\bibfnamefont{D.}~\bibnamefont{Dekkers}},
  \bibinfo{author}{\bibfnamefont{J.~A.} \bibnamefont{Geibel}},
  \bibinfo{author}{\bibfnamefont{R.}~\bibnamefont{Mermod}},
  \bibinfo{author}{\bibfnamefont{G.}~\bibnamefont{Weber}},
  \bibinfo{author}{\bibfnamefont{T.~R.} \bibnamefont{Willitts}},
  \bibinfo{author}{\bibfnamefont{K.}~\bibnamefont{Winter}},
  \bibinfo{author}{\bibfnamefont{B.}~\bibnamefont{Jordan}},
  \bibinfo{author}{\bibfnamefont{M.}~\bibnamefont{Vivargent}},
  \bibinfo{author}{\bibfnamefont{N.~M.} \bibnamefont{King}}, \bibnamefont{and}
  \bibinfo{author}{\bibfnamefont{E.~J.~N.} \bibnamefont{Wilson}},
  \bibinfo{journal}{Phys. Rev.} \textbf{\bibinfo{volume}{137}},
  \bibinfo{pages}{B962} (\bibinfo{year}{1965}).

\bibitem[{\citenamefont{{T. Eichten {\em et al.}}}(1972)}]{eichten}
\bibinfo{author}{\bibnamefont{{T. Eichten {\em et al.}}}},
  \bibinfo{journal}{Nucl. Phys. B} \textbf{\bibinfo{volume}{44}},
  \bibinfo{pages}{333} (\bibinfo{year}{1972}).

\bibitem[{\citenamefont{Lundy et~al.}(1965)\citenamefont{Lundy, Novey,
  Yovanovitch, and Telegdi}}]{lundy}
\bibinfo{author}{\bibfnamefont{R.~A.} \bibnamefont{Lundy}},
  \bibinfo{author}{\bibfnamefont{T.~B.} \bibnamefont{Novey}},
  \bibinfo{author}{\bibfnamefont{D.~D.} \bibnamefont{Yovanovitch}},
  \bibnamefont{and} \bibinfo{author}{\bibfnamefont{V.~L.}
  \bibnamefont{Telegdi}}, \bibinfo{journal}{Phys. Rev. Lett.}
  \textbf{\bibinfo{volume}{14}}, \bibinfo{pages}{504} (\bibinfo{year}{1965}).

\bibitem[{\citenamefont{Marmer et~al.}(1969)\citenamefont{Marmer, Reibel,
  Schwartz, Stevens, Winston, Wolfe, Rush, Phillips, Swallow, and
  Romanowski}}]{marmer}
\bibinfo{author}{\bibfnamefont{G.~J.} \bibnamefont{Marmer}},
  \bibinfo{author}{\bibfnamefont{K.}~\bibnamefont{Reibel}},
  \bibinfo{author}{\bibfnamefont{D.~M.} \bibnamefont{Schwartz}},
  \bibinfo{author}{\bibfnamefont{A.}~\bibnamefont{Stevens}},
  \bibinfo{author}{\bibfnamefont{R.}~\bibnamefont{Winston}},
  \bibinfo{author}{\bibfnamefont{D.}~\bibnamefont{Wolfe}},
  \bibinfo{author}{\bibfnamefont{C.~J.} \bibnamefont{Rush}},
  \bibinfo{author}{\bibfnamefont{P.~R.} \bibnamefont{Phillips}},
  \bibinfo{author}{\bibfnamefont{E.~C.} \bibnamefont{Swallow}},
  \bibnamefont{and} \bibinfo{author}{\bibfnamefont{T.~A.}
  \bibnamefont{Romanowski}}, \bibinfo{journal}{Phys. Rev.}
  \textbf{\bibinfo{volume}{179}}, \bibinfo{pages}{1294} (\bibinfo{year}{1969}).

\bibitem[{\citenamefont{Vorontsov et~al.}(1988)\citenamefont{Vorontsov,
  Safronov, Sibirtsev, Smirnov, and Trebukhovsky}}]{vorontsov}
\bibinfo{author}{\bibfnamefont{I.~A.} \bibnamefont{Vorontsov}},
  \bibinfo{author}{\bibfnamefont{G.~A.} \bibnamefont{Safronov}},
  \bibinfo{author}{\bibfnamefont{A.~A.} \bibnamefont{Sibirtsev}},
  \bibinfo{author}{\bibfnamefont{G.~N.} \bibnamefont{Smirnov}},
  \bibnamefont{and} \bibinfo{author}{\bibfnamefont{Y.~V.}
  \bibnamefont{Trebukhovsky}} (\bibinfo{year}{1988}),
  \bibinfo{note}{{ITEP-88-011}}.

\bibitem[{\citenamefont{Conrad et~al.}()\citenamefont{Conrad, Monroe, and
  Shaevitz}}]{fs}
\bibinfo{author}{\bibfnamefont{J.~M.} \bibnamefont{Conrad}},
  \bibinfo{author}{\bibfnamefont{J.~R.} \bibnamefont{Monroe}},
  \bibnamefont{and} \bibinfo{author}{\bibfnamefont{M.~H.}
  \bibnamefont{Shaevitz}}, \bibinfo{note}{in preparation}.

\bibitem[{\citenamefont{Abe et~al.}(1987)}]{abe}
\bibinfo{author}{\bibfnamefont{F.}~\bibnamefont{Abe}} \bibnamefont{et~al.},
  \bibinfo{journal}{Phys. Rev.} \textbf{\bibinfo{volume}{D36}},
  \bibinfo{pages}{1302} (\bibinfo{year}{1987}).

\bibitem[{\citenamefont{Mokhov et~al.}(1998)}]{mars}
\bibinfo{author}{\bibfnamefont{N.~V.} \bibnamefont{Mokhov}}
  \bibnamefont{et~al.} (\bibinfo{year}{1998}), \eprint{nucl-th/9812038}.

\bibitem[{\citenamefont{Ferrari et~al.}()\citenamefont{Ferrari, Sala, Fasso,
  and Ranft}}]{fluka}
\bibinfo{author}{\bibfnamefont{A.}~\bibnamefont{Ferrari}},
  \bibinfo{author}{\bibfnamefont{P.~R.} \bibnamefont{Sala}},
  \bibinfo{author}{\bibfnamefont{A.}~\bibnamefont{Fasso}}, \bibnamefont{and}
  \bibinfo{author}{\bibfnamefont{J.}~\bibnamefont{Ranft}},
  \emph{\bibinfo{title}{{FLUKA: A multi-particle transport code (Program
  version 2005)}}}, \bibinfo{note}{cERN-2005-010}.

\bibitem[{\citenamefont{{G. Bellettini {\em et al.}}}(1966)}]{bellettini}
\bibinfo{author}{\bibnamefont{{G. Bellettini {\em et al.}}}},
  \bibinfo{journal}{Nuc. Phys} \textbf{\bibinfo{volume}{79}},
  \bibinfo{pages}{609} (\bibinfo{year}{1966}).

\bibitem[{\citenamefont{{Aguilar-Arevalo, A. A. and others}}(2009)}]{boonenim}
\bibinfo{author}{\bibnamefont{{Aguilar-Arevalo, A. A. and others}}}
  (\bibinfo{collaboration}{MiniBooNE}), \bibinfo{journal}{Nucl.Instrum.Meth.A}
  \textbf{\bibinfo{volume}{599}}, \bibinfo{pages}{28} (\bibinfo{year}{2009}),
  \eprint{arXiv:0806.4201}.

\bibitem[{\citenamefont{Casper}(2002)}]{nuance}
\bibinfo{author}{\bibfnamefont{D.}~\bibnamefont{Casper}},
  \bibinfo{journal}{Nucl. Phys. Proc. Suppl.} \textbf{\bibinfo{volume}{112}},
  \bibinfo{pages}{161} (\bibinfo{year}{2002}), \eprint{hep-ph/0208030}.

\end{thebibliography}

\end{document}